\documentclass[reqno,11pt,a4paper,final,fleqn]{amsart}
\usepackage[a4paper,left=30mm,right=30mm,top=30mm,bottom=30mm,marginpar=20mm]{geometry} 
\usepackage{amsmath}
\usepackage{amssymb}
\usepackage{amsthm}
\usepackage{amscd}
\usepackage{textcomp}
\usepackage{stmaryrd}
\usepackage{cite}
\usepackage{bbold}
\usepackage{color}
\usepackage[english=american]{csquotes}
\usepackage{graphicx}
\usepackage{enumerate}
\usepackage{times}
\usepackage{accents}
\usepackage{bm}
\usepackage[normalem]{ulem}
\usepackage{subcaption}
\usepackage[bookmarks=true,bookmarksnumbered=true,hypertexnames=true]{hyperref}

\graphicspath{{pics/}}

\numberwithin{equation}{section}

\newtheoremstyle{thmlemcorr}{10pt}{10pt}{\itshape}{}{\bfseries}{.}{10pt}{{\thmname{#1}\thmnumber{ #2}\thmnote{ (#3)}}}
\newtheoremstyle{thmlemcorr*}{10pt}{10pt}{\itshape}{}{\bfseries}{.}{10pt}{{\thmname{#1}\thmnote{ (#3)}}}
\newtheoremstyle{defi}{10pt}{10pt}{\itshape}{}{\bfseries}{.}{10pt}{{\thmname{#1}\thmnumber{ #2}\thmnote{ (#3)}}}
\newtheoremstyle{remexample}{10pt}{10pt}{}{}{\bfseries}{.}{10pt}{{\thmname{#1}\thmnumber{ #2}\thmnote{ (#3)}}}
\newtheoremstyle{ass}{10pt}{10pt}{}{}{\bfseries}{.}{10pt}{{\thmname{#1}\thmnumber{ A#2}\thmnote{ (#3)}}}
\newtheoremstyle{hyp}{10pt}{10pt}{\itshape}{}{\bfseries}{.}{10pt}{{\thmname{#1}\thmnote{ (#3)}}}

\theoremstyle{thmlemcorr}
\newtheorem{theorem}{Theorem}
\numberwithin{theorem}{section}

\theoremstyle{thmlemcorr*}
\newtheorem*{theorem*}{Theorem}
\newtheorem*{lemma*}{Lemma}
\newtheorem*{corollary*}{Corollary}
\newtheorem*{proposition*}{Proposition}
\newtheorem*{problem*}{Problem}
\newtheorem*{conjecture*}{Conjecture}

\theoremstyle{defi}

\theoremstyle{remexample}
\newtheorem{remark}[theorem]{Remark}
\newtheorem{example}[theorem]{Example}

\theoremstyle{hyp}
\newtheorem{hypothesis}{Hypothesis}

\newcommand{\Erm}{\mathrm{E}}

\newcommand{\Lrm}{\mathrm{L}}

\newcommand{\Orm}{\mathrm{O}}

\newcommand{\Trm}{\mathrm{T}}

\newcommand{\Wrm}{\mathrm{W}}

\newcommand{\Bcal}{\mathcal{B}}

\newcommand{\Dcal}{\mathcal{D}}
\newcommand{\Ecal}{\mathcal{E}}

\newcommand{\Hcal}{\mathcal{H}}

\newcommand{\Lcal}{\mathcal{L}}

\newcommand{\Qcal}{\mathcal{Q}}
\newcommand{\Rcal}{\mathcal{R}}
\newcommand{\Scal}{\mathcal{S}}

\newcommand{\Wcal}{\mathcal{W}}

\newcommand{\Pfrak}{\mathfrak{P}}

\newcommand{\Sfrak}{\mathfrak{S}}

\newcommand{\pfrak}{\mathfrak{p}}

\newcommand{\Mbf}{\mathbf{M}}

\newcommand{\pbf}{\mathbf{p}}

\newcommand{\tbf}{\mathbf{t}}

\newcommand{\Ebb}{\mathbb{E}}
\newcommand{\Fbb}{\mathbb{F}}

\newcommand{\Tb}{\bm{T}}
\newcommand{\Sb}{\bm{S}}
\newcommand{\vb}{\bm{v}}
\newcommand{\taub}{\bm{\tau}}

\DeclareMathOperator{\graph}{graph}

\DeclareMathOperator{\dive}{div}
\DeclareMathOperator{\curl}{curl}

\DeclareMathOperator{\tr}{tr}
\DeclareMathOperator{\spn}{span}

\DeclareMathOperator{\cof}{cof}

\DeclareMathOperator{\Wedge}{{\textstyle\bigwedge}}

\newcommand{\ee}{\mathrm{e}}

\newcommand{\id}{\mathrm{id}}
\newcommand{\Id}{\mathrm{Id}}

\newcommand{\setb}[2]{\bigl\{\, #1 \ \ \textup{\textbf{:}}\ \ #2 \,\bigr\}}

\newcommand{\setBB}[2]{\biggl\{\, #1 \ \ \textup{\textbf{:}}\ \ #2 \,\biggr\}}

\newcommand{\norm}[1]{\mathopen{\|} #1 \mathclose{\|}}

\newcommand{\abs}[1]{\mathopen{|} #1 \mathclose{|}}

\newcommand{\absb}[1]{\bigl| #1 \bigr|}

\newcommand{\absBB}[1]{\biggl| #1 \biggr|}

\newcommand{\tv}[1]{\norm{ #1 }}

\newcommand{\altnorm}[1]{{\left\vert\kern-0.25ex\left\vert\kern-0.25ex\left\vert #1 \right\vert\kern-0.25ex\right\vert\kern-0.25ex\right\vert}}

\newcommand{\dpr}[1]{\mathopen{\langle} #1 \mathclose{\rangle}}	

\newcommand{\dprb}[1]{\bigl\langle #1 \bigr\rangle}

\newcommand{\dprBB}[1]{\biggl\langle #1 \biggr\rangle}

\newcommand{\dbr}[1]{\mathopen{\llbracket} #1 \mathclose{\rrbracket}}

\newcommand{\di}{\mathrm{d}}
\newcommand{\dd}{\;\mathrm{d}}
\newcommand{\DD}{\mathrm{D}}
\newcommand{\N}{\mathbb{N}}
\newcommand{\R}{\mathbb{R}}

\newcommand{\Z}{\mathbb{Z}}

\newcommand{\sym}{\mathrm{sym}}

\newcommand{\res}{\mathrm{res}}

\newcommand{\free}{\mathrm{fr}}
\newcommand{\linet}{\mathrm{lt}}

\newcommand{\toup}{\uparrow}
\newcommand{\todown}{\downarrow}

\newcommand{\hodge}{{\star}}

\newcommand{\sbullet}{\begin{picture}(1,1)(-0.5,-2.5)\circle*{2}\end{picture}}
\newcommand{\frarg}{\,\sbullet\,}
\newcommand{\BV}{\mathrm{BV}}

\newcommand{\eps}{\epsilon}

\newcommand{\GL}{\mathrm{GL}}
\newcommand{\SO}{\mathrm{SO}}
\newcommand{\SL}{\mathrm{SL}}
\newcommand{\Diss}{\mathrm{Diss}}

\newcommand{\ff}{\gg}

\newcommand{\T}{\mathsf{T}}

\DeclareMathOperator{\Lie}{Lie}
\DeclareMathOperator{\Var}{Var}

\DeclareMathOperator{\Exp}{Exp}

\DeclareMathOperator{\fr}{fr}

\makeatletter
\def\@tempa#1{\@xp\@tempb\meaning#1\@nil#1}
\def\@tempb#1>#2#3 #4\@nil#5{%
  \@xp\ifx\csname#3\endcsname\mathaccent
    \@tempc#4?"7777\@nil#5%
  \else
    \PackageWarningNoLine{amsmath}{%
      Unable to redefine math accent \string#5}%
  \fi
}
\def\@tempc#1"#2#3#4#5#6\@nil#7{%
  \chardef\@tempd="#3\relax\set@mathaccent\@tempd{#7}{#2}{#4#5}}

\DeclareFontFamily{U}{mathx}{\hyphenchar\font45}
\DeclareFontShape{U}{mathx}{m}{n}{
      <5> <6> <7> <8> <9> <10>
      <10.95> <12> <14.4> <17.28> <20.74> <24.88>
      mathx10
      }{}
\DeclareSymbolFont{mathx}{U}{mathx}{m}{n}
\DeclareFontSubstitution{U}{mathx}{m}{n}
\DeclareMathAccent{\widecheck}{0}{mathx}{"71}
\DeclareMathAccent{\wideparen}{0}{mathx}{"75}

\renewcommand{\hat}{\widehat}
\renewcommand{\check}{\widecheck}
\renewcommand{\tilde}{\widetilde}
\renewcommand{\phi}{\varphi}
\renewcommand{\rho}{\varrho}

\renewcommand*{\dot}[1]{\accentset{\mbox{\normalfont\large\bfseries .}}{#1}}

\def\XXint#1#2#3{{\setbox0= \hbox{$#1{#2#3}{\int}$} 
\vcenter{\hbox{$#2#3$}}\kern-.5\wd0}}

\usepackage{pict2e}
\makeatletter
\DeclareRobustCommand{\intprod}{%
  \mathbin{\mathpalette\int@prod{(0.1,0)(0.9,0)(0.9,0.8)}}}
\DeclareRobustCommand{\restrict}{%
  \mathbin{\mathpalette\int@prod{(0.1,0.8)(0.1,0)(0.9,0)}}}	
\newcommand{\int@prod}[2]{%
  \begingroup
  \sbox\z@{$\m@th#1+$}%
  \setlength\unitlength{\wd\z@}%
  \begin{picture}(1,1)
  \roundcap
  \polyline#2
  \end{picture}%
  \endgroup
}
\makeatother

\renewcommand{\eps}{\varepsilon}
\renewcommand{\phi}{\varphi}
\renewcommand{\tilde}{\widetilde}
\renewcommand{\hat}{\widehat}
\renewcommand{\bar}{\overline}

\begin{document}


\title[Small-distortion elasto-plasticity]{A small-distortion geometric model for\\elasto-plasticity in single crystals driven by the\\motion of dislocations}


\author{Filip Rindler}
\address{Mathematics Institute, University of Warwick, Coventry CV4 7AL, United Kingdom.}
\email{F.Rindler@warwick.ac.uk}


\hypersetup{
  pdfauthor = {Filip Rindler},
  pdftitle = {A small-distortion geometric model for elasto-plasticity in single crystals driven by the motion of dislocations}
}



\begin{abstract}
A central problem in nonlinear elasto-plasticity theory is to formulate a geometrically nonlinear, large-strain model of elasto-plasticity in single crystals in which plastic flow is driven directly by the motion of dislocations and which allows for a homogenization procedure from discrete dislocation lines to dislocation densities. In this work, such a model is introduced based on two hypotheses: (1)~The Small-Distortion Hypothesis posits that the total plastic distortion may be arbitrarily large but its dominant part admits a representation as a gradient, meaning that the distorting effect of dislocations is small relative to the specimen size. Concretely, this hypothesis is realized here through a ``multiplicative Helmholtz-type decomposition'', which splits an arbitrary matrix field into the product of a gradient and a matrix field with controlled curl. (2)~The Line-Tension Approximation treats all dislocations as infinitesimally thin lines whose local stress fields are individually negligible compared to specimen-scale stresses and only matter in the aggregate. The resulting model, termed the Small-Distortion Geometric Model, furthermore employs the so-called Space-Time Framework, which furnishes a geometric language to precisely describe the advection of dislocation lines. As such, it provides a physically grounded account of crystal plasticity in an idealized mesoscopic setting that bridges dislocation mechanics and continuum elasto-plasticity.
\vspace{4pt}




\noindent\textsc{Date:} \today{}.
\end{abstract}

\maketitle

\setcounter{tocdepth}{1} 
\tableofcontents


\section{Introduction}

The basic motivation for the present work is the following \emph{dislocation homogenization problem}: A theory of plasticity driven by dislocation \emph{fields} should be obtainable from a model describing individual dislocation lines in the limit of more and more lines with smaller and smaller weight (or, equivalently, a rescaling of the specimen length scales). This requirement is to be interpreted as a form of consistency requirement for the model to be developed since it entails that the \enquote{correct} amount of information is retained in the internal variables: Too much information (i.e., too extensive internal variables) introduces redundancy, while too little information kept in the (limiting) internal variables precludes the discrete-to-field limit passage for dislocations. Indeed, if the internal variables identify too many different dislocation configurations into the same representation (e.g., by only keeping track of \emph{average} dislocation densities), then it becomes impossible to keep track of differing effects (e.g., stresses) since these configurations are now indistinguishable in the field model.

In fact, the author posits that such a discrete-to-field limit passage should be carried out as a mathematically fully rigorous homogenization procedure. In this way, no question as to its validity remains (under the imposed technical assumptions, at least) and the mathematical analysis reveals much additional information about the structure of the equations. In fact, the limit passage from discrete to continuum dislocation densities involves so-called \enquote{weak} notions of convergence (in the mathematical sense), which involve a form of averaging. In general, weak limits do not commute with nonlinear operations and it is paramount that the model for which the homogenization procedure is to be performed is formulated in a way that is amenable to such weak limit passages.

Based on this motivation, the present work introduces what we call the Small-Distortion Geometric Model. While the present paper develops the full modeling framework, the rigorous homogenization procedure, at least in the small-strain regime, will be carried out in the mathematical companion work~\cite{BonicattoRindlerTurnbull26c?} (based on the preparatory technical articles~\cite{BonicattoRindlerTurnbull26,BonicattoRindlerTurnbull26b?}). The model is specifically designed to permit the discrete-to-field weak-limit procedure described above. At the end of Section~\ref{sc:PDE}, we compare the two models, particularly their treatment of independently moving Burgers-vector components and of averaging the associated plastic-flow contributions.

On top of the groundwork laid by~\cite{HudsonRindler22}, which contains a lower-level approach (and to which the reader is referred for an extensive review of existing approaches), the Small-Distortion Geometric Model imposes two main additional assumptions:

\begin{enumerate}[(1)]
  \item The \emph{Small-Distortion Hypothesis} assumes that while the total plastic distortion can be arbitrarily large, whereby we are in the (plastically) large-strain regime, most of the plastic distortion takes the form of a \emph{gradient}. This fact is expressed via what we call a \emph{Multiplicative Helmholtz-type Decomposition}, whereby an arbitrary matrix field is decomposed as a product of a gradient and a matrix field with controlled (small) curl.\smallskip

  \item The \emph{Line-Tension Approximation} posits that at the length scale of the model all dislocations may be considered as \emph{infinitesimally thin} lines. In particular, their local stress fields are assumed to have a length scale that is negligible with respect to the size of the specimens to which the model is applied.
\end{enumerate}

Together, these assumptions may be understood as comprising an extended form of the premise of \emph{continuum mechanics} that atomistic or near-atomistic effects are too small to be resolved individually and only matter in the aggregate. Evidently, these are modeling choices, which are clearly not satisfied at very small specimen sizes. On the other hand, they seem to be reasonable for medium-sized specimens, for instance the single-crystal grains within polycrystals. So, we work at a \emph{mesoscale}.

Let us stress that while at some point in the modeling we will linearize the elastic energy (as a consequence of the Small-Distortion Hypothesis), it is \emph{never} assumed that the plastic distortion itself is small. As a consequence, here plastic flow is supposed to commence at relatively small deformation, as is the case in most metals~\cite{Lubliner08}.

As in the preceding work~\cite{HudsonRindler22}, dislocations and their motion will be represented in what we call the Space-Time Framework, a geometric language to describe the transport of (low-regularity) geometric objects based on the theory of \emph{space-time currents}. In it, the evolution of a curve (representing a dislocation line) is represented by a \emph{space-time trajectory}, that is, the $2$-dimensional surface in $4$-dimensional space-time traced out by the moving curve. Despite its abstractness, this yields a versatile and flexible approach in which all relevant quantities can be directly and elegantly defined. The foundations of the Space-Time Framework were laid in~\cite{HudsonRindler22} and the mathematical theory was further developed in~\cite{Rindler23,Rindler25,BonicattoDelNinRindler24,BonicattoDelNinRindler25}. In particular, in~\cite{Rindler25} the author was able to establish the first existence theorem for fully nonlinear elasto-plasticity driven by the flow of dislocation lines using the Space-Time Framework.

While the present work contains no rigorous mathematics, all the objects described here in an intuitive fashion can in fact be given a precise meaning using the mathematical theory of \emph{integral and normal currents} in space-time~\cite{Rindler23,BonicattoDelNinRindler24,BonicattoDelNinRindler25}. Currents, at least in their \enquote{integral} flavor, are low-regularity generalizations of classical geometric structures such as curves, surfaces, and general manifolds~\cite{Federer69book,KrantzParks08book}. For instance, an \enquote{integral $2$-current} can be thought of as a $2$-dimensional oriented surface, or a \enquote{sum} of such surfaces, which may however have corners, kinks, holes, or fractally-distributed defects. Their theory turns out to be a perfect language to describe dislocations since they indeed may not have good regularity properties and topological changes, such as cancellations or intersections, are well known to occur during some of their motions. A prominent example of this is the Frank--Read Source~\cite{FrankRead50,HudsonRindlerRydell25}, a \enquote{micro-machine} facilitating the nucleation of dislocation lines via a complex oscillatory line motion that involves topological change (i.e., cutting as the result of intersection of oppositely-oriented dislocations). As another desirable feature, so-called \enquote{normal} currents can also directly express \emph{fields} (densities) of geometric objects such as curves or surfaces, which is required for the aforementioned homogenization procedure.

In fact, in a static (non-evolutionary) context, currents have been used for a long time to describe dislocations, see, for instance,~\cite{ContiGarroniOrtiz15,ContiGarroniMassaccesi15,KupfermanOlami20,ContiGarroniMarziani23}. On the other hand, parametric descriptions of dislocation motion are used in the theory of Discrete Dislocation Dynamics, see, e.g.,~\cite{BulatovCai06book} for an overview and also~\cite{FonsecaGinsterWojtowytsch21} for a recent mathematical work on this parametric approach. However, parametric approaches to describe evolutions of lines have several limitations, most notably that the parametrization may \enquote{artificially} deteriorate as the evolution progresses. Currents as intrinsic low-regularity geometric objects have no such problems.

Another important feature of the model of~\cite{HudsonRindler22} (which is also crucial for the mathematical treatment in~\cite{Rindler25}) is that while dislocation flow is described in a geometric way for the individual lines, their plastic effects are smeared out over a \enquote{core} region with non-negligible width. On the contrary, the present work derives a mesoscale model, where also the plastic effects of moving dislocations are infinitesimal.

The presentation of the Small-Distortion Geometric Model aims to be completely explicit in its imposed hypotheses, which are typographically offset from the remainder of the text. In this way it should be clear what is an assumption and what is a derivation. 

Let us finally mention that Acharya's Field Dislocation Mechanics (FDM), see~\cite{Acharya01,Acharya03,Acharya04,AroraAcharya20} and the references cited therein, constitutes a related model, which has several structural features in common with the Small-Distortion Geometric Model. We compare this model to the present one at the end of Section~\ref{sc:PDE}, focusing in particular on their different treatments of independently moving Burgers-vector components and of averaging the associated plastic-flow contributions.

This paper is organized as follows: We begin with a description of nonlinear elasto-plasticity in Section~\ref{sc:plasticity}. This is mostly standard material expressed in the notation and terminology of~\cite{HudsonRindler22}. The following Section~\ref{sc:spacetime_framework} contains an introduction to the Space-Time Framework. While many of the ideas presented here have been explored in the works cited above, the presentation is made more accessible and complete. This theory is then applied to dislocation motion and plastic flow in Section~\ref{sc:dislocations}. The core Section~\ref{sc:SDGM} introduces the Small-Distortion Hypothesis based on the Multiplicative Helmholtz-type Decomposition as well as the Line-Tension Approximation. After a summary of all relations derived up to this point in Section~\ref{sc:summary_relations}, in Section~\ref{sc:energetic} we formulate the flow rule, that is, the central dissipative force balance, in the so-called energetic formulation. In Section~\ref{sc:SDGM_linearization} we consider the linearized small-strain version of the model, which is then written as a system of PDEs via the Geometric (Lie) Transport Equation in Section~\ref{sc:PDE}.

\subsection*{Acknowledgements}

This project has received funding from the UKRI Frontier Research Guarantee (ERC guarantee) grant EP/Z000297/1 (ERC CONCENTRATE). The author would like to thank Amit Acharya, Paolo Bonicatto, Thomas Hudson, Alexander Mielke, Harry Turnbull, and Arghir Zarnescu for discussions related to this work.

\subsection*{AI disclosure}

AI was only used for proofreading, checking, and polishing the language, but not for the scientific content of this work.


\section{Geometrically non-linear elasto-plasticity} \label{sc:plasticity}

In this section we recall basic facts about geometrically nonlinear elasto-plasticity theory and fix notation.

\subsection{Basic kinematics}

Suppose that our single-crystal specimen occupies an open, bounded, connected \textbf{reference configuration} $\Omega \subseteq \R^3$. Under a time-dependent \textbf{(total) deformation} $y = (y^1,y^2,y^3) \colon [0,T] \times \Omega \to \R^3$, every \textbf{referential point} $x \in \Omega$ is mapped at time $t \in [0,T]$ to the \textbf{spatial (Eulerian) point} $y(t,x)$. We assume that the map $y$ and all other maps are as smooth as required and we will frequently suppress the arguments $t,x$ in our notation.

\begin{remark}
For the avoidance of confusion, let us explicitly state that no special assumptions are made on the reference configuration (e.g., stress-freeness); $\Omega$ is simply the shape of the specimen at a reference time $t = 0$.  
\end{remark}

We start from the following \textbf{Kr\"{o}ner--Lee decomposition}, as in most of elasto-plasticity theory in the large-strain regime:

\begin{hypothesis}
The deformation gradient $\nabla y$ decomposes multiplicatively into \textbf{elastic} and \textbf{plastic distortions} $E = E(t,x) \in \R^{3 \times 3}$, $P = P(t,x) \in \R^{3 \times 3}$ as
\[
  \nabla y = E P.
\]
\end{hypothesis}

This assumption has many derivations. For instance, in~\cite{HudsonRindler22} it is derived by considering slips acting on the so-called \emph{crystal scaffold}, which represents the inter-atomic bonds in the crystal lattice. Many other derivations also exist, see for instance~\cite{Kroner60,LeeLiu67FSEP,Lee69EPDF,GreenNaghdi71,CaseyNaghdi80,GurtinFriedAnand10book,ReinaConti14,KupfermanMaor15, ReinaContiSchlomerkemper16,ReinaDjodomOrtizConti18,EpsteinKupfermanMaor20}.

In general, $P$ is not the gradient of any map $f \colon \Omega \to \R^3$. At least on simply connected domains (those without \enquote{topological holes}) this is equivalent to the differential condition
\[
  \curl P \neq 0.
\]
The curl here is defined row-wise, i.e., for a matrix field $F = [F^i_j]^i_j \colon \Omega \to \R^{3 \times 3}$ (with the upper index $i$ enumerating the rows and the lower index $j$ enumerating the columns),
\begin{align*}
  \curl F
  &:= \bigl[ \nabla \times F^i \bigr]^i \\
  &\phantom{:}= \biggl[ \frac{\partial F^i_{j+2}}{\partial x_{j+1}} - \frac{\partial F^i_{j+1}}{\partial x_{j+2}}\biggr]^i_j,
\end{align*}
where in the second line we have used a cyclic $j$-index, that is,
\[
  j \colon 1 \overset{+1}{\longrightarrow} 2 \overset{+1}{\longrightarrow} 3 \overset{+1}{\longrightarrow} 1.
\]
Indeed, a non-zero curl is the necessary consequence of modeling plasticity driven by dislocations, as we will see below. So, if $\curl P \neq 0$, then there is no \enquote{plastic deformation}, which would have $P$ as its gradient, and hence we only speak of $E$ and $P$ as the elastic and plastic \emph{distortions}, respectively. 

We also make the following \textbf{plastic incompressibility assumption}:

\begin{hypothesis}
The plastic distortion is volume-preserving, that is, $\det P \equiv 1$.
\end{hypothesis}

This expresses a local conservation of mass under plastic flow. In the context of dislocations, this corresponds to the \enquote{glide motions} being dominant, where the Burgers vector (i.e., the direction in which atoms move when a dislocation passes through) lies in the slip plane, and dislocation \enquote{climb} is forbidden (or, at least, negligible).

Note that we will treat $P$ as one of the \textbf{internal variables} of the system, that is, it describes some aspect of the specimen's state that is not directly observable. Our model involves an explicit evolution law for the plastic distortion $P$, which will be discussed in Section~\ref{sc:plastic_flow}. Therefore, the \enquote{uniqueness problem} for $P$, that is, the question which invariances should be required of $E$ and $P$ in order to make the Kr\"{o}ner--Lee decomposition $\nabla y = EP$ unique, which is much discussed in the literature~\cite{GreenNaghdi71,Rice71,Mandel73,NematNasser79,Dafalias87,LubardaLee81,Naghdi90,Zbib93,CaseyNaghdi80,Mielke03a}, is avoided.

\subsection{Elastic energy} \label{sc:energy}

We next make the following \textbf{hyperelasticity} assumption:

\begin{hypothesis}
  The elastic properties of the material are described by the \textbf{elastic energy}
\[
  \Wcal_e := \int_\Omega W_e(\nabla y P^{-1}) \dd x,
\]
where $W_e \colon \GL^+(3) \to \R$ is the \textbf{elastic energy density}, which satisfies the \textbf{objectivity (frame-indifference)} condition
\begin{equation} \label{eq:frame_indiff}
  W_e(RE) = W_e(E)  \qquad\text{for all $R \in \SO(3)$, $E \in \GL^+(3)$,}
\end{equation}
as well as the property that $W_e$ takes its minimum value $0$ on $\SO(3)$ and nowhere else.
\end{hypothesis}

Here we recall that
\begin{align*}
  \GL(3) &:= \setb{A \in \R^{3 \times 3}}{ \det A \neq 0 }, \\
  \GL^+(3) &:= \setb{A \in \GL(3)}{ \det A > 0 }, \\
  \SL(3) &:= \setb{A \in \GL(3)}{ \det A = 1 }, \\
  \Orm(3) &:= \setb{A \in \GL(3)}{ A^{-1} = A^\T }, \\
  \SO(3) &:= \setb{A \in \Orm(3)}{ \det A = 1 }
\end{align*}
denote the \textbf{general linear group} (containing all invertible matrices) as well as the groups of \textbf{orientation-preserving matrices}, \textbf{special linear matrices}, \textbf{orthogonal matrices}, and \textbf{special orthogonal matrices}, respectively.

Often, $W_e$ also reflects additional material symmetries, such as $W_e(ES) = W_e(E)$ for all $S$ from a \textbf{(point) symmetry group} $\Sfrak$ of the crystal and all $E \in \GL^+(3)$.

The above shape of the energy functional is in line with nonlinear elasticity theory as presented for instance in~\cite{Ciarlet88book}. In contrast to pure nonlinear elasticity, however, the dependence of the energy on $\nabla y P^{-1}$, and not just on $\nabla y$, gives rise to significantly different behavior. In particular, when $\curl P \neq 0$, no complete elastic unloading is possible.

\begin{remark} \label{rem:We_compressible}
If we do not make the plastic incompressibility assumption, then we should instead define
\[
  \Wcal_e := \int_\Omega W_e(\nabla y P^{-1}) \, \abs{\det P} \dd x.
\]
The additional term $\abs{\det P}$ here accounts for the plastic volume change: If $\abs{\det P} > 1$, there has been plastic volume change, which corresponds to additional material, and this has to be accounted for in the elastic energy; likewise, for $\abs{\det P} < 1$ material has been removed (transported away by the plastic flow). In crystal plasticity, a change in the quantity of material around a point only occurs through dislocation climb, so this additional determinant term is neglected here, in accordance with the plastic incompressibility assumption.
\end{remark}

\begin{remark} \label{rem:We_alternatives}
There are various ways of writing $W_e(E)$ as depending on quantities other than $E$. For instance, one may express $W_e(E) = \hat{W}_e(E^\T E)$. Indeed, by the polar decomposition of matrices, we can write any matrix $E$ with $\det E > 0$ as 
\[
  E = R\sqrt{E^\T E}
\]
for some $R\in \SO(3)$. Then, from~\eqref{eq:frame_indiff} we have 
\[
  W_e(E)
  = W_e(R\sqrt{E^\T E})
  = W_e(\sqrt{E^\T E})
  =: \hat{W}_e(E^\T E).
\]
In particular, if we had that $E = \nabla y$ (so $P \equiv \Id$), then
\[
  E^\T E = (\nabla y)^\T \nabla y.
\]
In elasticity theory, the expression $(\nabla y)^\T \nabla y$ is called the \textbf{(right) Cauchy--Green strain tensor}. Moreover, writing $y = \id + u$ with the \textbf{displacement} $u \colon \Omega \to \R^3$, we may expand this as
\[
  E^\T E
  = (\Id + \nabla u)^\T (\Id + \nabla u)
  = \Id + \nabla u + (\nabla u)^\T + (\nabla u)^\T \nabla u.
\]
One calls
\begin{equation} \label{eq:GSV_strain}
  G := \frac12 \bigl(\nabla u + (\nabla u)^\T + (\nabla u)^\T \nabla u \bigr)
\end{equation}
the \textbf{Green--St Venant strain tensor}. Evidently,
\[
  E^\T E = \Id + 2G,
\]
so we may alternatively write $W_e$ as a function of $G$. This is quite useful when expressing the dependence of $W_e$ on various natural material constants; see~\cite{Ciarlet88book} for more information on this.
\end{remark}

\subsection{Energy linearization} \label{sc:We_linearize}

Later, we will need to linearize the elastic energy $W_e$ around the identity matrix $\Id$, corresponding to small \emph{elastic} strain (note that this does not preclude large \emph{plastic} strain). So, assume that $E = \Id + \nabla u$ for a small displacement $u \colon \Omega \to \R^3$ (with small gradient as well). We then appeal to Remark~\ref{rem:We_alternatives} to write
\[
  W_e(E) = \hat{W}_e(\Id + 2G),  \qquad G := \frac12 \bigl(\nabla u + (\nabla u)^\T + (\nabla u)^\T \nabla u \bigr),
\]
with $G$ the Green--St Venant strain tensor that was defined in~\eqref{eq:GSV_strain}. Taylor-expanding $\hat{W}_e$ (with respect to $E^\T E$) around $\Id$,
\[
  W_e(E) \approx \hat{W}_e(\Id) + \DD\hat{W}_e(\Id)[2G] + \frac{1}{2} \Ebb G : G.
\]
Here, the fourth-order \textbf{(linearized) elastic tensor} $\Ebb = \Ebb^{ijkl}$ is defined as
\[
  \Ebb := 4 \, \DD^2 \hat W_e(\Id),
\]
which acts on a matrix $A = (A^i_j) \in \R^{3 \times 3}$ via
\[
  \Ebb A := \bigl[ \Ebb^{ijkl} A^k_l \bigr]^i_j \in \R^{3 \times 3},
\]
with implied summation over all repeated indices. From the assumptions on $W_e$ above, $\hat{W}_e(\Id) = 0$. Furthermore, one supposes $\DD \hat{W}_e(\Id) = 0$ since the material is considered not to be pre-stressed (see~\cite{Ciarlet88book} or~\cite{GurtinFriedAnand10book} for details). As the multiple of a second derivative, the tensor $\Ebb$ is symmetric in the sense that
\[
  \Ebb^{ijkl} = \Ebb^{klij}.
\]

Now, for $u$ a small deformation, the Green--St Venant strain tensor can be approximated as
\[
  G \approx \epsilon(u) := \frac12 \bigl(\nabla u + (\nabla u)^\T \bigr),
\]
that is, the \textbf{symmetrized gradient} $\epsilon(u)$ of $u$, which is also called the \textbf{infinitesimal (linearized) strain tensor}. Since $W_e$ can only depend on $G \approx \epsilon(u)$ by Remark~\ref{rem:We_alternatives}, we may suppose that also $\Ebb$ depends only on the symmetric part of its argument and that the resulting matrix after application of $\Ebb$ is symmetric (any skew-symmetric part would cancel when multiplied by the argument again in the bilinear form). Thus, we posit also the symmetry relations
\[
  \Ebb^{ijkl} = \Ebb^{jikl},  \qquad
  \Ebb^{ijkl} = \Ebb^{ijlk}.
\]

In the following, we will write
\[
  \abs{A}_\Ebb^2 := \Ebb A : A,
\]
even though this expression really only depends on the symmetric part $\sym A := (A + A^\T)/2$ of $A \in \R^{3 \times 3}$. So, $\abs{W}_\Ebb^2 = 0$ for any skew-symmetric $W \in \R^{3 \times 3}$ (hence, despite the notation, $\abs{\frarg}_\Ebb$ is not a norm on all of $\R^{3 \times 3}$). In conclusion,
\[
  W_e(E) \approx \frac{1}{2} \abs{\epsilon(u)}_\Ebb^2.
\]

Finally, to reflect that any non-identity deformation costs energy, it is realistic to require that
\[
  \abs{A}_\Ebb^2 \geq c \abs{\sym A}^2
\]
for a constant $c > 0$. 

\begin{remark}
In fact, alternatively one may define $\Ebb$ as
\[
  \Ebb = \DD^2 W_e(\Id).
\]
To see this, let $C(E):=E^\T E$, so that $W_e=\hat{W}_e\circ C$. For $A,B\in\R^{3\times 3}$, we have
\[
  \DD C(\Id)[A]=A+A^\T.
\]
Thus, the chain rule together with $\DD\hat{W}_e(\Id)=0$ gives
\[
  \DD^2 W_e(\Id)[A,B] = \DD^2\hat{W}_e(\Id)[A+A^\T,B+B^\T].
\]
Hence,
\begin{align*}
  \DD^2 W_e(\Id)[A,B]
  &= \DD^2\hat{W}_e(\Id)[A+A^\T,B+B^\T] \\
  &= 4\,\DD^2\hat{W}_e(\Id)[\sym A,\sym B] \\
  &= \Ebb \, \sym A:\sym B,
\end{align*}
where $\Ebb=4\,\DD^2\hat{W}_e(\Id)$. Thus, $\Ebb = \DD^2 W_e(\Id)$ on symmetric matrices.
\end{remark}

\begin{remark}
If one additionally assumes that $W_e$ is \textbf{isotropic}, that is,
\[
  W_e(ER) = W_e(E)  \qquad\text{for all $R \in \SO(3)$,}
\]
one may show, see, e.g., Theorems~3.7-1,~3.8-1 in~\cite{Ciarlet88book}, that
\[
  \Ebb A = \lambda (\tr A) \Id + 2 \mu A,  \qquad A \in \R^{3 \times 3}_\sym,
\]
where $\lambda, \mu$ are the \textbf{Lam\'{e} constants} of the material (satisfying $\mu > 0$, $3\lambda+2\mu > 0$). In this case,
\[
  \abs{A}_\Ebb^2 = \lambda (\tr A)^2 + 2 \mu \, \abs{\sym A}^2.
\]
\end{remark}

\section{Space-time framework} \label{sc:spacetime_framework}

In this section we give an introduction to the Space-Time Framework that was introduced in~\cite{HudsonRindler22,Rindler23,Rindler25}. The basic modeling idea is to describe the dynamics of dislocation lines via the $2$-dimensional surfaces in space-time traced out by their motions, the so-called \emph{slip trajectories}. This description employs the language of multilinear algebra, which is quite well known in mathematics, but less so in mechanics. While multilinear algebra requires a bit of terminology, which may seem non-intuitive at first sight, it ultimately provides a very robust and efficient formalism to reason about curves and surfaces together with their transformations by linear and non-linear maps.

Let us mention already here that the use of multilinear algebra goes beyond a mere formalism: The solution in~\cite{BonicattoRindlerTurnbull26c?}
 of the homogenization problem for the model presented here crucially depends on this approach since the Space-Time Framework induces a different kind of \emph{weak topology} than the classical weak topology for vector-valued functions (or measures). In this topology, the discrete-to-field limit passage can then be performed for dislocations without the appearance of \enquote{commutators}; Example~\ref{ex:wedge_better} will contain a more precise illustration of this observation.

\subsection{Geometry of planes} \label{sc:planes}

We first discuss the geometry of planes in $3$-dimensional space $\R^3$ and also in $4$-dimensional (Galilean) \textbf{space-time}, that is,
\[
  \R^{1+3} = \spn\{\ee_0,\ee_1,\ee_2,\ee_3\} \cong \R \times \R^3 \cong \R^4.
\]
In $\R^{1+3}$ the first component takes the role of time and the remaining components take the role of space. Here, $\ee_1$, $\ee_2$ and $\ee_3$ are the three canonical \textbf{space-like} vectors and $\ee_0 = (1,0,0,0)$ is the additional \textbf{time-like} vector pointing in the (positive) time direction. We will implicitly consider vectors in $\R^3$ as space-time vectors in $\R^{1+3}$ by extending them by zero in the $\ee_0$ direction, i.e., we tacitly identify $(x_1,x_2,x_3) \in \R^3 $ with $(0,x_1,x_2,x_3) \in \R^{1+3}$.

\begin{figure}[tb]
  \centering
  \includegraphics{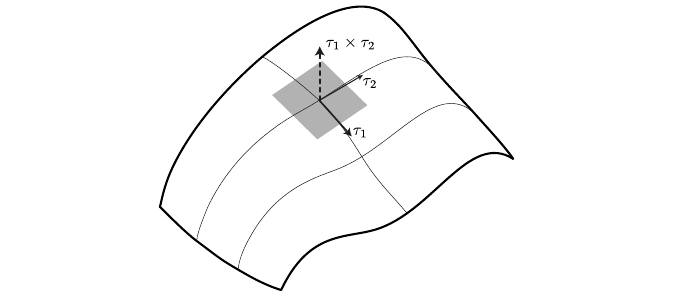}
  \caption{An oriented surface with tangent vectors.} 
  \label{fig:surface}
\end{figure}

A classical way to specify the \textbf{orientation} of a $2$-dimensional surface $S$ in $3$-dimensional space is to choose a continuous unit normal vector field defined at every point $x \in S$. The $2$-dimensional vector space orthogonal to this normal vector is called the \textbf{tangent space} to the surface at $x$ and denoted by $\Trm_x S$; see Fig.~\ref{fig:surface} for an illustration. The choice of orientation of the normal vector in fact turns $\Trm_x S$ into an \textbf{oriented vector space} as follows: If the vectors in the ordered pair $(v_1, v_2)$ span the tangent space $\Trm_x S$, then we say that $v_1, v_2$ are positively oriented (or 
\enquote{in the right-hand sense}) if the chosen normal vector at $x$ points in the same direction as the cross-product $v_1 \times v_2$ (and not in the direction of $v_2 \times v_1 = -v_1 \times v_2$).

However, this approach to orienting tangent vector spaces via a normal vector field is highly dependent on the ambient dimension being (at most) $3$. In particular, $2$-dimensional planes in $4$ dimensions (like in space-time $\R^{1+3}$) cannot be described by one normal vector only, but would require \emph{two} normal vectors. This is rather impractical and necessitates a more efficient description of planes and, more generally, vector sub-spaces.

\begin{figure}[tb]
  \centering
  \includegraphics{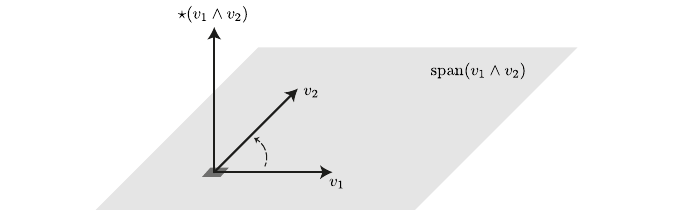}
  \caption{A $2$-vector describing a plane.} 
  \label{fig:planes}
\end{figure}

The basic idea is to equip $\R^N$ ($N = 2,3,4,\ldots$) with a new operation, the \textbf{wedge}, which is denoted by \enquote{$\wedge$}. It takes two vectors, say $v_1, v_2 \in \R^N$ and combines them into a \textbf{\mbox{$2$-(multi-)vector}}, which is denoted by $v_1 \wedge v_2$. Multi-vectors represent oriented, and possibly \enquote{weighted}, planes; see Fig.~\ref{fig:planes} for an illustration. In practice the actual definition of multi-vectors is largely irrelevant and one only needs to work with the computation rules, which are given below. While these rules tend to be a bit unusual to start with, eventually one obtains a very flexible, elegant, and efficient description of plane geometry. In the following, we give an intuitive introduction to multi-vectors; proofs can be found in Section~1.4 of~\cite{KrantzParks08book}.

More formally, it can be proved that for any $\R^N$ (here always space $\R^3$ or space-time $\R^{1+3}$) and all $k \in \N$ there exists a vector space
\[
  \Wedge_k \R^N
\]
of \textbf{$k$-(multi-)vectors} together with an operation
\[
  \wedge \colon \underbrace{\R^N \times \cdots \times \R^N}_{\text{$k$ times}} \to \Wedge_k \R^N 
\]
that is linear in every entry and such that for $v_1, \ldots, v_k, v_1' \in \R^N$, $i,j = 1,\ldots,k$, and $\lambda \in \R$ the following \textbf{Gaussian rules} hold:
\begin{align*}
  v_1 \wedge \cdots \wedge (\lambda v_i) \wedge \cdots \wedge v_j \wedge \cdots \wedge v_k &= v_1 \wedge \cdots \wedge v_i \wedge \cdots \wedge (\lambda v_j) \wedge \cdots \wedge v_k, \\
  v_1 \wedge \cdots \wedge v_i \wedge \cdots \wedge v_j \wedge \cdots \wedge v_k &= v_1 \wedge \cdots \wedge (v_i+\lambda v_j) \wedge \cdots \wedge v_j \wedge \cdots \wedge v_k, \\
  v_1 \wedge \cdots \wedge v_i \wedge \cdots \wedge v_j \wedge \cdots \wedge v_k &= v_1 \wedge \cdots \wedge (-v_j) \wedge \cdots \wedge v_i \wedge \cdots \wedge v_k, \\
  \lambda(v_1 \wedge \cdots \wedge v_k) &= (\lambda v_1) \wedge \cdots \wedge v_k, \\
  v_1 \wedge v_2 \wedge \cdots \wedge v_k + v_1' \wedge v_2 \wedge \cdots \wedge v_k &= (v_1 + v_1') \wedge v_2 \wedge \cdots \wedge v_k.
\end{align*}
The best way to remember these computation rules is to observe that they correspond precisely to the transformation rules for the rows of a matrix that do not change the determinant of the matrix (which is of course no coincidence). Additionally, we will always require that $v_1 \wedge \cdots \wedge v_k \neq 0$ (in $\Wedge_k \R^N$) if and only if the vectors $v_1, \ldots, v_k$ are linearly independent. It turns out that the above rules uniquely determine $\Wedge_k \R^N$ and the wedge (up to rescalings, which will be fixed later when we define \enquote{unit} $k$-vectors).

One can show that $\Wedge_1 \R^N$ can be canonically identified with $\R^N$ itself and $\Wedge_k \R^N  = \{0\}$ if $k > N$, as can be checked by the above rules. We also set $\Wedge_0 \R^N := \R$.

If an element $\xi \in \Wedge_k \R^N $ can be written as an element of the form $v_1 \wedge \cdots \wedge v_k$, then the $k$-vector $\xi$ is called \textbf{simple}. Not all vectors are simple, however. For instance, the space-time $2$-vector $\ee_0 \wedge \ee_1 + \ee_2 \wedge \ee_3 \in \Wedge_2 \R^{1+3}$ is not simple (which can be shown using the computation rules).

It is also useful to define the \textbf{wedge product} (or \textbf{exterior product}) of a $k$-vector $\xi \in \Wedge_k \R^N $ and an $l$-vector $\eta \in \Wedge_l \R^N$ by linear extension: For a simple $k$-vector $\xi = v_1 \wedge \cdots \wedge v_k$ and a simple $l$-vector $\eta = w_1 \wedge \cdots \wedge w_l$, the $(k+l)$-vector $\xi \wedge \eta$ is defined by \enquote{wedge-concatenation},
\[
  \xi \wedge \eta := v_1 \wedge \cdots \wedge v_k \wedge w_1 \wedge \cdots \wedge w_l \in \Wedge_{k+l} \R^N.
\]
This associative operation is then extended to non-simple vectors distributively, that is, for $\xi_1, \xi_2 \in \Wedge_k \R^N$ and $\eta_1, \eta_2 \in \Wedge_l \R^N$ simple we set
\[
  (\xi_1 + \xi_2) \wedge (\eta_1 + \eta_2)
  := \xi_1 \wedge \eta_1 + \xi_1 \wedge \eta_2 + \xi_2 \wedge \eta_1 + \xi_2 \wedge \eta_2
\]
and so on. It can be proved that this is well-defined (by the above computation rules).

Let $\xi \in \Wedge_k \R^N $, let $\eta, \rho \in \Wedge_l \R^N$ (where $k,l \in \{0,\ldots,d\}$), and let $\lambda \in \R$. Then we have the following further rules, which follow by relatively straightforward computations from the above identities:
\begin{enumerate}[(i)]
  \item \textbf{Linearity:} $\xi \wedge (\eta + \rho) = \xi \wedge \eta + \xi \wedge \rho$ and $(\lambda \xi) \wedge \eta = \xi \wedge (\lambda \eta) = \lambda (\xi \wedge \eta)$.
  \item \textbf{Skew-commutativity:} $\xi \wedge \eta = (-1)^{k l} \eta \wedge \xi$; in particular $\xi \wedge \xi = 0$ if $k$ is odd.
  \item \textbf{Nilpotence:} $\xi \wedge \xi = 0$ if $\xi$ is simple; in particular $v \wedge v = 0$ for $v \in \R^N$.
\end{enumerate}

We define the \textbf{span} of a simple $k$-vector $\xi = v_1 \wedge \cdots \wedge v_k \in \Wedge_k \R^N $ by
\[
  \spn \xi := \spn \{ v_1, \ldots, v_k \}.
\]
For non-simple vectors we take the span of the spans of any decomposition into simple parts (which is always possible and turns out to be well-defined). 

Multi-vectors in fact describe \emph{weighted} planes and it can be shown that for simple $k$-vectors $\xi, \eta \in \Wedge_k \R^N$ it holds that
\[
  \spn \xi = \spn \eta  \qquad\text{if and only if}\qquad
  \text{$\eta = \lambda \xi$ for some $\lambda \neq 0$.}
\]
As for the geometric interpretation of the weighting, call a simple $k$-vector $\xi$ a \textbf{unit} if it can be expressed as $\xi = v_1 \wedge \cdots \wedge v_k$ with the $\{v_i\}_i$ forming an orthonormal basis of $\spn \xi$ (that is, $v_i \cdot v_j = 1$ if and only if $i = j$, and $0$ otherwise). For $\xi = v_1 \wedge \cdots \wedge v_k$ a simple $k$-vector, assume that $\xi$ can be written as $\xi = \lambda \xi_1$ with a unit $k$-vector $\xi_1$ and $\lambda > 0$; note that the positivity can always be achieved by flipping one of the $v_i$'s sign. It can be proved that this $\lambda$ is the ($k$-dimensional) volume of the parallelotope (i.e., a \enquote{parallelogram} in $\R^2$ and a \enquote{parallelepiped} in $\R^3$) spanned by the vectors $v_j$ for $j = 1,\ldots,k$. In this context note that all the above computation rules for $k$-vectors do not contradict this interpretation. For instance, one rule said that adding a multiple of one vector to another in a (simple) wedge product leaves this wedge product invariant. Geometrically, this corresponds to shearing the parallelotope, which, as is well-known from elementary geometry, does not change its volume. In particular, by repeatedly applying the rules one may reduce to the case of a cuboid, from which the existence of the representation $\xi = \lambda \xi_1$ follows.

For two simple $k$-vectors $\xi = v_1 \wedge \cdots \wedge v_k$, $\eta = w_1 \wedge \cdots \wedge w_k$ the \textbf{scalar product} is defined as
\[
  (\xi, \eta) := \det\left(
    \begin{matrix}
      v_1 \cdot w_1 & \cdots & v_1 \cdot w_k\\
      \vdots &\ddots & \vdots \\
      v_k \cdot w_1 &\cdots &v_k \cdot w_k
    \end{matrix}\right),
\]
which is extended to general $k$-vectors by linearity. This means that if $\xi = \xi_1 + \xi_2$ and $\eta = \eta_1 + \eta_2$ with $\xi_1, \xi_2, \eta_1, \eta_2$ simple, then by \enquote{expanding the product},
\[
  (\xi, \eta) := (\xi_1, \eta_1) + (\xi_1, \eta_2) + (\xi_2, \eta_1) + (\xi_2, \eta_2),
\] 
with the individual products defined as above. Fixing a basis and first defining the scalar product for wedge products of the basis vectors, one can show that this is well-defined, meaning that the value does not depend on which decomposition of $\xi$ and $\eta$ one chooses.

The scalar product then also gives rise to a \textbf{(Euclidean) norm} of $k$-vectors, namely
\begin{equation} \label{eq:multivector_Euclidean_norm}
  \abs{\xi} := (\xi, \xi)^{1/2},  \qquad \xi \in \Wedge_k \R^N.
\end{equation}
One can check that if $\xi = v_1 \wedge \cdots \wedge v_k$ for an orthonormal system of vectors $v_1, \ldots, v_k \in \R^N$, then $\abs{\xi} = 1$. Combining this with the previous paragraph, we conclude that $\abs{v_1 \wedge \cdots \wedge v_k}$ is the volume of the parallelotope spanned by the vectors $v_1, \ldots, v_k$.

Another way to describe an oriented $k$-plane $R$ is by providing a $(N-k)$-dimensional unit \textbf{normal plane} $U := \spn \xi$, where $\xi \in \Wedge_{N-k} \R^N$ is simple, and setting $R := U^\perp$, the orthogonal complement of $U$ (with an \enquote{orthogonal orientation}). For the case of ordinary $2$-planes in $\R^3$ (that is, $N = 3$, $k = 2$), this orthogonal space is spanned by just one unit vector, the \textbf{unit normal vector}, which is determined uniquely up to the choice of sign. For the orientation of the resulting plane we may use the reasoning via the ordering of the cross product from the beginning of the section.

This approach is systematized as follows: For a $k$-vector $\eta \in \Wedge_k \R^N $, we define the \textbf{Hodge dual} $\hodge \eta \in \Wedge_{N-k} \R^N$ as the unique $(N-k)$-vector satisfying for all $\xi \in \Wedge_k \R^N$ that
\[
  \xi \wedge \hodge \eta = (\xi, \eta) \, \Erm,  \qquad\text{where}\qquad
  \Erm := \ee_1 \wedge \cdots \wedge \ee_N.
\]
Here, as before $\ee_1, \ldots, \ee_N$ denote the unit vectors (in their canonical order) in $\R^N$ (that is, $\Erm := \ee_1 \wedge \ee_2 \wedge \ee_3$ in $\R^3$ and $\Erm := \ee_0 \wedge \ee_1 \wedge \ee_2 \wedge \ee_3$ in $\R^{1+3}$).

In particular, in the case $N=3$ we have the following geometric interpretation of the Hodge dual: For $\eta \in \Wedge_2 \R^3$, the \mbox{($1$-)vector} $\hodge \eta$ is the oriented normal vector to the $2$-dimensional hyperplane with orientation $\eta$, that is,
\[
  \hodge \eta \perp \spn \eta.
\]
See Fig.~\ref{fig:planes} for a visualization of the Hodge dual. Note that no similar representation of a $2$-plane by a normal vector holds in $\R^4$ since the Hodge dual of a $2$-vector in $\R^4$ is again a $2$-vector.

The wedge product and the vector cross product are closely related in $\R^3$. Indeed, for vectors $v, w \in \Wedge_1 \R^3 \cong \R^3$ the identities
\begin{equation}\label{eq:hodge_wedge}
  \hodge(v \times w) = v \wedge w,  \qquad
  \hodge (v \wedge w) = v \times w
\end{equation}
hold, where \enquote{$\times$} denotes the classical vector \textbf{cross product} in $\R^3$. In fact, in large parts of the engineering literature, the symbol \enquote{$\wedge$} stands for the vector cross product and the above formulas show that this is consistent with our language \enquote{up to Hodge duality}. However, the wedge product and Hodge formalism easily extend to any dimension, in particular to $\R^{1+3}$, whereas the cross product is limited to $2$ or $3$ ambient dimensions.

In the following, we will also need to extend a linear map $L \colon \R^N \to \R^M$ to $k$-vectors. We do this by \enquote{pulling out the wedge}, that is, by setting
\[
  L(v_1 \wedge \cdots \wedge v_k) := (Lv_1) \wedge \cdots \wedge (Lv_k), \qquad v_1, \ldots, v_k \in \R^N,
\]
and then extending this definition by (multi-)linearity to all of $\Wedge_k \R^N$ (that is, $L(\xi + \eta) := L\xi + L\eta$ and so on). We call this so-extended $L$ the $k$-times \textbf{wedge product of a linear map}. In the literature the wedge product of a linear map $L$ is sometimes written as $\Wedge^k L$, but we here denote it just by $L$ again to avoid notational clutter.

One can then derive a formula for the transformation of a Hodge dual of a $2$-vector under an invertible linear map $L \colon \R^3 \to \R^3$, which is identified with its matrix representation $L \in \R^{3 \times 3}$: For $\eta := v \wedge w \in \Wedge_2 \R^3$ the Hodge dual $\hodge (L\eta)$ of $L\eta$ satisfies 
\begin{equation} \label{eq:hodge_transform2}
  \hodge (L\eta) = (\det L) \, L^{-\T} (\hodge \eta).
\end{equation}
Note that if $L$ is orthogonal and $\det L = 1$, this agrees with the transformation rule for normal vectors under orthogonal transformations of a surface. Furthermore, for any vector $v \in \R^3$ one obtains the dual transformation rule
\begin{equation} \label{eq:hodge_transform1}
  \hodge (Lv) = (\det L) \, L^{-\T} (\hodge v).
\end{equation}

We note that for the inverse of the Hodge dual it holds that
\begin{equation}\label{eq:hodge_inverse}
  \hodge^{-1} = (-1)^{k(N-k)} \, \hodge,
\end{equation}
in particular $\hodge^{-1} = \hodge$ when applied to \mbox{($1$-)vectors} in $\R^3$.

Finally, the Hodge dual operation also preserves the (Euclidean) norm, that is, $\abs{\hodge \xi} = \abs{\xi}$ for all $\xi \in \Wedge_k \R^N$.

\subsection{From surfaces to currents} \label{sc:surfaces_to_currents}

Having developed an efficient language to describe planes, we now move on to develop some theory for geometric objects, most importantly curves and surfaces. We start from the prototypical case of a parametrized $2$-dimensional surface
\[
  S = F(Z) \subseteq \R^N
\]
for a single injective, smooth, and smoothly invertible (on its image) \textbf{chart} $F \colon Z \to \R^N$, which is defined on a bounded coordinate domain $Z \subseteq \R^2$ (actually, it is the inverse $F^{-1}$ of $F$ that is customarily called a \enquote{chart}, but here this is less convenient); see Fig.~\ref{fig:surface} for an illustration.

Classical vector analysis tells us that the tangent space $\Trm_x S$ to $S$ at a point $x = F(z) \in S$ is given as the span of the vectors
\begin{equation} \label{eq:xix_xiy}
  \tau_1(x) := \frac{\partial F}{\partial z_1}(F^{-1}(x)), \qquad
  \tau_2(x) := \frac{\partial F}{\partial z_2}(F^{-1}(x)).
\end{equation}
The (oriented) tangent space (together with the orientation induced by $(\partial_{z_1}F,\partial_{z_2}F)$) is completely specified by the \textbf{orienting $2$-vector}
\begin{equation} \label{eq:vecS_surface}
  \vec{S}(x) := \frac{\tau_1 \wedge \tau_2}{\abs{\tau_1 \wedge \tau_2}} \in \Wedge_2 \R^N.
\end{equation}
Note that we have normalized this $2$-vector by the (Euclidean) norm defined in~\eqref{eq:multivector_Euclidean_norm} since, at this point at least, we do not want a \enquote{weighted} tangent space. It is intuitively clear that $\vec{S}(x)$ does not depend on the choice of parametrization since it is determined solely by the shape of $S$ around $x$. One can also prove this parametrization-invariance formally using the computation rules for $2$-vectors from the previous section.

The formalism with the orienting $2$-vector $\vec{S}$ works in any ambient space $\R^N$, in particular in space-time $\R^{1+3}$. The description of a surface through a surface normal, on the other hand, cannot be used in $\R^{1+3}$, since we would need \emph{two} normals to describe a $2$-surface in $4$-dimensional space.

A fundamental idea from the field of geometric measure theory is to identify an (oriented) surface with the \emph{integration} over this surface. While at first sight this may seem like a formalistic contortion, it turns out that in this way one can \emph{compute} with surfaces just like one can compute with vectors and tensors. In particular, this formalism will allow us to add, subtract and scalar-multiply surfaces. Furthermore, \emph{geometric} operations, such as taking the boundary or slicing by a hyperplane, may be expressed by \emph{algebraic} formulas in this language.

If we want to consider $S$ as an integral of some sort, it is a natural requirement that this integral is \emph{independent of the chosen parametrization} of the surface (our $F$ above), so the \enquote{integrands} we want to consider should only take $\vec{S}(x)$ as input at a point $x$. In order to obtain a linear operation with respect to the surface (a basic requirement to be an \enquote{integral}), these integrands should naturally be given, at any point $x \in S$, as the application of a linear functional on $\Wedge_2 \R^N$.

For this, we need multi-vectors based on the dual space $(\R^N)^*$ of $\R^N$, that is, linear functionals on $\R^N$ (linear maps from $\R^N$ to $\R$). Of course, $(\R^N)^*$ could be identified with $\R^N$ itself, but we will not do this since $(\R^N)^*$ has a different \enquote{data type} than $\R^N$ and the distinction will become important.

The $k$-vectors over the dual space $(\R^N)^*$ are called \textbf{$k$-(multi-)covectors} (over $\R^N$) and are denoted by
\[
  \Wedge^k \R^N := \Wedge_k (\R^N)^*.
\]
Once again we set $\Wedge^0 \R^N := \R$ (or, more precisely, the \emph{dual space} to $\R$) and define the wedge product on covectors similarly to before. The standard dual basis is denoted by $\di x^1, \ldots, \di x^N$, as usual. Even though in coordinates, these are identical to the unit vectors $\ee_1, \ldots, \ee_N$, it is quite important to keep multi-vectors and multi-covectors apart, hence we will never identify $\ee_i$ and $\di x^i$.

In fact, the space $\Wedge^k \R^N$ can be considered as the dual space to $\Wedge_k \R^N$ (which is a linear space). The duality pairing between a simple $k$-vector $\xi = v_1 \wedge \cdots \wedge v_k$ and a simple $k$-covector $\alpha = w_1 \wedge \cdots \wedge w_k$ is given via
\begin{equation} \label{eq:multi_duality_pairing}
  \dprb{\xi,\alpha} := \det \, [v_i \cdot w_j]^i_j.
\end{equation}
Here, $[v_i \cdot w_j]^i_j$ is the matrix with entry $v_i \cdot w_j$ (the scalar product between $v_i$ and $w_j$ in coordinates) in the $i$'th row and $j$'th column. This is indeed a \emph{linear} pairing because we are considering $\xi, \alpha$ to be paired and not the individual vectors (in which this pairing is only \emph{separately} linear). Furthermore, $\Wedge^k \R^N$ could also be identified with the space of $k$-multilinear (linear in every argument) and alternating (switching sign when exchanging two arguments) functions on $(\R^N)^k$, but we will have little need for this.

Smooth (i.e., infinitely often differentiable) maps, defined on $\R^N$ and with values in $\Wedge^k \R^N$, are called \textbf{(smooth) differential $k$-forms}.

Given a $2$-dimensional surface $S$ in $\R^N$, we can thus define for any (smooth) differential $2$-form $\omega \colon S \to \Wedge^2 \R^N$ the \textbf{integral} as
\[
  \int_S \omega := \int_S \dprb{\vec{S}(x), \omega(x)} \dd \Hcal^2(x).
\]
Here, $\dpr{\vec{S}(x), \omega(x)}$ is the duality pairing between the orienting $2$-vector $\vec{S}(x)$ and the $2$-covector $\omega(x)$ as defined in~\eqref{eq:multi_duality_pairing}. Moreover, $\Hcal^2$ denotes the surface element, which is also called the $2$-dimensional Hausdorff measure, which measures $2$-dimensional area.

If $S = F(Z)$ is a surface parametrized by a single chart, we can unwind the definition of the above integral as follows: It is shown in elementary vector analysis that if the ambient space is $\R^2$ or $\R^3$ then for any integrand $\psi \colon S \to \R$,
\[
  \int_S \psi(x) \dd \Hcal^2(x)
  = \iint_Z \psi(F(z)) \; \abs{\tau_1(F(z)) \times \tau_2(F(z))} \dd z,
\]
with $\tau_1, \tau_2$ defined in~\eqref{eq:xix_xiy}. Now, using that the Hodge dual preserves the multi-vector norm, and the computation rules of~\eqref{eq:hodge_wedge}, we have that
\[
  \abs{\tau_1 \times \tau_2} = \abs{\hodge(\tau_1 \wedge \tau_2)} = \abs{\tau_1 \wedge \tau_2}.
\]
Consequently, using also the definition of $\vec{S}$ in~\eqref{eq:vecS_surface},
\begin{align}
  \int_S \omega
  &= \iint_Z \dprb{\vec{S}(F(z)), \omega(F(z))} \; \abs{\tau_1(F(z)) \wedge \tau_2(F(z))} \dd z  \notag\\
  &= \iint_Z \dprb{\tau_1(F(z)) \wedge \tau_2(F(z)), \omega(F(z))} \dd z.  \label{eq:intS_xi}
\end{align}
This agrees with the usual definition of surface integrals for differential forms, see, e.g.,~\cite{Lee13book}.

Now we (re-)define $S$ as a linear functional on (smooth) differential $2$-forms $\omega \colon S \to \Wedge^2 \R^N$ as follows:
\[
  S[\omega] := \int_S \omega = \int_S \dprb{\vec{S}, \omega} \dd \Hcal^2.
\]
This definition can be easily extended to situations where $S$ is not parametrized by just one chart, but by many, once one establishes that the definition of the integral does not depend on the choice of charts. Locally, that is, in a suitably small ball around any point, one can reduce to the case of one chart.

With this definition, we can now even make sense of \emph{linear} combinations of surfaces $S_1, S_2$ in this way, setting for any $\alpha, \beta \in \R$,
\[
  (\alpha S_1 + \beta S_2)[\omega] 
  := \alpha S_1[\omega] + \beta S_2[\omega] 
  = \alpha \int_{S_1} \omega + \beta \int_{S_2} \omega.
\]
Furthermore, given surfaces $S_n$ with \textbf{multiplicities} $m_n \in \N$, we call the linear combination
\[
  S = \sum_n m_n S_n  
\]
a \textbf{(integral) \mbox{$2$-current}}. Here, the index set may be finite or countable. This linear combination has a sense as a linear functional on (smooth) differential $2$-forms $\omega \colon \R^N \to \Wedge^2 \R^N$ in the above sense, that is,
\[
  S[\omega] 
  := \sum_n m_n \, \dprb{S_n,\omega} 
  = \sum_n m_n \int_{S_n} \omega
  = \sum_n m_n \int_{S_n} \dprb{\vec{S}_n, \omega} \dd \Hcal^2.
\]
The above definition in particular shows that currents corresponding to parametrized surfaces are \emph{independent} of the surface's parametrization.  

A mathematical introduction to the theory of currents can be found in~\cite{KrantzParks08book} (or the encyclopedic~\cite{Federer69book}). However, as the mathematical treatment is quite technical and since for the modeling we do not in fact require the intricacies of the full theory, we here develop all the notions in an intuitive way.

For the expression $S[\omega]$ we also introduce the following convenient notation:
\[
  \int \dprb{\vec{S}, \omega} \dd \tv{S} := S[\omega].
\]
This expresses that we are integrating the duality product of $\vec{S}$ and $\omega$ over the \textbf{mass measure} $\tv{S}$, which adds up all the $2$-dimensional Hausdorff measures restricted to the surfaces $S_n$, together with their multiplicities $m_n$ (correctly treating overlaps, e.g., canceling oppositely-oriented overlapping pieces). To emphasize the splitting of the \mbox{$2$-current} $S$ into the orienting $2$-vector field $\vec{S}$ and the surface \enquote{measure} $\tv{S}$, we will often write $S$ via the so-called \textbf{Radon--Nikodym decomposition}
\[
  S = \vec{S} \, \tv{S}.
\]
One observes that this notation is not ambiguous, because $\vec{S}$ is well-defined with respect to $\tv{S}$: If two surfaces $S_n,S_m$ from the collection making up $S$ overlap over a set $U$, then the orienting $2$-vector $\vec{S}$ is \emph{uniquely} determined on $U$, up to a \enquote{negligible} (lower-dimensional) part, since the tangent spaces agree (up to orientation). If they did not agree, the surfaces would cross and could not overlap in more than a curve. As $\tv{S}$ incorporates the multiplicities (whose sign may be flipped in tandem with $\vec{S}$), the \enquote{additivity} of $S$ is thus correctly represented by the above formula.

We clearly can add or scalar-multiply currents, thus turning \mbox{$2$-currents} into a linear space: If
\[
  S = \sum_n m_n S_n,  \qquad
  \tilde{S} = \sum_n \tilde{m}_n \tilde{S}_n
\]
and $\alpha, \beta \in \R$, we set
\[
  \alpha S + \beta \tilde{S}
  = \sum_n (\alpha m_n) S_n + \sum_n (\beta \tilde{m}_n) \tilde{S}_n.
\]
One special case is that
\[
  -S = (-\vec{S}) \, \tv{S} = \sum_n (-m_n) S_n.
\]
In fact, currents can even be extended to \enquote{densities} of surfaces, where integrals replace sums; these currents are called \enquote{normal} (if the boundary mass is also finite).

Finally, nothing of the above derivation used the $2$-dimensionality in an essential way. Everything works just the same way for $k$-dimensional (hyper-)surfaces, yielding the notion of a \textbf{(rectifiable) \mbox{$k$-current}} $T = \vec{T} \, \tv{T}$ with $\vec{T}(x) \in \Wedge_k \R^N$ for every $x$ with respect to the measure $\tv{T}$. In particular, for curves a \mbox{$1$-current} is a linear combination of a family of curves together with multiplicities.

\begin{remark}
There is some relation of our framework to the so-called \enquote{geometrical language of continuum mechanics}~\cite{CiarletGratieMardare09,Epstein10book,EpsteinKupfermanMaor20}. One noteworthy point in this respect is that the dislocation density of continuously distributed (and grain-homogenized) dislocations can be identified with the torsion tensor of an affine material connection~\cite{Kondo55,Nye53,BilbyBulloughSmith55,Noll58,KronerSeeger59,Kroner60,Kroner61,Wang67,Lazar00,Katanaev05,Malyshev07,HehlObukhov07,LazarAnastassiadis08,LazarAnastassiadis09,KupfermanMaor15,EpsteinKupfermanMaor20,KupfermanOlami20} (which can be seen as the limit of discrete dislocations~\cite{KupfermanMaor15,EpsteinKupfermanMaor20,KupfermanOlami20}). This is in a precise sense dual to the approach with currents;  see the appendix of~\cite{HudsonRindler22} for a comparison.
\end{remark}

\subsection{Exterior differential and boundaries} \label{sc:boundary}

We were deliberately vague in the previous section whether our $2$-dimensional surface $S$ has a \enquote{boundary} or not. The geometric notion of boundary is intuitively clear -- it is those points \enquote{where the surface ends}. If $S = F(Z)$ as above then the boundary $\partial S$ of $S$ is the image of the boundary of $Z$.

\begin{figure}[tb]
  \centering
  \includegraphics{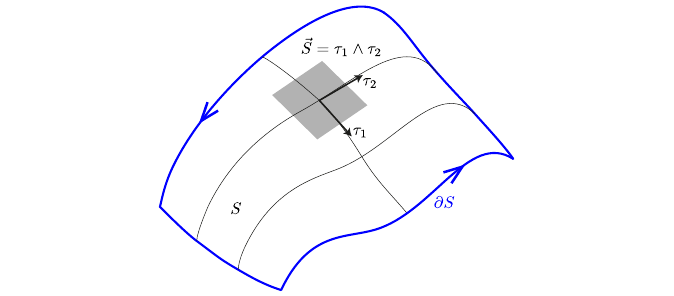}
  \caption{An integral current and its boundary.} 
  \label{fig:current_boundary}
\end{figure}

In three ambient dimensions, the boundary $\partial S$ of $S$ is $1$-dimensional (at least if it is not too \enquote{ragged}) and in fact it can be oriented using the \textbf{boundary orientation}: The boundary (unit) orientation $\tau$ is the one such that if $n$ is the unit \emph{inner} normal, then $\tau \times n$ equals the oriented normal to the surface, that is, $\hodge \vec{S}$. Equivalently, an imagined pedestrian with head in direction of the oriented normal to the surface (so, in direction $\hodge \vec{S}$) traverses the boundary in positive direction if the surface always lies to the left of the pedestrian. See Fig.~\ref{fig:current_boundary} for an illustration of the boundary and its orientation. 

To define boundaries in full generality, we need the following notion: The \textbf{exterior differential} $\di \omega$ of a $k$-form $\omega \colon \R^N \to \Wedge^k \R^N$ is defined as follows. For a $0$-form $f$ we set
\[
  \di f := \frac{\partial f}{\partial x^1} \; \di x^1 + \cdots + \frac{\partial f}{\partial x^N} \; \di x^N.
\]
For a $k$-form $\omega$ that can be written as $\omega = f \, \di x^{i_1} \wedge \cdots \wedge \di x^{i_k}$ for some indices $i_1, \ldots, i_k \in \{1,\ldots,N\}$, we set
\[
  \di\omega := \di f \wedge \di x^{i_1} \wedge \cdots \wedge \di x^{i_k}.
\]
If $\omega$ is not of the above simple form, then it is a linear combination of such simple forms and we extend this definition by linearity, e.g., for $\omega = \omega_1 + \omega_2$ with $\omega_1, \omega_2$ only taking \emph{simple} $k$-covectors as values, we set $\di \omega := \di\omega_1 + \di\omega_2$, and so on.

The usual vector analysis operations are contained in the exterior calculus, which, however, is more systematic. For instance, we have only one \textbf{Leibniz rule} for a $k$-form $\omega$ and an $l$-form $\eta$, namely
\[
  \di(\omega \wedge \eta) = \di\omega \wedge \eta + (-1)^k \omega \wedge \di\eta.
\]
and it always holds that
\begin{equation} \label{eq:d2_0}
  \di^2 \omega = \di(\di\omega) = 0,
\end{equation}
as a computation shows.

The \textbf{boundary} $\partial T$ of a generic \mbox{$k$-current} $T$ is then defined so that the generalized \textbf{Stokes theorem} holds, that is,
\begin{equation} \label{eq:boundary_Stokes}
  (\partial T)[\omega] := T[\di\omega]
\end{equation}
for any differential $(k-1)$-form $\omega$ (for which $\di\omega$ is a $k$-form).

Via~\eqref{eq:d2_0}, we obtain the important relation
\[
  \partial (\partial S) = 0,
\]
a fact we will often abbreviate to \enquote{$\partial^2 = 0$}.

For a $1$-current representing a curve $\gamma$ in $\R^N$, that is,
\[
  T = \vec{T} \, \Hcal^1 \restrict \gamma,
\]
with $\gamma \colon [0,1] \to \R^N$ a curve and $\vec{T}$ its tangent, the boundary is the $0$-current given by
\[
  \partial T = \delta_{\gamma(1)} - \delta_{\gamma(0)},
\]
that is, a point mass at the end minus a point mass at the start of the curve. This can be proved via the above general definition of the boundary and using the fact that for a $0$-form $f$ it holds that $\di f$ (see above) represents a gradient, so that
\[
  \dprb{\partial T, f}
  = \int_\gamma \nabla f \cdot \vec{T} \dd \Hcal^1
  = f(\gamma(1)) - f(\gamma(0)).
\]

\subsection{Operations on currents} \label{sc:operations_on_currents}

Consider a surface $S = F(Z) \subseteq \R^N$ with a (single) chart $F \colon Z \to \R^N$, which is defined on an open coordinate domain $Z \subseteq \R^2$. Recall that the \mbox{$2$-current} $S$ acts on a (smooth) differential $2$-form $\omega \colon \R^N \to \Wedge^2 \R^N$ as
\begin{equation} \label{eq:S_act_coordinates}
  S[\omega]
  = \iint_Z \dprBB{\frac{\partial F}{\partial z_1}(z) \wedge \frac{\partial F}{\partial z_2}(z), \; \omega(F(z))} \dd z.
\end{equation}
Here, we have written the expression $\tau_1(F(z)) \wedge \tau_2(F(z))$ in~\eqref{eq:intS_xi} (with $\tau_1, \tau_2$ defined in~\eqref{eq:xix_xiy}) as
\[
  \tau_1(F(z)) \wedge \tau_2(F(z))
  = \frac{\partial F}{\partial z_1}(z) \wedge \frac{\partial F}{\partial z_2}(z).
\]
Define the \textbf{differential} $\DD F(z)$ of $F$ at $z \in Z$ as the linear map
\[
  \DD F(z) := \frac{\partial F}{\partial z_1}(z) \dd x_1 + \frac{\partial F}{\partial z_2}(z) \dd x_2
\]
with $\di x_1, \di x_2$ the functionals that take a vector $v \in \R^2$ and return its first and second component, respectively. This means that for $v = (v_1,v_2)$ we have
\[
  \DD F(z)[v] = \frac{\partial F}{\partial z_1}(z) \, v_1 + \frac{\partial F}{\partial z_2}(z) \, v_2.
\]

According to our rules, $\DD F$ is extended to $\Wedge_k \R^2$ by \enquote{pulling out the wedge}, that is,
\[
  \DD F(z)[v \wedge w] := \DD F(z)[v] \wedge \DD F(z)[w].
\]
In particular,
\[
  \DD F(z)[\ee_1 \wedge \ee_2]
  = \DD F(z)[\ee_1] \wedge \DD F(z)[\ee_2]
  = \frac{\partial F}{\partial z_1}(z) \wedge \frac{\partial F}{\partial z_2}(z)
\]
since $\di x^i(\ee_j) = 1$ if and only if $i = j$ and $0$ otherwise. So, comparing this expression to~\eqref{eq:S_act_coordinates}, our surface $S$, considered as a \mbox{$2$-current}, acts on a differential $2$-form $\omega$ via
\begin{equation} \label{eq:S_as_pushforward}
  S[\omega] = \iint_Z \dprb{\DD F(z)[\ee_1 \wedge \ee_2], \omega(F(z))} \dd z.
\end{equation}

If we transform the surface $S$ via a smooth map $\theta \colon \R^N \to \R^M$ (or even just $\theta \colon S \to \R^M$), we want to describe the \mbox{$2$-current} corresponding to the resulting surface $\theta(S)$. Let again $S = F(Z)$ as above. Then, the \mbox{$2$-current} corresponding to this transformed surface, which is called the \textbf{pushforward of $S$ under $\theta$} and denoted by $\theta_* S$, can be expressed via~\eqref{eq:S_as_pushforward} for a smooth differential $2$-form (now defined on $\R^M$) as
\[
  (\theta_* S)[\omega] = \iint_Z \dprb{\DD (\theta \circ F)(z)[\ee_1 \wedge \ee_2], \omega((\theta \circ F)(z))} \dd z.
\]
Using the chain rule, we obtain for the differential $\DD (\theta \circ F)$ of the concatenation $(\theta \circ F)(z) = \theta(F(z))$ the expression
\[
  \DD (\theta \circ F)(z)[v] = \DD \theta(F(z)) [\DD F(z)[v]]
\]
for any vector $v$. Extending this definition again to multi-vectors by \enquote{pulling out the wedge}, we may thus write
\begin{align}
  (\theta_* S)[\omega]
  &= \iint_Z \dprb{\DD \theta(F(z)) [\DD F(z)[\ee_1 \wedge \ee_2]], \omega((\theta \circ F)(z))} \dd z  \notag\\
  &= \int \dprb{\DD \theta[\vec{S}], \omega \circ \theta} \dd \tv{S}.  \label{eq:pushforward_def}
\end{align}
In this way we see that the pushforward does not actually depend on the parametrization and then can be extended from parametrized surfaces to all \mbox{$2$-currents} by the above formula.

From the above formulation one may also derive (with some effort) the important property that
\begin{equation} \label{eq:partial_pushforward}
  \partial (\theta_* S) = \theta_* (\partial S),
\end{equation}
that is, pushforwards commute with the boundary operation.

The next operation we want to consider concerns the cutting-out of parts of a \mbox{$2$-current}: The \textbf{restriction} $S \restrict U$ of a surface $S$ to a set $U$ is just defined by intersecting $S$ with $U$. For a general \mbox{$2$-current} this simply means that we restrict the integral, that is,
\[
  (S \restrict U)[\omega]
  := \int_U \dprb{\vec{S}, \omega} \dd \tv{S}.
\]
This operation does not affect the orientation. Note that if $U$ is open, then the result can still be understood as a surface (with or without boundary). On the other hand, a general intersection may have zero area and then we just identify the result with the zero \mbox{$2$-current} (which yields zero when integrated against any differential $2$-form).

As we will see in the next section, the Space-Time Framework has as its central objects $2$-surfaces in space-time $\R^{1+3}$. For additional convenience, we define the \textbf{time projection} $\tbf$ and the \textbf{spatial projection} $\pbf$ via
\begin{equation} \label{eq:tbf_pbf_def}
  \tbf(t,x_1,x_2,x_3) := t \qquad\text{and}\qquad
  \pbf(t,x_1,x_2,x_3) := (x_1,x_2,x_3),
\end{equation}
respectively. These linear maps act on vectors (which at least conceptually should be distinguished from points, even though they share the same coordinate representation) via their \emph{pushforwards} as follows:
\[
  \tbf v := v_0,  \quad
  \pbf v := (v_1,v_2,v_3) \qquad\text{for}\qquad
  v = (v_0,v_1,v_2,v_3) \in \R^{1+3}.
\]
Technically, the pushforward $\pbf_*$ of $\pbf$ applied to a vector $v \in \R^{1+3}$ is defined to be
\[
  \pbf(t,x)_* v := \DD \pbf(t,x) v
\]
with the differential $\DD \pbf$ the linear map given as
\[
  \DD \pbf(t,x) = \bigl[ 0 \; \Id_{3 \times 3} \bigr] \in \R^{3 \times 4}.
\]
However, since $\DD \pbf$ does not actually depend on $(t,x)$ we simply write $\pbf v$ for the application to $v$; likewise for $\tbf v$.

\begin{figure}[tb]
  \centering
  \includegraphics{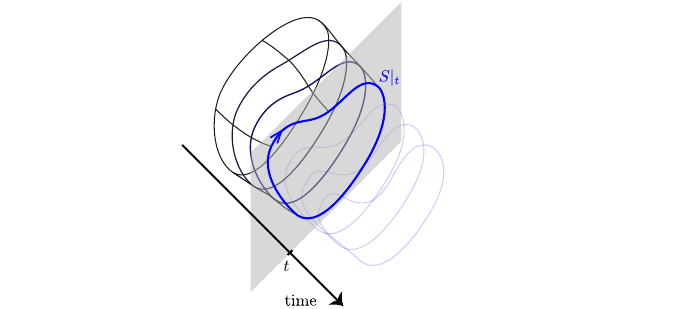}
  \caption{The slicing of a surface in space-time.}
  \label{fig:slip_trajectory_slice}
\end{figure}

We can also define another form of \enquote{restriction}: Consider a $2$-dimensional surface $S$ in space-time $\R^{1+3}$ that has the property that every intersection
\[
  S \restrict \{\tbf = t\},  \qquad\text{where}\qquad
  \{\tbf = t\} := \{t\} \times \R^3 = \setb{ (t,\hat{x}) }{ \hat{x} \in \R^3 },
\]
for any $t \in \R$, does not contain a positive-area piece of $S$, meaning that no part of the area of $S$ is space-like. Then, $S \restrict \{\tbf = t\}$ is in fact a curve and we would like to consider it as a \mbox{$1$-current}. To do this, we need an orientation. This can be achieved by defining the \textbf{time-slice} $S|_t$ of the surface $S$ at time $t$ as the \mbox{$1$-current}
\begin{equation} \label{eq:slice_by_cylinder}
  S|_t := \partial(S \restrict \{\tbf < t\}) - (\partial S) \restrict \{\tbf < t\},
\end{equation}
with the boundary operator $\partial$ defined in the previous section; see Fig.~\ref{fig:slip_trajectory_slice} for an illustration of the slicing operation.

The Radon--Nikodym decomposition of the slice $S|_t$ will be written as
\[
  S|_t = \vec{S}|_t \, \tv{S|_t},
\]
with orienting \mbox{($1$-)vector} $\vec{S}|_t$ and the \enquote{curve measure} $\tv{S|_t}$ (representing integration over the curve), which lies in the plane $\{\tbf = t\} \subseteq \R^{1+3}$. To be notationally precise, we could have written the orienting vector $\vec{S}|_t$ in the more cumbersome way $\overrightarrow{S|_t}$ (restriction first, \emph{then} the overset arrow) to emphasize that it is the orienting \mbox{($1$-)vector} of $S|_t$, but one may in fact interpret $\vec{S}|_t$ as a type of \enquote{restriction} of $\vec{S}$, so we will in the following always use $\vec{S}|_t$ for convenience.

Note that $S|_t$ for a fixed time $t$ is entirely contained in the plane $\{\tbf = t\}$. Often it will be convenient to \enquote{forget} the $t$-component and just translate $S|_t$ to $\R^3$. We can do this by considering the pushforward under the space projection $\pbf$, which was defined in~\eqref{eq:tbf_pbf_def}, to set
\[
  S(t) := \pbf_*(S|_t).
\]
Unwinding the definitions, this means that $S(t)$ acts on a (smooth) differential $1$-form $\omega \colon \R^3 \to \Wedge^1 \R^3$ via
\[
  S(t)[\omega] = \int \dprb{\pbf\vec{S}|_t, \omega \circ \pbf} \dd \tv{S|_t}.
\]
For the avoidance of confusion, let us fix that the symbol $\vec{S}(t)$ will always denote the orienting $2$-vector of $S$, as a function on the plane $\{\tbf = t\}$, and will never be used to denote the orienting \mbox{($1$-)vector} $\overrightarrow{S(t)}$ of $S(t)$.

It can be checked very easily from the definitions that boundaries, pushforwards, restrictions, and slices are linear in the currents, so that linear operations \enquote{pull out} of the operation, e.g.,
\[
  \partial (\alpha S + \beta \tilde{S}) = \alpha \partial S + \beta \partial \tilde{S}
\]
and likewise for the other operations.

\subsection{Smooth currents} \label{sc:smooth_currents}

In this section we want to consider currents and their operations in the more familiar case of (smooth) vector and multi-vector fields. This builds a bridge from the geometric theory seen so far to classical vector analysis and also relates to dislocation \emph{fields}.

So, we consider \mbox{$k$-currents} in $\R^N$ ($k = 0,\ldots,N$) that can be written as
\[
  S = \vec{S}(x) \, \Lcal^N(\di x),
\]
where $\vec{S} \colon \R^N \to \Wedge_k \R^N$ and \enquote{$\Lcal^N(\di x)$} is the volume (Lebesgue) measure on $\R^N$. This means that $S$ acts on a $k$-form $\omega \colon \R^N \to \Wedge^k \R^N$ that is non-zero only on a bounded set, by integration, that is,
\[
  S[\omega] = \int \dprb{\vec{S}(x), \omega(x)} \dd x.
\]

First, we investigate the pushforward $\theta_* T$ of a smooth \mbox{$1$-current} $T = \vec{T} \, \Lcal^N$ under a smooth map $\theta \colon \R^N \to \R^M$. By definition, see~\eqref{eq:pushforward_def},
\[
  (\theta_* T)[\omega] = \int \dprb{\DD \theta(x)[\vec{T}(x)], \omega(\theta(x))} \dd x
\]
for any $1$-form $\omega$ on $\R^M$. Let $\theta$ be invertible with smooth inverse. Then, we can change variables to $\tilde{x} = \theta(x)$, for which
\[
  \di \tilde{x} 
  = \abs{\det \nabla \theta(x)} \; \di x,
  \qquad\text{or, equivalently,}\qquad
  \di x = \abs{\det \nabla \theta^{-1}(\tilde x)} \; \di \tilde{x},
\]
to rewrite this as
\[
  (\theta_* T)[\omega] = \int \dprb{\DD \theta(\theta^{-1}(\tilde{x}))[\vec{T}(\theta^{-1}(\tilde{x}))], \omega(\tilde{x})} \, \abs{\det \nabla \theta^{-1}(\tilde x)} \dd \tilde{x}.
\]
Hence, also $\theta_* T$ is a smooth current, namely
\begin{equation} \label{eq:smooth_pushforward}
  \theta_* T = \abs{\det \nabla \theta^{-1}} \, (\DD \theta \circ \theta^{-1})[\vec{T} \circ \theta^{-1}] \; \Lcal^N.
\end{equation}
The same formula also holds for a smooth \mbox{$2$-current} $S$, extending the differential $\DD \theta$ as usual by \enquote{pulling out the wedge}.

If $T = \vec{T} \, \Lcal^N$ is a smooth \mbox{$1$-current}, we can use the Stokes theorem above to compute as follows for a $0$-form $\omega \colon \R^N \to \R$ (that is, a function $\omega = \omega(x)$):
\begin{align*}
  (\partial T)[\omega]
  &= T[\di \omega] \\
  &= \int \dprb{\vec{T}(x), \di\omega(x)} \dd x \\
  &= \int \vec{T}(x) \cdot \nabla \omega(x) \dd x.
\end{align*}
Then, assuming that $\omega$ has bounded support, we use the divergence theorem to conclude
\[
  (\partial T)[\omega] = \int - \dive \vec{T}(x) \cdot \omega(x) \dd x.
\]
So, $\partial T$ is a smooth $0$-current given as
\[
  \partial T = (-\dive \vec{T}) \, \Lcal^N.
\]
In this way we see that for \mbox{$1$-currents} the boundary operator satisfies
\begin{equation} \label{eq:partial_div}
  \partial = -\dive
\end{equation}
and one can think of the boundary operator as a \enquote{geometric divergence}.

Finally, we also compute for a \mbox{$2$-current} $S = \vec{S} \, \Lcal^3$ that
\[
  S(t) = \pbf_*(S|_t) = \vec{S}(t,x) \, \Lcal^3(\di x),
\]
which motivated the notation \enquote{$S(t)$} in the first place.

\section{Dislocations and plastic flow} \label{sc:dislocations}

We now describe dislocations and their trajectories via the Space-Time Framework developed in the previous section. Then we will consider how the plastic distortion $P$ changes as a consequence of dislocation flux.

\subsection{Dislocation systems} \label{sc:dislocation_systems}

We will need to work with different \enquote{types} of vectors, for which we need to introduce a bit of terminology: Each referential point $x \in \Omega$ comes with a tangent space $\Trm_x \Omega$, which by definition contains all vectors occurring as tangent vectors of curves lying entirely in $\Omega$ and passing through $x$. Since $\Omega$ was assumed to be open, it contains a small ball around any point $x \in \Omega$ and so all directions in $\R^3$ are possible as tangent vectors, whereby we may identify $\Trm_x \Omega$ with $\R^3$. Vectors in the tangent space $\Trm_x \Omega$ are called \textbf{referential vectors} (attached at the point $x$ in the reference configuration $\Omega$).

Moreover, in a simple cubic crystal we may index the position of the atoms around a referential point $x \in \Omega$ via abstract \textbf{lattice coordinates} from $\Z^3$, with the origin corresponding to the atom at the point $x$. In these coordinates, the vectors $\pm \ee_1$, $\pm \ee_2$ and $\pm \ee_3$ (where $\ee_1, \ee_2, \ee_3$ denote the canonical unit vectors in the three axes) represent a translation to the nearest neighbor atoms in the lattice. To get to a next-to-nearest neighbor atom, we would have to go a distance $\pm 2 \ee_i$ or $\pm \ee_i \pm \ee_j$ (with $i,j \in \{1,2,3\}$), and so on. Vectors in the lattice coordinate system are called \textbf{lattice vectors}. For non-cubic crystals we use lattice vectors in $\R^3$; see~\cite{HullBacon11book,AndersonHirthLothe17book} for basic crystallography.

For a given material (and physical state), only a few possibilities for the Burgers vector of a dislocation are observed experimentally, so we now make the following \textbf{single-crystal} assumption:

\begin{hypothesis}
There is a fixed finite set $\Bcal = \bigl\{ \pm b_1, \ldots, \pm b_n \} \subseteq \R^3 \setminus \{0\}$ containing all possible Burgers vectors, which are all given as lattice vectors.
\end{hypothesis}

Since lattice vectors reflect the lattice structure of the material, the set $\Bcal$ should be assumed to be invariant under any point symmetries of the crystal.

The following is a well-known property of dislocations:

\begin{hypothesis}
Dislocation defects always occur as \emph{lines}, that is, curves going through the crystal that are either loops or end at the boundary of the specimen (respectively, the boundary of a grain in a polycrystal). The Burgers vector $b$ is constant along every dislocation line.
\end{hypothesis}

Let us denote the \textbf{lattice length scale}, which is on the order of the average distance between atoms, by $\eps > 0$. We collect all dislocations with Burgers vector $b$ into a formal sum
\begin{equation} \label{eq:Tb_formal_sum}
  T^b 
  = \sum_k T^b_k
  = \sum_k \vec{T}^b_k \, \eps^2 m^b_k \gamma^b_k,
\end{equation}
where $\gamma^b_k$ is an oriented curve in the reference configuration, $\vec{T}^b_k$ is its tangent vector, and $m^b_k \in \Z \setminus \{0\}$ is a multiplicity (technically, $\eps^2 m^b_k$ is the multiplicity).

The reason for the $\eps^2$-weighting is that one should think of every line as an abstraction of a cylinder with radius on the order of the lattice length scale, that is, $\eps / \sqrt{\pi}$, in the referential coordinate system, giving a total weight of $\eps^2$ (which is then multiplied with the length).

The multiplicity $m^b_k$ determines \enquote{how many times} the curve is to be counted. Negative multiplicities correspond to reversed curves. Indeed, reversing a curve flips the sign of $\vec{T}^b_k$, so in the formal sum for $T^b$ it has the same effect as flipping the sign of $m^b_k$. 

For every dislocation curve with Burgers vector $b$, the reversed curve is a dislocation curve with Burgers vector $-b$. Consequently,
\begin{equation} \label{eq:Tb_sym}
  T^{-b} = -T^b  \qquad\text{for all $b \in \Bcal$.}
\end{equation}

As part of the above hypothesis, all curves in $T^b$ are either closed loops or end at the boundary of the crystal. We express this statement formally as
\[
  \partial T^b \restrict \Omega = 0  \qquad\text{for all $b \in \Bcal$.}
\]
Here, $\partial T^b$ denotes the boundary of $T^b$, which is the formal sum of the point masses at the end points of the curves in $T^b$ with their sign induced by the orientation of the respective curve ($+1$ at the \enquote{end} and $-1$ at the \enquote{start}); see Section~\ref{sc:boundary} for details.

We also denote by $\vec{T}^b$ the (referential) unit \textbf{tangent vector} along a curve in $T^b$; this is well-defined for almost every point in the curves $T^b$ since even if there are many lines in $T^b$, they either lie on top of each other (hence giving the same tangent) or cross in a \emph{single} point. This fact can be made rigorous within the theory of currents, but we do not pursue this further here.

The totality of all dislocations is given via a (referential) \textbf{dislocation system}
\[
  \Tb = (T^b)_{b\in\Bcal},
\]
where every $T^b$ is the set of dislocations with Burgers vector $b$, as defined above. Without going into the mathematical intricacies, the object $T^b$ in~\eqref{eq:Tb_formal_sum} is called a \enquote{\mbox{$1$-current}}. For the moment, however, the above purely formal expression will suffice. Note that by~\eqref{eq:Tb_sym} every physical dislocation line appears \emph{twice} in $\Tb$, once for Burgers vector $b$ and once for $-b$.

Assume for the sake of the following argument that $T^b$ represents only a \emph{single} dislocation loop. The type of the dislocation at a referential point $x$ is described by the relationship between $\vec{T}^b(x)$ and the Burgers vector $b$, but these vectors are of a different nature: $b$ is a lattice vector whereas $\vec{T}^b(x)$ is a referential vector. The correct way to compare them is therefore to map $\vec{T}^b$ into the lattice coordinates via the plastic distortion $P$, giving the curve tangent relative to the lattice. Moreover, $b$ is constant along the dislocation, but $P\vec{T}^b$ varies with the point, so a dislocation can have different type at different points.

\begin{figure}
  \centering
  \includegraphics{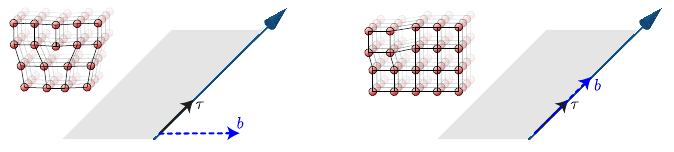}
  \caption{Edge and screw dislocations, where $\tau := P\vec{T}^b(x)$.}
  \label{fig:dislocation_types}
\end{figure}

We have the following terminology:
\begin{enumerate}[(i)]
  \item If $P\vec{T}^b(x)$ and $b$ are orthogonal at a point $x \in \Omega$, then $T^b$ is an \textbf{edge dislocation} at $x$; see the left of Fig.~\ref{fig:dislocation_types} for an illustration.
  \item If $P\vec{T}^b(x)$ and $b$ are parallel at a point $x \in \Omega$, then $T^b$ is a \textbf{screw dislocation} at $x$; see the right of Fig.~\ref{fig:dislocation_types} for an illustration.
  \item Otherwise the dislocation is of \textbf{intermediate type} at $x \in \Omega$.
\end{enumerate}

Since the Burgers vector is constant along the dislocation, the type varies with the tangent vector, so a dislocation will often be of edge-type in some parts and of screw-type in others. Note also that the movement of other ambient dislocations may change the type of a fixed dislocation loop since this movement changes $P$.

We will also consider dislocation systems which are not discrete. In this case we have instead
\begin{equation} \label{eq:Tb_density}
  T^b = \vec{T}^b \, \rho^b,
\end{equation}
where we now have a \emph{density} $\rho^b$ of continuously-distributed dislocations with tangent vector field $\vec{T}^b$. The notion of dislocation system is then suitably adapted. The density is to be understood in the reference coordinate system, which means that a single line corresponds to a unit-density cylinder with radius $\eps / \sqrt{\pi}$, which is the same as an abstract singular line with weight $\eps^2$. So, a single dislocation line could alternatively be thought to have the \emph{singular} density $\vec{T}^b_k \, \eps^2 m^b_k \gamma^b_k$, which technically is only a measure, not a vector field. From now on we will not make a distinction between continuous and singular densities.

One may combine the dislocations for all Burgers vectors into the \textbf{Nye--Kr\"{o}ner dislocation density tensor}
\begin{equation} \label{eq:alpha}
  \alpha := \frac12 \sum_{b \in \Bcal} b \otimes T^b.
\end{equation}

Note that the symmetry condition~\eqref{eq:Tb_sym} entails that every dislocation occurs exactly twice in $\Tb = (T^b)_{b\in\Bcal}$, and hence we introduce a factor of $\frac12$ in the sum to cancel this double counting. If we are in the situation of discrete dislocations, that is,~\eqref{eq:Tb_formal_sum}, then
\[
  \alpha = \frac12 \sum_{b \in \Bcal} \sum_k b \otimes \vec{T}^b_k \, \eps^2 m^b_k \gamma^b_k,
\]
whereas in the situation of dislocation densities, that is,~\eqref{eq:Tb_density}, we have
\begin{equation} \label{eq:alpha_density}
  \alpha = \frac12 \sum_{b \in \Bcal} b \otimes \vec{T}^b \, \rho^b.
\end{equation}

\begin{remark} \label{rem:sign}
Much of the existing literature instead uses the opposite sign convention for the Burgers vector. However, when we look at the dynamics of dislocations below we will see that our sign convention (where the Burgers vector in our sense equals the shear direction) avoids a proliferation of minus signs.
\end{remark}

\begin{remark}
It is a key difference between the approach presented in this work and more classical approaches that in our choice of internal variables we in fact do \emph{not} combine dislocations with different Burgers vectors into the Nye--Kr\"{o}ner dislocation density tensor $\alpha$. The reason for this choice is that in general there is no uniqueness for the decomposition of a matrix with rank more than one into dyadic products (all of which have rank one). Using only $\alpha$ as an internal variable, one cannot reconstruct the splitting along Burgers vectors on the right-hand side of~\eqref{eq:alpha_density}, at least in the case of densities (for discrete dislocations, the decomposition can be performed geometrically). This would then create an ambiguity in the definition of the motion of dislocation fields. It also turns out that the evolutionary system does not \enquote{factor through $\alpha$}; also see Remark~\ref{rem:commutators}.
\end{remark}

\subsection{Slip trajectories} \label{sc:slip_trajectories}

\begin{figure}[tb]
  \centering
  \includegraphics{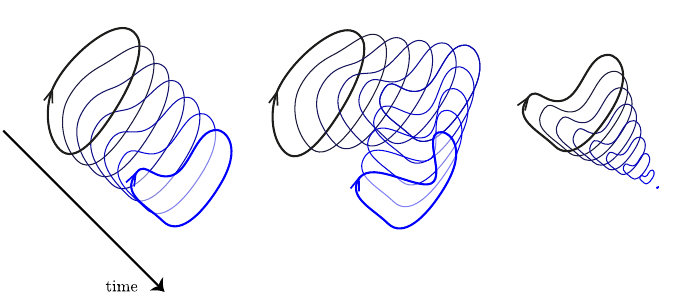}
  \caption{Slip trajectories.} 
  \label{fig:slip_trajectories}
\end{figure}

We now have all the geometric language available to describe dislocation motion and the associated plastic slip. Consider one fixed dislocation line that at time $t$ lies on a curve $s \in [0,1] \mapsto \gamma(t,s)$ with tangent vector $\tau = \tau(t,x)$. For each $t$, the curve $s \mapsto (t,\gamma(t,s))$ lies in the plane $\{\tbf = t\} = \{t\} \times \R^3$ and $\tau$ is space-like, i.e., $\tau \in \{0\} \times \R^3$. Gluing all these lines for $t \in \R$ together yields a $2$-dimensional surface $S$ in (Galilean) space-time $\R^{1+3}$. We consider this surface as a \mbox{$2$-current} and call it the \textbf{slip trajectory} associated with the dislocation motion; see Fig.~\ref{fig:slip_trajectories} for some examples.

Conversely, consider a $2$-dimensional surface $S$ in space-time $\R^{1+3}$. To get back to the dislocation curves $\gamma(t,\frarg)$, interpreted as \mbox{$1$-currents}, we use the slicing operation from the preceding section, setting
\begin{equation} \label{eq:gamma_St}
  \gamma(t) = S(t) := \pbf_*(S|_t),
\end{equation}
whenever this is well-defined. Here recall that $S|_t$ is the slice of $S$ by the plane $\{\tbf = t\}$ and the pushforward $\pbf_*(S|_t)$ then \enquote{forgets} the $t$-coordinate, so that we obtain a curve lying in $\R^3$ (and not in $\{\tbf = t\} = \{t\} \times \R^3$). Slip trajectories and their slices are illustrated in Fig.~\ref{fig:slip_trajectory_slice}.

This change in the point of view, namely that the slip trajectory is the \emph{fundamental} object and the dislocation lines are the \emph{derived} objects, has a number of benefits, as we will see shortly.

\begin{figure}[tb]
  \centering
  \includegraphics{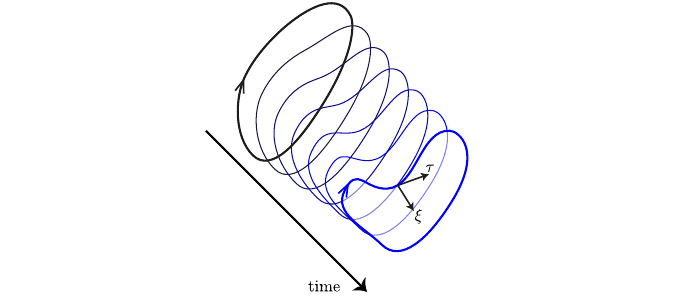}
  \caption{The tangent vectors at a slip trajectory.} 
  \label{fig:slip_trajectory_vectors}
\end{figure}

However, starting with slip trajectories and not the individual curves brings with it the issue that not every $2$-dimensional surface in $\R^{1+3}$ is admissible: In order for the slices $S|_t$ to be well-defined as \mbox{$1$-currents}, we need to ensure that the slip trajectory $S$ has no space-like parts, that is, non-zero area contained in a time-slice $\{\tbf = t\}$. For this, we first remark that at each $(t,x) \in S$ the tangent space $\Trm_{(t,x)}S$ to $S$ is spanned by two vectors. Since every point $(t,x)$ lies on precisely one curve $\gamma(t)$, we may take the curve tangent, denoted by $\tau = \tau(t,x)$, as one of these spanning vectors. Denoting the unique \textbf{displacement vector} that completes $\tau$ to an orthonormal basis of $\Trm_{(t,x)}S$ by $\xi = \xi(t,x)$, we thus have
\begin{equation} \label{eq:Tan_surface}
  \Trm_{(t,x)}S = \spn\bigl\{ \tau(t,x), \xi(t,x) \bigr\},  \qquad \xi \cdot \tau = 0, \; \abs{\tau} = \abs{\xi} = 1.
\end{equation}
See Fig.~\ref{fig:slip_trajectory_vectors} for an illustration.

Recalling the definition of time projection $\tbf$ and the spatial projection $\pbf$ from~\eqref{eq:tbf_pbf_def}, we then henceforth require the condition
\begin{equation} \label{eq:xi_forward}
  \tbf\xi = \xi \cdot \ee_0 > 0.
\end{equation}
This ensures that the slices $S|_t$ are indeed $1$-dimensional currents and we may use~\eqref{eq:gamma_St} to \emph{define} the dislocation curves $\gamma(t)$.

We now generalize from individual dislocation slip trajectories to sums thereof. Since a Burgers vector is attached to the whole of a dislocation line and does not change as it moves, we need one dislocation \mbox{$2$-current}, denoted by $S^b$, for each Burgers vector $b \in \Bcal$. By the definition of $2$-currents, this $S^b$ is a sum of \emph{all} slip trajectories for the dislocation lines with this Burgers vector. In fact, we will also call $S^b$ a \enquote{slip trajectory}.

Then define the \textbf{(referential) slip trajectory system} as the collection
\[
  \Sb = (S^b)_{b\in\Bcal},
\]
of \mbox{$2$-currents}
\[
  S^b = \vec{S}^b \, \tv{S^b}
\]
in $\R^{1+3}$ with the property that all the $S^b$ are contained in $[0,T] \times \Omega$, where $\Omega$ is our specimen domain. The dislocation system $\Tb = (T^b(t))_{b \in \Bcal}$ at time $t$ can then be found via the slicing operation from the preceding section, so that
\[
  T^b(t) = S^b(t) = \pbf_*(S^b|_t),
\]
which generalizes~\eqref{eq:gamma_St} in a natural way.

As before, we require~\eqref{eq:xi_forward} for every $S^b$. Even though there are now many surfaces in $S^b$, this expression is not ambiguous since the tangent space of $S^b$ is well-defined via the span of the orienting $2$-vector $\vec{S}^b$, at least if $\vec{S}^b$ is simple, that is $\vec{S}^b = \xi \wedge \tau$. 

We also assume
\begin{equation} \label{eq:Sb_sym}
  S^{-b} = -S^b  \qquad\text{for all $b\in\Bcal$}
\end{equation}
and that $S^b$ has no boundary on the interior of the space-time cylinder, i.e.
\begin{equation} \label{eq:Sb_bdry}
  (\partial S^b) \restrict (0,T) \times \Omega = 0.
\end{equation}
From~\eqref{eq:Sb_sym} we also immediately derive the symmetry condition
\[
  S^{-b}(t) = -S^b(t)  \qquad\text{for all $b \in \Bcal$.}
\]
For the boundaries of the dislocations, it then follows via the identity $\partial(S|_t) = -(\partial S)|_t$ that
\[
  \partial S^b(t) \restrict \Omega = 0, 
\]
which expresses the requirement that dislocations are composed of loops inside $\Omega$ or end at the boundary $\partial \Omega$.

The slip trajectory $S^b$ can also be understood as \enquote{acting} on $S^b(t_0)$ to transport it to $S^b(t)$ for any times $t_0 < t$ as follows:
\begin{equation} \label{eq:S_forward}
  S^b(t_0) + \pbf_*\partial \bigl( S^b\restrict\{t_0<\tbf<t\} \bigr)
  = S^b(t_0) + \pbf_* \bigl( S^b|_t - S^b|_{t_0} \bigr)
  = S^b(t).
\end{equation}
Here we have used the definition of slices from~\eqref{eq:slice_by_cylinder} and the assumption~\eqref{eq:Sb_bdry}.

If we constrain all slip trajectories $S^b$ to be collections of surfaces (and not densities thereof), we speak of a \textbf{discrete dislocation model}. In this case, the $S(t)$ are collections of \emph{curves} for all times $t$. Otherwise, we speak of a \textbf{field dislocation model}. 

\begin{remark}
Some works, such as~\cite{ScalaVanGoethem19}, have used time-indexed \mbox{$1$-currents} $(T^b_t)_{t \geq 0}$ to model dislocation motion (without their plastic effects, however). Unfortunately, time-indexed families of geometric objects lack a lot of information that is required to describe the plastic effect of dislocation motion. Most notably, it is difficult to compute the (normal) velocity of a moving curve since it is not clear which \enquote{points} on the curve should be compared in a difference quotient. Moreover, it is unclear what regularity of curves (both in space and time) is required to make such computations valid. Additionally, the dissipation \enquote{distance} between two curves, which is roughly proportional to the traversed area (but without canceling patches traversed both forward and backward), can only be defined implicitly (via a minimization) in this description.  
\end{remark}

\begin{remark}
Moreover, one can (and should) ask whether the space-time formulation has the same expressive power as working directly with time-indexed \mbox{$1$-currents} $(T^b_t)_{t \geq 0}$, without considering them to be the slices of space-time trajectory $S^b$. In more mathematical terms, this is essentially the question whether one can always glue a collection $(T^b_t)_{t>0}$ together into a $2$-surface $S^b$ such that $S^b(t) = T^b_t$ for all $t \geq 0$. This turns out to be true, as is shown by the Rademacher-type differentiability theorems of~\cite{BonicattoDelNinRindler25}, up to dealing with a number of regularity issues (which are, however, more on the level of excluding pathologies than genuine restrictions). 
\end{remark}

\subsection{Plastic flow} \label{sc:plastic_flow}

The formulation of dislocation motion via slip trajectories in space-time allows us to define the velocity of dislocations in a direct way: Given a slip trajectory $S^b$ consisting of only one $2$-dimensional surface, recall the definition of the displacement vector $\xi^b(t,x)$ via~\eqref{eq:Tan_surface},~\eqref{eq:xi_forward}; also see Fig.~\ref{fig:slip_trajectory_vectors} for an illustration. We then define the \textbf{(referential, normal) dislocation velocity} of $S^b$ at a point $(t,x)$ via
\begin{equation} \label{eq:geometric_derivative}
  \frac{\DD}{\DD t} S^b \bigg|_t(x)
  := \frac{\pbf\xi^b(t,x)}{\tbf\xi^b(t,x)} \in \R^3.
\end{equation}
Indeed, noting that $\pbf\xi^b$ is the spatial displacement of the dislocation per $\tbf\xi^b$ units of time, the above formula represents the velocity of the dislocation. To ensure that $\frac{\DD}{\DD t} S^b$ is well-defined, we need to require the regularity condition~\eqref{eq:xi_forward} that we already needed to define the slices. The quantity $\frac{\DD}{\DD t} S^b$ is called the \enquote{geometric derivative} of $S^b$ in~\cite{BonicattoDelNinRindler25}.

The dislocation velocity is orthogonal to the dislocation tangent $\tau^b$, or, more precisely to $\hat{\tau}^b$ if we write $\tau^b = (0,\hat{\tau}^b)^\T$. Indeed, by construction, $\xi^b$ and $\tau^b$ are orthogonal, whereby
\[
  0 = \xi^b \cdot \tau^b = \pbf\xi^b \cdot \hat{\tau}^b = \tbf\xi^b \cdot \biggl( \frac{\DD}{\DD t} S^b \cdot \hat{\tau}^b \biggr).
\]
From this the orthogonality of $\frac{\DD}{\DD t} S^b$ and $\hat{\tau}^b$ follows since $\tbf\xi^b > 0$ by~\eqref{eq:xi_forward}. This orthogonality expresses that the dislocation velocity is \emph{normal}, that is, it does not have a component along the tangent of the curve (which would have no physical meaning). In the following we identify $\tau^b$ and $\hat{\tau}^b$.

We can directly extend the definition of the dislocation velocity (geometric derivative) from surfaces to \mbox{$2$-currents} $S^b$, at least if they have a \emph{simple} orienting $2$-vector $\vec{S}^b = \xi^b \wedge \tau^b$. Indeed, if $\tau^b = (0,\pbf\tau^b)$ is a space-like tangent vector to $S^b$, which is unique up to choice of sign, then $\xi^b$ is uniquely determined, again up to choice of sign. The sign choice is in fact also determined by our requirement that $\tbf\xi^b > 0$ by~\eqref{eq:xi_forward} since $\xi^b$ and $\tau$ flip their sign in tandem. It is possible to write a concrete expression for $\xi^b$ as a function of $\vec{S}^b$ and $\tau^b$, but this requires additional notation, so it is omitted here.

The \textbf{(referential) slip normal rate} $g^b \colon \Omega \to \R^3$ is defined to be the vector field that is normal (in the referential frame) to the dislocation slip motion, with magnitude proportional to the velocity and density of flowing dislocations, that is, the \emph{slip flux}. As we have seen above, in the Space-Time Framework the dislocation velocity is given by $\frac{\DD}{\DD t} S^b|_t$, which was defined in~\eqref{eq:geometric_derivative}. Since the infinitesimal slip plane is spanned by the velocity $v^b$ and the tangent direction (with multiplicity) $\tau^b$, one obtains
\begin{equation} \label{eq:gb_cross_currents}
  g^b(t)
  = \bigl( v^b(t) \times \tau^b(t) \bigr) \, \tv{S^b(t)}
  = \biggl(\frac{\DD}{\DD t} S^b \bigg|_t \times \tau^b(t) \biggr) \, \tv{S^b(t)}.
\end{equation}
As above, $\tau^b(t) \in \R^3$ here denotes the tangent vector to the slice $S^b(t)$, and $\tv{S^b(t)}$ is the \enquote{measure} of all the curves together with their multiplicities. Note that here we have omitted the $x$-arguments for readability (and because these formulas only really make rigorous sense in a measure-theoretic way).

The following assumption describes the evolution of the plastic distortion (from the Kr\"{o}ner--Lee decomposition $\nabla y = EP$). It is derived in detail in~\cite{HudsonRindler22}, but it is also classically known, see Remark~\ref{rem:choice_of_quantities} below.

\begin{hypothesis}
The plastic distortion evolves according to the (pointwise) differential equation
\begin{equation} \label{eq:plastic_flow}
  \dot{P}
  = \frac12 \sum_b b \otimes g^b,
\end{equation}
with $b \otimes g^b := b \, (g^b)^T$ the dyadic product of the Burgers vector $b$ and the associated slip normal rate $g^b$.
\end{hypothesis}

If the drift is concentrated on a surface, the above formula must be understood with $\dot{P}$ and $g^b$ being \emph{measures} in the mathematical sense. This means that $\dot{P}$ assigns a compound drift matrix to any (measurable) subset of $\Omega$. Unlike vector fields, this allows one to assign non-zero values of drift and plastic distortion to surfaces (which have zero three-dimensional volume and are thus \enquote{negligible} from the view of fields). 

We will now express the right-hand side of~\eqref{eq:plastic_flow} via the slip trajectories $S^b$. For this we first rewrite~\eqref{eq:gb_cross_currents} using the (spatial) Hodge operator \enquote{$\hodge$} as
\[
  g^b(t) = \hodge \biggl[\frac{\DD}{\DD t} S^b \bigg|_t \wedge \tau^b(t) \biggr] \, \tv{S^b(t)},
\]
which holds since $\hodge (v \wedge w) = v \times w$ by~\eqref{eq:hodge_wedge}. Then we plug in the definition of $\frac{\DD}{\DD t} S^b$ and use that
\[
  \tau^b(t) = \pbf(\vec{S}^b|_t),
\]
with $\vec{S}^b|_t$ the orienting vector of the time-slice of $S^b$ at $t$, to compute
\[
  \frac{\DD}{\DD t} S^b \bigg|_t \wedge \tau^b = \frac{\pbf\xi^b}{\tbf\xi^b} \wedge \pbf(\vec{S}^b|_t).
\]
Thus, we arrive at
\begin{equation} \label{eq:gb_geometric}
  g^b(t) = \hodge \biggl[\frac{\pbf\xi^b}{\tbf\xi^b} \wedge \pbf(\vec{S}^b|_t)\biggr] \, \tv{S^b(t)}.
\end{equation}
Here we have liberally omitted arguments to keep the formulas concise. The preceding equality should really have been written in \enquote{measure-theory notation} as
\[
  g^b(t, \di x) = \hodge \biggl[\frac{\pbf\xi^b}{\tbf\xi^b} \wedge \pbf(\vec{S}^b|_t)\biggr] \, \tv{S^b(t)}(\di x)
\]
since $\tv{S^b(t)}$ is a measure that integrates over $x$. However, also in the following we avoid these contortions and omit all arguments not absolutely required by the notation (such as the $t$ in \enquote{$S|_t$} or \enquote{$S(t)$}, which cannot be left out without distorting the meaning) to keep the formulas maximally readable.

Combining~\eqref{eq:gb_geometric} with the plastic flow equation~\eqref{eq:plastic_flow}, which was recalled above,
\begin{equation} \label{eq:dotP_geometric}
  \dot{P}(t) = \frac12 \sum_{b \in \Bcal} b \otimes \hodge \biggl[\frac{\pbf\xi^b}{\tbf\xi^b} \wedge \pbf(\vec{S}^b|_t)\biggr] \, \tv{S^b(t)}
\end{equation}
The expression inside the sum should be understood as follows: For each fixed $t$, $\tv{S^b(t)}$ is a collection of curves in space (loops inside $\Omega$), or more precisely the \enquote{measure} representing integration over these curves. This measure is multiplied by a field of space-like vectors, namely $\hodge \bigl[\frac{\pbf\xi^b}{\tbf\xi^b} \wedge \pbf(\vec{S}^b|_t)\bigr]$, and then tensored from the left with the Burgers vector $b$. This yields a \enquote{vector measure} on $\R^3$, which, if given by a density, can be identified with a field valued in the $\R^{3 \times 3}$-matrices.

Now we integrate~\eqref{eq:dotP_geometric} from $0$ to $t$ to obtain
\begin{align}
  P(t)
  &= P(0) + \int_0^t \dot{P}(s) \dd s  \notag\\
  &= P(0) + \frac12 \sum_{b \in \Bcal} b \otimes \hodge \int_0^t \pbf\xi^b \wedge \pbf(\vec{S}^b|_s) \, \tv{S^b(s)} \; \frac{\di s}{\abs{\tbf\xi^b}}. \label{eq:P_via_S_pre}
\end{align}
Here note that $\tbf\xi^b = \abs{\tbf\xi^b}$ by~\eqref{eq:xi_forward}. To transform this expression further, we first observe the algebraic identity
\begin{equation} \label{eq:P_via_S_part1}
  \pbf\xi^b \wedge \pbf(\vec{S}^b|_s)
  = \pbf [\xi^b \wedge \vec{S}^b|_s]
  = \pbf \vec{S}^b,
\end{equation}
where we have used that $\vec{S}^b = \xi^b \wedge \vec{S}^b|_s$ and also employed the extension of $\pbf$ to $2$-vectors by \enquote{pulling out the wedge}.

We furthermore observe
\begin{equation} \label{eq:P_via_S_part2}
  \int_0^t \pbf \vec{S}^b \, \tv{S^b(s)} \; \frac{\di s}{\abs{\tbf\xi^b}}
  = \pbf_* \!\left( \int_0^t \vec{S}^b \, \tv{S^b|_s} \; \frac{\di s}{\abs{\tbf\xi^b}} \right).
\end{equation}
by the definition of the pushforward and the linearity of the integral. This can be made rigorous by following the point of view of Section~\ref{sc:surfaces_to_currents}, namely to understand \mbox{$2$-currents} as linear functionals acting on differential $2$-forms $\omega = \omega(t,x)$, and to carefully write out both sides of the equality.

Finally, it can be shown that
\begin{equation} \label{eq:P_via_S_part3}
  \int_0^t \vec{S}^b \, \tv{S^b|_s} \; \frac{\di s}{\abs{\tbf\xi^b}} = S^b \restrict \{0<\mathbf t<t\}
\end{equation}
This can be proved via the so-called \emph{coarea formula}, but it can at least be explained intuitively here: The left integral is the \enquote{slice-by-slice} view on $S^b$, so it stands to reason that it is just a Fubini-type way to write $S^b$. Moreover, the factor $\abs{\tbf\xi^b}^{-1}$ gives the \enquote{steepness} of the space-time surface traced out by the $S^b|_s$: If the (local) \enquote{steepness} of the surface is high, $\abs{\tbf\xi^b}$ is small since $\xi^b$ is close to being space-like. So, $\abs{\tbf\xi^b}^{-1}$ is large, corresponding to the fact that $S^b$ concentrates a large surface area in a short time interval. On the other hand, if the local \enquote{steepness} of the surface is low, then an analogous argument shows that not much area of $S^b$ is concentrated in a short time interval.

Plugging~\eqref{eq:P_via_S_part1}--\eqref{eq:P_via_S_part3} into~\eqref{eq:P_via_S_pre}, we conclude
\begin{align}
  P(t)
  &= P(0) + \frac12 \sum_{b \in \Bcal} b \otimes \hodge \int_0^t \pbf \vec{S}^b \, \tv{S^b(s)} \; \frac{\di s}{\abs{\tbf\xi^b}}  \notag\\
  &= P(0) + \frac12 \sum_{b \in \Bcal} b \otimes \hodge \pbf_* \!\left( \int_0^t \vec{S}^b \, \tv{S^b|_s} \; \frac{\di s}{\abs{\tbf\xi^b}} \right) \notag\\
  &= P(0) + \frac12 \sum_{b \in \Bcal} b \otimes \hodge \pbf_*(S^b \restrict \{0 < \tbf < t\}). \label{eq:P_forward}
\end{align}
Since $\pbf_*(S^b \restrict \{0 < \tbf < t\})$ amounts to what is usually called the \textbf{slip surface}, that is, the surface in space over which dislocation motion has occurred in the time interval $(0,t)$, this formula expresses that the plastic distortion changes directly via the shear caused by the slip surface. One can think of~\eqref{eq:P_forward} as paralleling~\eqref{eq:S_forward} in that it describes the evolution by means of how $S$ \enquote{acts} on $P(0)$.

Also observe that the formula~\eqref{eq:P_forward} has the \textbf{semigroup property}, that is, for all times $t_0 < t$ it holds that
\[
  P(t) = P(t_0) + \frac12 \sum_{b \in \Bcal} b \otimes \hodge \pbf_*(S^b \restrict \{t_0 < \tbf < t\}).
\]
This means that we could \enquote{reinitialize} the flow at time $t_0$ and we would get the same dynamics, starting from $P(t_0)$ and with dislocation system $(S^b(t_0))_{b \in \Bcal}$. In conjunction with the plastic consistency condition, which is the topic of the next section, this shows that the start time $t = 0$ is not special and any other time could be chosen as well, without changing the dynamics. We call this the \textbf{reinitialization property} of the flow.

For later reference, we also record the formulas
\begin{equation} \label{eq:gb_geometric_nowedge}
  g^b(t) = \frac{\hodge \pbf \vec{S}^b}{\tbf\xi^b} \, \tv{S^b(t)}
\end{equation}
and
\begin{equation} \label{eq:dotP_geometric_nowedge}
  \dot{P}(t) = \frac12 \sum_{b \in \Bcal} b \otimes \biggl(\frac{\hodge \pbf \vec{S}^b}{\tbf\xi^b} \, \tv{S^b(t)}\biggr).
\end{equation}
For this, combine~\eqref{eq:gb_geometric} and~\eqref{eq:dotP_geometric} with~\eqref{eq:P_via_S_part1}.

\section{Space-time variation and dissipation} \label{sc:spacetime_variation}

\begin{figure}
  \centering
  \includegraphics{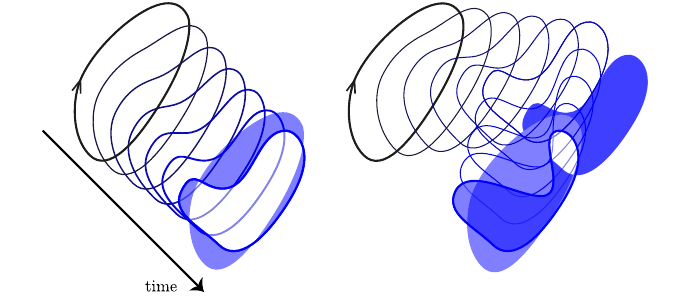}
  \caption{The variation of a slip trajectory in a simple and in a complex motion (with some parts counted twice).}
  \label{fig:variations}
\end{figure}

Let $S = \vec{S} \, \tv{S}$ be a slip trajectory, i.e., a \mbox{$2$-current} in $\R^{1+3}$. Then, the \textbf{space-time variation} (or just \textbf{variation}) of $S$ in the interval $[s,t]$ is defined to be
\begin{equation} \label{eq:variation}
  \Var(S;[s,t]) := \int_{[s,t] \times \Omega} \abs{\pbf\vec{S}} \dd \tv{S}.
\end{equation}
This formula was introduced in~\cite{Rindler23} and is motivated as follows: The pushforward $\pbf_* S$ of $S$ onto $\R^3$ can be expressed via~\eqref{eq:pushforward_def} as
\[
  \dprb{\pbf_* S, \omega} = \int \dprb{\pbf\vec{S}, \omega \circ \pbf} \dd \tv{S}.
\]
However, the pushforward $\pbf_* S$ has the property that if the space-time trajectory traverses an area and then again in the opposite direction, the two contributions \emph{cancel}. The variation adds a norm around the \enquote{oriented area element} $\pbf\vec{S}$ in order to count area \emph{absolutely}, without cancellation. In this way, the space-time variation measures the (absolute) area traversed by a moving dislocation. Figure~\ref{fig:variations} contains an illustration of the variation for two different motions.

The (absolute) area traversed by the dislocation is proportional to the bonds broken and reconnected (assuming they are spaced evenly), so the variation as defined above is an approximate measure of the dissipation. Of course, as the specimen deforms and the lattice is distorted, how one measures area needs to be adapted to the changing geometry of the specimen, but at least for small deformations, this may be neglected.

Next, we can use the coarea formula in a similar fashion as in~\eqref{eq:P_via_S_part3} to rewrite the space-time variation as
\begin{equation} \label{eq:variation_slices}
  \Var(S;[s,t]) = \int_s^t \int \abs{\pbf\vec{S}} \dd \tv{S(\tau)} \; \frac{\di \tau}{\abs{\tbf\xi}}.
\end{equation}
This of course assumes that we do not have any area of $S$ contained in any plane $\{\tbf = \tau\}$. Decomposing the $2$-vector $\vec{S}$ as before via
\[
  \vec{S} = \xi \wedge \vec{S}|_\tau,  \qquad \xi \perp \vec{S}|_\tau,  \qquad \abs{\vec{S}|_\tau} = \abs{\xi} = 1,  \qquad \tbf\xi > 0,
\]
we further see that
\begin{align*}
  \Var(S;[s,t]) 
  &= \int_s^t \int \absBB{\frac{\pbf\xi \wedge \pbf\vec{S}|_\tau}{\tbf\xi}} \dd \tv{S(\tau)} \dd \tau \\
  &= \int_s^t \int \absBB{\frac{\DD}{\DD \tau} S \wedge \pbf\vec{S}|_\tau} \dd \tv{S(\tau)} \dd \tau.
\end{align*}
Since $\frac{\DD}{\DD \tau} S$ and $\pbf\vec{S}|_\tau$ are orthogonal, the norm has the property that for $v,w$ orthogonal $\abs{v \wedge w} = \abs{v} \cdot \abs{w}$, as well as the fact that $\abs{\pbf\vec{S}|_\tau} = 1$, we arrive at
\[
  \Var(S;[s,t]) = \int_s^t \int \absBB{\frac{\DD}{\DD \tau} S} \dd \tv{S(\tau)} \dd \tau.
\]
In this formulation, instead of directly measuring the traversed area, one takes an \enquote{infinitesimal} point of view and looks at the \emph{rate} of traversing area (understood as the rate of breaking bonds). This involves the (normal) dislocation velocity integrated over the length of the dislocation line and then further integrating up this instantaneous area traversal over time. This should also correspond (again, approximately) to the dissipation and our argument deducing~\eqref{eq:variation_slices} from~\eqref{eq:variation} shows precisely this equality.

In~\cite{HudsonRindler22}, the \textbf{rate-independent dissipation} in a time interval $[s,t]$ was derived to be
\begin{equation} \label{eq:Diss1}
  \Diss_1([s,t]) = \frac12 \sum_{b\in\Bcal} \int_s^t \int_\Omega R^b_1(P^{-\T} g^b) \dd x \dd \tau,
\end{equation}
where $R^b_1 \colon \R^3 \to [0,\infty]$ is the \textbf{rate-independent dissipation potential}, which is always assumed to be positively $1$-homogeneous and convex. In our situation, for a single \mbox{$2$-current} and assuming that $R^b_1 = \abs{\frarg}$ (the simplest possible $1$-homogeneous convex function), as well as $P \approx \Id$ (close to the identity plastic distortion), this corresponds to the expression
\[
  \int_s^t \int \absBB{\frac{\hodge \pbf \vec{S}}{\tbf\xi}} \dd \tv{S(\tau)} \dd \tau
  = \int_s^t \int \absBB{\frac{\DD}{\DD \tau} S} \dd \tv{S(\tau)} \dd \tau.
\]
Here we have used similar arguments as before,~\eqref{eq:gb_geometric_nowedge}, and $\abs{\hodge v} = \abs{v}$. The right-hand side we have already identified above as $\Var(S;[s,t])$. Thus, we see that under the given simplifying assumptions the variation indeed expresses the dissipation.

\begin{remark}
For the sake of illustration, we will now see that the space-time variation really is an analogue of the classical variation of a differentiable function $u \colon [s,t] \to \R$. Indeed, for such an $u$, the associated \mbox{$1$-current} $S_u$ is the curve corresponding to the graph
\[
  \graph u := \setb{ (\tau,u(\tau)) }{ \tau \in [s,t] },
\]
which is oriented with the forward-pointing tangent field
\[
  \vec{S}_u = \frac{1}{\sqrt{1+\abs{\dot{u}}^2}} \begin{bmatrix} 1 \\ \dot{u} \end{bmatrix}.
\]
Then,
\[
  \pbf \vec{S}_u = \frac{\dot{u}}{\sqrt{1+\abs{\dot{u}}^2}},
\]
and so, invoking the classical area formula for graphs,
\[
  \int_s^t \abs{\dot{u}} \dd \tau
  = \int_{\graph u \cap ([s,t] \times \R)} \frac{\abs{\dot{u}}}{\sqrt{1+\abs{\dot{u}}^2}} \dd \Hcal^1 
  = \int_{\graph u \cap ([s,t] \times \R)} \abs{\pbf\vec{S}_g} \dd \Hcal^1.
\]
The first expression is the classical variation of $u$ and the last expression is the analogue of our space-time variation for the \mbox{$1$-current} $S_u$ (whose slices are points), which are thus seen to be equal.
\end{remark}

\subsection{Consistency}

Suppose that, starting from the identity plastic distortion, a dislocation with Burgers vector $b$ sweeps out a referential \textbf{slip surface} $H$ with oriented unit normal $N$, whose oriented boundary is a curve $\gamma$ with tangent vector field $\tau$. The resulting increment of the plastic distortion is
\[
  \Delta P = b \otimes N \, \Hcal^2 \restrict H,
\]
so that, in this idealized singular setting, $P=\Id+\Delta P$. Taking the distributional row-wise curl and choosing the orientation of $\tau$ induced by $N$, we obtain
\[
  \curl \Delta P
  = b \otimes \tau \, \Hcal^1 \restrict \gamma.
\]
Generalizing this to the dislocation system at time $t$, namely
\[
  \Sb(t) := (S^b(t))_b,
\]
where
\[
  S^b(t)=\tau^b(t)\,\tv{S^b(t)},
\]
we obtain the \textbf{plastic consistency condition}
\begin{equation} \label{eq:curlP_consistency}
  \curl P(t)
  = \frac12 \sum_{b\in\Bcal} b\otimes S^b(t)
  = \frac12 \sum_{b\in\Bcal}
    b\otimes\tau^b(t)\,\tv{S^b(t)}
  = \alpha(t).
\end{equation}
Here, $\alpha(t)$ is the Nye--Kr\"{o}ner dislocation density tensor defined in~\eqref{eq:alpha}, now generalized to the time-dependent plastic flow. Taking the time derivative of the consistency condition and using~\eqref{eq:dotP_geometric_nowedge} and~\eqref{eq:gb_cross_currents}, we obtain
\begin{align*}
  \dot\alpha(t)
  &= \curl\dot P(t) \\
  &= \frac12\sum_{b\in\Bcal} \curl\bigl(b\otimes g^b(t)\bigr) \\
  &= \frac12\sum_{b\in\Bcal} \curl \Bigl(b \otimes\bigl[(v^b(t)\times\tau^b(t))\,\tv{S^b(t)}\bigr] \Bigr) \\
  &= \frac12\sum_{b\in\Bcal} b \otimes \curl\bigl[(v^b(t)\times\tau^b(t))\,\tv{S^b(t)}\bigr],
\end{align*}
where the row-wise curl is understood distributionally and acts on the complete tensor-valued measure. Note that there is no \emph{single} velocity that advects $\alpha$; rather, a velocity $v^b$ is associated with each slip trajectory $S^b$, $b\in\Bcal$. In fact, one can show from the above that~\eqref{eq:curlP_consistency} is preserved along the flow; see~\cite{HudsonRindler22}.

\begin{remark}
The reason we need \emph{both} the plastic distortion $P(t)$ and the dislocation system $\Sb(t) = (S^b(t))_b$ as internal variables, despite them being so closely related by the above consistency relation~\eqref{eq:curlP_consistency}, is that neither variable contains enough information to recover the other: While the curl of $P$ can be computed from $\Sb(t)$, the plastic distortion $P(t)$ itself cannot be computed in this way as it is only determined up to a gradient. Conversely, the knowledge of $P(t)$ determines only the Nye--Kr\"{o}ner dislocation density tensor $\alpha$, but not the individual dislocations. This is not enough to describe the full dynamics, as can be seen from the following example.
\end{remark}

\begin{example}
Consider a material with a hexagonal close-packed lattice (which occurs, for instance, in titanium or zinc~\cite{Lubliner08}). Slip overwhelmingly occurs only on planes with normal $n=(0,0,1)$, but in 6 different slip directions, that is, for the Burgers vectors
\[
  \bigl(\pm1,0,0 \bigr),
  \bigl(\pm\tfrac12,\pm\tfrac{\sqrt{3}}2,0 \bigr),
  \bigl(\mp\tfrac12,\pm\tfrac{\sqrt{3}}2,0 \bigr).
\]
As we have more Burgers vectors than dimensions, different combinations of active dislocations could produce the same slip. For instance, the motion of a dislocation with Burgers vector $(1,0,0)$ over a plane with normal $(0,0,1)$ could alternatively be accomplished by the motion of a dislocation with Burgers vector $(\tfrac12,\tfrac{\sqrt{3}}2,0)$ and one with Burgers vector $(\tfrac12,-\tfrac{\sqrt{3}}2,0)$, both moving over the same plane as before. The point is that the resulting net slip, i.e., the change to $P$ and the change to the Nye--Kr\"{o}ner dislocation density tensor $\alpha$ are the same in all scenarios, but the evolution for $\Sb$ is different. In particular, the velocities and dissipation may well be different. In conclusion, the Nye--Kr\"{o}ner dislocation density tensor $\alpha$ contains too little information to build a well-defined geometric theory of dislocation motion and its associated elasto-plastic evolution.
\end{example}

\subsection{Modes of dislocation motion}

Consider a dislocation line $\gamma$ with tangent vector $\tau$ that moves with the referential velocity $v$, over a referential slip plane $H$ with (weighted) normal $N = v \times \tau$. The corresponding lattice slip plane is denoted by $H'$ and it has the (weighted) normal vector $N' = Pv \times P\tau = P^{-\T} N$ (assuming that $\det P \equiv 1$). One distinguishes the following two (pure) situations:

\begin{figure}[tb]
  \centering
  \includegraphics{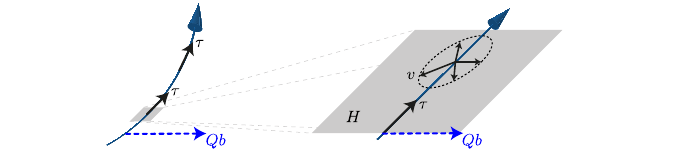}
  \caption{Allowed directions for edge dislocation glide.} 
  \label{fig:edge_glide}
\end{figure}

\begin{enumerate}[(i)]
  \item If the (lattice) Burgers vector $b$ lies in the plane $H'$, which corresponds to $N' \cdot b = 0$ (equivalently, $N \cdot (P^{-1}b) = 0$), the atoms can slide over each other fairly freely. This slip motion is called \textbf{glide}; see Fig.~\ref{fig:edge_glide} for an illustration.
  \item If $b$ is orthogonal to the plane $H'$, which corresponds to $N' \times b =  0$, then during slip the atoms need to be pushed up or down relative to the slip plane, which is much more expensive in terms of dissipated energy. This slip motion is called \textbf{climb}.
\end{enumerate}
In real materials, climb motion is much more energetically expensive, essentially because whole rows of atoms have to be moved as opposed to a shearing motion for glides. Mixed motions may also occur.

Notice that at points where our dislocation has screw character as defined in Section~\ref{sc:dislocation_systems}, $P\tau$ and $b$ are parallel, whereby (using the triple product rule)
\[
  N' \cdot b = (Pv \times P\tau) \cdot b = Pv \cdot (P\tau \times b) = 0.
\]
Hence, all motions of pure screw dislocations are glides. 

For the complementary case of a dislocation with an edge character, or at least non-zero edge component, it holds that $P\tau \times b \neq 0$. If the motion is a glide, it is necessary that
\[
  0 = N' \cdot b = (Pv \times P\tau) \cdot b = Pv \cdot (P\tau \times b).
\]
But this can only hold if $Pv$ lies in the plane spanned by both $P\tau$ and $b$. So, there is only \emph{one} velocity direction in which the motion of an edge dislocation is a glide; in all other directions the motion has at least partial climb character.

Consider now a \emph{curved} dislocation line that lies wholly in a (lattice) plane $H'$ and assume that its Burgers vector also lies in this plane. Then, the above arguments show that the velocity vector $v$ is constrained to lie in this plane $H$ wherever the dislocation has edge character. Since the dislocation is curved, it has edge character in many places and thus dislocation motion is \enquote{trapped} in the slip plane $H$, at least if the force does not exceed the threshold for dislocation climb.

If in the situation of a dislocation line lying in a slip plane $H$, however, a straight part of the dislocation that has screw character can move out of the slip plane $H$. This phenomenon is called \textbf{cross slip}. In practice, this also occurs for parts of the dislocation line where it has only \emph{almost} screw character; the very small edge component needs to climb, but this only contributes a minor amount to the dissipation.

\begin{remark} \label{rem:choice_of_quantities}
Let us compare our description of plastic flow to the one in~\cite{GurtinFriedAnand10book}. The (referential) slip normal rate for one Burgers vector $b$ can be expressed as
\[
  g^b = \nu^b \rho^b N^b,
\]
where $\rho^b = \rho^b(t,x)$ is the \textbf{slip field density} for the Burgers vector $b$, $N^b$ is the \textbf{(referential) slip plane normal}, and $\nu^b$ is the \textbf{slip rate}, that is, the speed at which the slip field is transported. All of these quantities are given in the referential frame. Note that $g^b$ can be directly expressed in the Space-Time Framework, both for individual dislocation lines and for densities thereof, whereas $\nu^b$, $N^b$, $\rho^b$ cannot be easily separated. The associated \textbf{Schmid tensor} is defined to be the dyadic product
\[
  b \otimes (\nu^b \rho^b {N^b}')
\]
with ${N^b}'$ the \emph{lattice (structural)} slip plane normal. Then, the structural plastic flow equation reads as
\[
  \dot{P}P^{-1} = L^p = \frac12 \sum_b b \otimes (\nu^b \rho^b {N^b}'),
\]
see, e.g., Section~106 in~\cite{GurtinFriedAnand10book}. It holds that $P^\T {N^b}' = N^b$ by the transformation rule for normals since $P$ transforms referential vectors to structural vectors, whereby $P^\T$ transforms structural normals to referential ones (see also~\eqref{eq:hodge_transform1}). Thus, we may multiply this relation by $P$ from the right to compute
\[
  \dot{P}
  = \frac12 \sum_b b \otimes \bigl[ (\nu^b \rho^b P^\T {N^b}') \bigr]
  = \frac12 \sum_b b \otimes \bigl[ (\nu^b \rho^b N^b) \bigr]
  = \frac12 \sum_b b \otimes g^b.
\]
Hence, the classical formulation in~\cite{GurtinFriedAnand10book} is equivalent to~\eqref{eq:plastic_flow} above.
\end{remark}

\section{Small-distortion geometric model} \label{sc:SDGM}

We now come to the second main theme of this work, namely how to modify the energy functionals according to the Small-Distortion Hypothesis and the Line-Tension Approximation.

\subsection{Multiplicative Helmholtz-type decomposition} \label{sc:multiplicative_Helmholtz}

Before we can precisely state the Small-Distortion Hypothesis, we need to consider multiplicative decompositions of the plastic distortion $P$ into a gradient and a residual non-gradient distortion. For this, we write
\begin{equation} \label{eq:multiplicative_Helmholtz_pre}
  P = Z_P \, \nabla y_p,
\end{equation}
with an orientation-preserving smoothly invertible \textbf{plastic deformation} $y_p \colon \Omega \to y_p(\Omega)$ (i.e., satisfying $\det \nabla y_p > 0$), and a matrix field $Z_P \colon \Omega \to \R^{3 \times 3}$. We denote the inverse map of $y_p$ by $y_p^{-1} \colon y_p(\Omega) \to \Omega$ and its gradient by $\nabla y_p^{-1}$. By the formula for the gradient of the inverse map we have
\begin{equation} \label{eq:gradient_inverse}
  \nabla y_p^{-1} \circ y_p = (\nabla y_p)^{-1},  \qquad
  (\nabla y_p \circ y_p^{-1})^{-1} = \nabla y_p^{-1}.
\end{equation}
The field $Z_P$ is unique because $\nabla y_p$ is an invertible matrix, and hence the splitting~\eqref{eq:multiplicative_Helmholtz_pre} is determined by the choice of $y_p$ only. Note further that a splitting as in~\eqref{eq:multiplicative_Helmholtz_pre} always exists (just choose $y_p = \id$ and $Z_P = P$), but below we will choose $\nabla y_p$ \enquote{maximally}, so that we isolate as much of the gradient part of $P$ as possible.
 
It is useful to rewrite the decomposition~\eqref{eq:multiplicative_Helmholtz_pre} as follows: Define the \textbf{residual plastic distortion} $P_\res \colon y_p(\Omega) \to \R^{3 \times 3}$ via
\[
  P_\res := (P \circ y_p^{-1}) \, \nabla y_p^{-1},
\]
meaning that $P_\res(z) := P(y_p^{-1}(z)) \, \nabla y_p^{-1}(z)$ for $z \in y_p(\Omega)$. Using~\eqref{eq:gradient_inverse} and rearranging, we obtain
\begin{equation} \label{eq:multiplicative_Helmholtz_canonical_pre}
  P = (P_\res \circ y_p) \, \nabla y_p
\end{equation}
that is,~\eqref{eq:multiplicative_Helmholtz_pre} for $Z_P := P_\res \circ y_p$. This is a more natural form of the splitting~\eqref{eq:multiplicative_Helmholtz_pre} in that it mimics the chain rule (which would apply if $P$ was the gradient of the concatenation of two maps).

The crucial advantage of~\eqref{eq:multiplicative_Helmholtz_canonical_pre} over~\eqref{eq:multiplicative_Helmholtz_pre} is that the curl of $P_\res$ has a natural expression (unlike the curl of $Z_P$). To see this, we use first that $\curl = \partial \circ \, \hodge$ in $\R^3$ (this is proved in Section~7.7 of~\cite{HudsonRindler22} in the form $\curl \circ \hodge = \partial$, from which the stated equality follows via~\eqref{eq:hodge_inverse}) to write
\[
  \curl P_\res = \partial \, \hodge \bigl[ (P \circ y_p^{-1}) \, \nabla y_p^{-1} \bigr],
\]
as an equation of $\R^{3 \times 3}$-valued fields. We recall that by convention our curl acts row-wise on matrices. Hence also the Hodge operator must act individually on the rows, which contain \mbox{($1$-)vectors}. So, $\hodge [(P \circ y_p^{-1}) \, \nabla y_p^{-1}]$ is a column of $2$-vector fields. Each entry we may then interpret as a smooth \mbox{$2$-current} as in Section~\ref{sc:smooth_currents}. In this way, the boundary operator $\partial$ becomes applicable (row-wise, that is, to each entry separately), yielding a column of \mbox{$1$-currents}. By the formula $\curl = \partial \circ \, \hodge$ we know that this concatenation of $\hodge$ and $\partial$ yields the row-wise curl of the original matrix.

For row vectors, the transformation rule~\eqref{eq:hodge_transform1} for the Hodge operator reads as
\[
  \hodge (vL) = (\det L) \, (\hodge v) L^{-\T}
\]
with $L \in  \R^{3 \times 3}$ an invertible matrix. One can either derive this directly (using that the scalar product for row vectors $v,w$ is $v \cdot w = vw^\T$) or by applying~\eqref{eq:hodge_transform1} to transposed vectors. Our situation corresponds to $v := P \circ y_p^{-1}$ and $L := \nabla y_p^{-1}$, so that we obtain (with the convention that the Hodge dual binds stronger than the matrix multiplication to avoid a proliferation of parentheses)
\begin{align}
  \curl P_\res
  &= \partial \bigl( \det \nabla y_p^{-1} \cdot \hodge (P \circ y_p^{-1}) \, (\nabla y_p^{-1})^{-\T} \bigr)  \notag\\
  &= \partial \bigl( \det \nabla y_p^{-1} \cdot \hodge (P \circ y_p^{-1}) \, (\nabla y_p \circ y_p^{-1})^\T \bigr). \label{eq:curl_Ptilde}
\end{align}
Here we have also used~\eqref{eq:gradient_inverse} for the last equality.

Multiplying in~\eqref{eq:curl_Ptilde} from the right by $(\nabla y_p)^\T$ at a point $z \in y_p(\Omega)$ is precisely the action of the differential $\DD y_p(x)$ on a \emph{row} vector (as it is the transpose of the gradient applied to a column vector). So, the pushforward $y_{p*} h$ (meaning $(y_p)_* h$) of a \emph{row} vector field $h \colon \Omega \to \R^3$, interpreted as a (row) \mbox{$1$-current}, is given by an analogous formula to~\eqref{eq:smooth_pushforward}, namely
\begin{align*}
  y_{p*} h
  &= \abs{\det \nabla y_p^{-1}} \, (\DD y_p \circ y_p^{-1})[h \circ y_p^{-1}] \\
  &= \abs{\det \nabla y_p^{-1}} \, (h \circ y_p^{-1}) \, (\nabla y_p \circ y_p^{-1})^\T.  
\end{align*}
This definition of the pushforward extends to column vector fields $H = (h^1,h^2,h^3)^\T$ by applying it to all rows separately. If for $H := \curl P = \partial \hodge P$ we compare this to~\eqref{eq:curl_Ptilde} (here we also use the orientation-preserving condition and the linearity of the boundary operator $\partial$), we see that
\[
  \curl P_\res = \partial (y_{p*} [\hodge P]).
\]
Now we use that the boundary operator commutes with pushforwards, see~\eqref{eq:partial_pushforward}, and again that $\partial \hodge P = \curl P$, to see that
\begin{equation} \label{eq:curlP_yp_pushforward}
  \curl P_\res = y_{p*} [\curl P].
\end{equation}
In this way we have directly related the curl of $P_\res$ to the curl of $P$. This is decisive since $\curl P$ corresponds to the dislocations via~\eqref{eq:curlP_consistency}.

Recall that we are also imposing the plastic incompressibility assumption $\det P \equiv 1$. We want this to be reflected in our splitting of $P$ by imposing that
\[
  \det \nabla y_p = \det P_\res \equiv 1.
\]

As mentioned before, it is desirable to choose $\nabla y_p$ \enquote{maximally}, so that we split off as much of a gradient in $P$ as possible. There are several ways of defining what maximality should mean, but here we shall rely on a physically meaningful approach: Define $y_p$ to be the minimizer of
\begin{equation} \label{eq:multiplicative_Helmholtz_minimum}
  \hat{y}_p \mapsto \int_\Omega W_e(\nabla \hat{y}_p P^{-1}) \dd x,
\end{equation}
over all admissible $\hat{y}_p \colon \Omega \to \R^3$, which in particular includes the requirement $\det \nabla \hat{y}_p \equiv 1$ as well as possible boundary conditions. This choice corresponds to the optimal elastic relaxation for a given plastic distortion $P$ (as expressed in the hyperelasticity assumption of Section~\ref{sc:energy}). Thus, we are looking for $y_p \colon \Omega \to \R^3$ and $P_\res \colon y_p(\Omega) \to \R^{3 \times 3}$ solving
\begin{equation} \label{eq:multiplicative_Helmholtz}
  P = (P_\res \circ y_p) \, \nabla y_p
  \quad\text{with}\quad
  \left\{\begin{aligned}
    &\det \nabla y_p \equiv 1,\\
    &\text{$y_p$ minimizes $\hat{y}_p \mapsto \int_\Omega W_e(\nabla \hat{y}_p P^{-1}) \dd x$,} \\
    &\curl P_\res = y_{p*}(\curl P).
  \end{aligned}\right.
\end{equation}
We call such a splitting, if it exists, a \textbf{multiplicative Helmholtz-type decomposition}; this terminology will be explained below. Observe that $y_p, P_\res$ as above may exist even if $y_p$ is not globally invertible since the inverse of $y_p$ does not actually feature in~\eqref{eq:multiplicative_Helmholtz}. However, since $\det \nabla y_p \equiv 1$, the map $y_p$ must at least be locally invertible by the inverse function theorem, that is, on a small open ball $B$ around any point $x_0 \in \Omega$, and for every such ball and local map $y_p$ it must hold that $P_\res(z) = P(y_p^{-1}(z)) \, \nabla y_p^{-1}(z)$ for $z \in y_p(B)$.

Clearly,~\eqref{eq:multiplicative_Helmholtz_minimum} involves highly nonlinear quantities and this makes the minimization problem difficult to analyze. However, it can be proved that a solution exists under reasonable assumptions (via the usual polyconvexity methods, see Chapter~6 of~\cite{Rindler26book}, together with their extension to integrands involving a fixed $P$, see~\cite{MainikMielke09,FrancfortMielke06}), but uniqueness of the minimizer (or \emph{global} invertibility) is a much harder question. In fact, it is well-known that even the functionals of pure (nonlinear) elasticity, that is, the case $P \equiv \Id$, do not have a unique minimizer in general~\cite{Ciarlet88book}.

In the remainder of this section we will discuss simplifications and variants of~\eqref{eq:multiplicative_Helmholtz}, which will also ultimately explain the relationship to the classical Helmholtz decomposition.

First, to compute a solution of~\eqref{eq:multiplicative_Helmholtz}, one may formally introduce a Lagrange multiplier field $p \colon \Omega \to \R$ and consider the Lagrangian
\[
  \Lcal(\hat{y}_p,p) := \int_\Omega W_e(\nabla \hat{y}_p P^{-1}) + p \, \bigl(\det \nabla \hat{y}_p - 1\bigr) \dd x.
\]
Here, $\hat{y}_p \colon \Omega \to \R^3$ ranges over the admissible deformations, which in particular satisfy $\det \nabla \hat{y}_p>0$ and any prescribed boundary conditions. The constrained problem is then formally represented by the saddle-point problem
\[
  \inf_{\hat{y}_p}\sup_p \Lcal(\hat{y}_p,p).
\]
Stationarity with respect to $p$ imposes the pointwise constraint $\det \nabla y_p=1$. The corresponding formal Euler--Lagrange equations are
\[
  \left\{
  \begin{aligned}
    -\dive \bigl[\DD W_e(\nabla y_p P^{-1})P^{-\T} + p\,\cof \nabla y_p\bigr] &= 0, \\
    \det \nabla y_p &= 1.
  \end{aligned}
  \right.
\]
The validity of this PDE system for minimizers is unknown under realistic growth conditions in nonlinear elasticity~\cite{Ball02}, so one must in general remain in the variational framework.

However, if we a priori know (or assume so, as in the next section) that for the minimizer $y_p$ of~\eqref{eq:multiplicative_Helmholtz} it holds that $\nabla y_p P^{-1} \approx \Id$ (so that $P$ is \enquote{nearly} a gradient), we may appeal to Section~\ref{sc:We_linearize} to obtain the linearization of~\eqref{eq:multiplicative_Helmholtz_minimum} as the minimization problem for the functional
\begin{equation} \label{eq:multiplicative_Helmholtz_minimum_linearized}
  \hat{y}_p \mapsto \int_\Omega \frac12 \absb{\nabla \hat{y}_p P^{-1} - \Id}_\Ebb^2 \dd x
\end{equation}
over all admissible $\hat{y}_p \colon \Omega \to \R^3$, which in particular includes the requirements that $\hat{y}_p$ is invertible. In this linearized setting we may also neglect the constraint $\det \nabla y_p \equiv 1$, which by assumption is at least approximately satisfied since $\det \nabla y_p \approx \det P \equiv 1$. Moreover, the splitting $P = (P_\res \circ y_p) \, \nabla y_p$ itself necessarily implies that $\det \nabla y_p \neq 0$ (since the left-hand side satisfies $\det P \equiv 1$ and $\det (AB) = \det A \cdot \det B$). Hence, any $y_p$ in the multiplicative Helmholtz-type splitting ~\eqref{eq:multiplicative_Helmholtz} must at least be locally invertible by the inverse function theorem.

The functional in~\eqref{eq:multiplicative_Helmholtz_minimum_linearized} is convex, so a minimizer exists (see Section~1.8 as well as Examples~2.12 and~3.17 in~\cite{Rindler26book}) and satisfies the Euler--Lagrange equation
\begin{equation} \label{eq:multiplicative_Helmholtz_minimum_linearized_EL}
  \dive [\Fbb(P^{-1}) \nabla y_p] = \dive H(P^{-1}),
\end{equation}
where
\[
  \Fbb(Q)^{ijkl} := \Ebb^{imkn} Q^j_m Q^l_n,  \qquad
  H(Q)^i_j := \Ebb^{imkk} Q^j_m,
\]
and
\[
  \bigl[ \dive [\Fbb(Q) \nabla z] \bigr]^i = \partial_j \bigl[ \Fbb(Q)^{ijkl} \partial_l z^k \bigr],  \qquad
  \bigl[ \dive H(Q) \bigr]^i = \partial_j H(Q)^i_j
\]
for $Q \in \R^{3 \times 3}$ and $z \colon \Omega \to \R^3$ smoothly differentiable. Here, we have again summed over all repeated indices (e.g., $\Ebb^{imkk} Q^j_m = \sum_{m,k} \Ebb^{imkk} Q^j_m$). These expressions can be checked by computing the variations for~\eqref{eq:multiplicative_Helmholtz_minimum_linearized} and integrating by parts as follows: Let $\psi \colon \Omega \to \R^3$ be zero in a strip close to the boundary (i.e., $\psi$ has compact support). We also observe that
\[
  \Fbb(\Id)^{ijkl} = \Ebb^{ijkl},  \qquad
  H(\Id)^i_j = \Ebb^{ijkk}.
\]
Then, set $\hat{y}_p := y_p + h \psi$ in~\eqref{eq:multiplicative_Helmholtz_minimum_linearized} and take the derivative with respect to $h$, to compute
\begin{align*}
  &\frac{\di}{\di h} \int_\Omega \frac12 \abs{\nabla (y_p + h\psi) Q - \Id}_\Ebb^2 \dd x \bigg|_{h=0} \\
  &\qquad = \int_\Omega \Ebb (\nabla y_p Q - \Id) : (\nabla \psi Q) \dd x  \\
  &\qquad = \int_\Omega \Ebb^{imkn} (\partial_l y_p^k Q^l_n) (\partial_j \psi^i Q^j_m) - \Ebb^{imkn} \delta^k_n (\partial_j \psi^i Q^j_m) \dd x  \\
  &\qquad = \int_\Omega \Ebb^{imkn} \bigl[ (\partial_l y_p^k Q^j_m Q^l_n) - \delta^k_n Q^j_m \bigr] \, \partial_j \psi^i \dd x  \\
  &\qquad = - \int_\Omega \partial_j \bigl[ (\Ebb^{imkn} Q^j_m Q^l_n) \partial_l y_p^k - \Ebb^{imkk} Q^j_m \bigr] \psi^i \dd x.
\end{align*}
Here, as usual, $\delta^k_n = 1$ if $k = n$ and $0$ otherwise. The left-hand side has to equal zero since $y_p$ is assumed to be a minimizer, whereby
\[
  h \mapsto \int_\Omega \frac12 \abs{\nabla (y_p + h\psi) Q - \Id}_\Ebb^2 \dd x
\]
has a minimum at $h = 0$. Hence,
\[
  \int_\Omega \partial_j \bigl[ (\Ebb^{imkn} Q^j_m Q^l_n) \partial_l y_p^k - \Ebb^{imkk} Q^j_m \bigr] \psi^i \dd x = 0
\]
for all $\psi$ as above. Since $\psi$ was arbitrary, this yields precisely the system~\eqref{eq:multiplicative_Helmholtz_minimum_linearized_EL} of PDEs.

In conclusion, the above arguments combine to yield that in the case $\nabla y_p P^{-1} \approx \Id$ instead of~\eqref{eq:multiplicative_Helmholtz} we may alternatively use the following approximation:
\begin{equation} \label{eq:multiplicative_Helmholtz_linearized}
  P = (P_\res \circ y_p) \, \nabla y_p
  \qquad\text{with}\qquad
  \left\{\begin{aligned}
    \dive \bigl[ \Fbb(P^{-1}) \nabla y_p \bigr] &= \dive H(P^{-1}), \\
    \curl P_\res &= y_{p*}(\curl P).
  \end{aligned}\right.
\end{equation}
While one could add a Lagrange multiplier as above in order to enforce the constraint $\det \nabla y_p \equiv 1$, this would make the PDE nonlinear in $\nabla y_p$. 

\begin{remark}
In the plastically small-strain regime, that is, if $P = \Id + p$ with $p$ small, we make the ansatz $y_p = \id + u_p$ for a small plastic displacement $u_p \colon \Omega \to \R^3$. Then, by a similar procedure to the one in Section~\ref{sc:We_linearize}, the minimization in~\eqref{eq:multiplicative_Helmholtz_minimum_linearized} linearizes further to the minimization of
\[
  \hat{u}_p
  \mapsto \int_\Omega \frac12 \absb{(\Id + \nabla \hat{u}_p)(\Id - p) - \Id}_\Ebb^2 \dd x
  = \int_\Omega \frac12 \absb{\nabla \hat{u}_p - p - \nabla \hat{u}_p \, p}_\Ebb^2 \dd x.
\]
We may neglect the term $\nabla \hat{u}_p \, p$ since it is much smaller than the others and so we consider the small-strain minimization problem
\[
  \hat{u}_p
  \mapsto \int_\Omega \frac12 \absb{\nabla \hat{u}_p - p}_\Ebb^2 \dd x.
\]
The Euler--Lagrange equation satisfied by a minimizer $u_p$ is
\[
  \dive \bigl[ \Ebb(\nabla u_p - p) \bigr] = 0,
\]
augmented by suitable boundary conditions. By the symmetries of $\Ebb$, a solution $u_p$ is only determined up to a rigid deformation, that is a map $v(x) = v_0 + Wx$ with $v_0 \in \R^3$ and $W \in \R^{3 \times 3}$ satisfying $W^\T = -W$ (skew-symmetry). For any such solution $u_p$, setting $\sigma := p - \nabla u_p$ we obtain instead of~\eqref{eq:multiplicative_Helmholtz_linearized} the fully linearized splitting
\[
  p = \nabla u_p + \sigma
  \qquad\text{with}\qquad
  \left\{\begin{aligned}
    \dive \Ebb \sigma  &= 0, \\
    \curl \sigma &= \curl p.
  \end{aligned}\right.
\]
This is the classical Helmholtz decomposition of $p$ in the case of symmetric matrices and using the scalar product associated to $\Ebb$. This explains why by analogy we called~\eqref{eq:multiplicative_Helmholtz} a \emph{multiplicative} Helmholtz-type decomposition. Note, however, that any additive rigid deformation is not seen by $\Ebb \sigma$, so the above splitting is well-defined only up to rigid deformations, that is, also $p = (\nabla u_p + W) + (\sigma - W)$ for any $W \in \R^{3 \times 3}$ satisfying $W^\T = -W$. 
\end{remark}

\subsection{Small-Distortion Hypothesis} \label{sc:small_distortion_hypothesis}

The \textbf{Small-Distortion Hypothesis} says that the non-gradient part of the plastic distortion, in the sense of the multiplicative Helmholtz decomposition~\eqref{eq:multiplicative_Helmholtz}, is small:

\begin{hypothesis}
The plastic distortion splits via the multiplicative Helmholtz-type decomposition, see~\eqref{eq:multiplicative_Helmholtz}, as
\[
  P = (P_\res \circ y_p) \, \nabla y_p
\]
with a residual plastic distortion $P_\res = P_\res(t) \colon y_p(\Omega) \to \R^{3 \times 3}$ that is close to the identity field $\Id$, so that
\[
  P_\res = \Id + r
\]
with $r \colon y_p(\Omega) \to \R^{3 \times 3}$ small, and an orientation-preserving smoothly invertible plastic deformation $y_p = y_p(t) \colon \Omega \to \R^3$ with $\det \nabla y_p \equiv 1$.
\end{hypothesis}

The idea here is that most of the plastic distortion is given by a gradient field, with only a small non-gradient distortion. If $y_p$ is chosen as the minimizer (or an approximate minimizer) of~\eqref{eq:multiplicative_Helmholtz_minimum} or~\eqref{eq:multiplicative_Helmholtz_minimum_linearized}, it is a reasonable assumption that if $\curl P$ is small, then $\nabla y_p$ can cancel \enquote{most of} $P^{-1}$, so that $P_\res$ may be assumed to be close to the identity field. We stress once more that we do not assume that $P$ itself is small, so we are still in a (plastically) large-strain regime.

Since $\curl P$ represents the dislocation density, the Small-Distortion Hypothesis expresses that \emph{dislocations do not dominate the crystal}. This is a physically realistic assumption since in defect-dominated crystals also damage and fracture phenomena need to be taken into consideration, which is not the regime we consider here.

Given a total deformation $y \colon \Omega \to \R^3$, we further define
\[
  y_e(z) := y(y_p^{-1}(z)),  \qquad z \in y_p(\Omega),
\]
which is well-defined since $y_p$ was assumed invertible. Hence,
\begin{equation} \label{eq:y_ye_yp}
  y = y_e \circ y_p, \qquad
  \nabla y = (\nabla y_e \circ y_p) \, \nabla y_p
\end{equation}
as well as
\[
  \nabla y P^{-1}
  = (\nabla y_e \circ y_p) \, \nabla y_p \, (\nabla y_p)^{-1} \, (P_\res \circ y_p)^{-1}
  = (\nabla y_e \circ y_p) \, (P_\res \circ y_p)^{-1}.
\]
For the elastic energy for a potentially not volume-preserving plastic distortion $P$ we justified in Remark~\ref{rem:We_compressible} the expression
\[
  \Wcal_e[y,P;\Omega] = \int_\Omega W_e(\nabla y P^{-1}) \, \abs{\det P} \dd x.
\]
For the new variable $z = y_p(x)$ and the new plastic distortion $P_\res$ we have
\[
  \abs{\det P_\res} \dd z
  = \abs{\det P_\res \circ y_p} \cdot \abs{\det \nabla y_p} \dd x
  = \abs{\det P} \dd x
\]
by the product rule for the determinant. So, changing variables, we obtain
\begin{align}
  \Wcal_e[y,P;\Omega]
  &= \int_\Omega W_e(\nabla y P^{-1}) \, \abs{\det P} \dd x  \notag\\
  &= \int_\Omega W_e \bigl( (\nabla y_e \circ y_p) \, (P_\res \circ y_p)^{-1} \bigr) \, \abs{\det P} \dd x  \notag\\
  &= \int_{y_p(\Omega)} W_e(\nabla y_e P_\res^{-1}) \, \abs{\det P_\res} \dd z  \notag\\
  &= \Wcal_e[y_e,P_\res;y_p(\Omega)].  \label{eq:We_transform}
\end{align}
Here, $y_p(\Omega)$ means $y_p(t,\Omega)$ according to our convention of leaving out arguments when this does not cause confusion.

Now, as $P_\res$ is assumed to be close to the identity field,
\begin{equation} \label{eq:r}
  r(z) := P_\res(z) - \Id,  \qquad z \in y_p(\Omega),
\end{equation}
is small. Using the approximation $(\Id + X)^{-1} \approx \Id - X$ for $X$ small, we then get
\[
  P^{-1} = \bigl[(\Id + r \circ y_p) \, \nabla y_p \bigr]^{-1} \approx (\nabla y_p)^{-1} \, (\Id - r \circ y_p).
\]
As $P_\res^{-1} \approx \Id - r$ and $\det P_\res \equiv 1$, we thus may transform~\eqref{eq:We_transform} further to arrive at
\begin{equation} \label{eq:small_distortion_We_expand}
  \Wcal_e[y,P;\Omega] \approx \int_{y_p(\Omega)} W_e \bigl( \nabla y_e \, (\Id - r) \bigr) \dd z.
\end{equation}

Let us furthermore assume the following \textbf{elastic coercivity hypothesis}:

\begin{hypothesis}
If $P$ is \enquote{close} to the identity field $\Id$, then a minimizing deformation $y$ of
\[
  \Wcal_e = \int_\Omega W_e(\nabla y P^{-1}) \dd x
\]
is \enquote{close}, modulo a rigid deformation, to the identity map. More precisely, there exist $Q \in \SO(3)$ and $c \in \R^3$ such that $y$ is close to the map $x \mapsto Qx+c$.
\end{hypothesis}

This requirement expresses that a low amount of lattice distortion will only require a low amount of energy that cannot be elastically relaxed away. The formulation is deliberately kept vague as to the meaning of \enquote{close}. A precise meaning can only be given within the rigorous theory. For now, it shall suffice to mean
\[
  \norm{P^{-1}-\Id}, \qquad \norm{Q^\T\nabla y-\Id} \ll 1
\]
with respect to suitable norms and for some $Q \in \SO(3)$. Prescribed boundary conditions may fix the rigid motion and, in particular, may enforce $Q=\Id$ and $c=0$.

Applying this coercivity hypothesis on $y_p(\Omega)$ to $P_\res$ and $y_e$ in~\eqref{eq:We_transform}, we deduce from the smallness of $r$, whereby $P_\res \approx \Id$, that $y_e$ is close to a rigid deformation. Thus, there exist $Q \in \SO(3)$ and $c \in \R^3$ such that
\[
  \bar y_e(z) := Q^\T (y_e(z) - c)
\]
is close to the identity map. Correspondingly, define the co-rotated total deformation by
\[
  \bar y(x) := Q^\T(y(x) - c).
\]
Then $\bar y=\bar y_e\circ y_p$, and frame indifference gives
\[
  W_e\bigl(\nabla y_e(\Id-r)\bigr)
  = W_e\bigl(\nabla\bar y_e(\Id-r)\bigr).
\]
We may therefore work in this co-rotated frame without changing the elastic energy. To avoid introducing additional notation, we henceforth write $y$ and $y_e$ again in place of $\bar y$ and $\bar y_e$. Equivalently, the rigid motion has now been normalized so that $Q=\Id$ and $c=0$.

It follows that
\begin{equation} \label{eq:ye_tilde_ue}
  \nabla y_e = \Id+\nabla\tilde{\eta}
\end{equation}
with a small elastic displacement $\tilde{\eta} \colon y_p(\Omega) \to \R^3$. Since multiplying two small quantities gives a negligible quantity, at least on the order of approximation that we consider, we get $\nabla\tilde{\eta}\,r\approx0$, so that
\begin{equation} \label{eq:nabla_ye_R}
  \nabla y_e(\Id-r)
  \approx(\Id+\nabla\tilde{\eta})(\Id-r)
  \approx\Id+\nabla\tilde{\eta}-r.
\end{equation}
We then use~\eqref{eq:nabla_ye_R} in~\eqref{eq:small_distortion_We_expand} and linearize the elastic energy, similarly to~\eqref{eq:multiplicative_Helmholtz_minimum_linearized} in the previous section, to obtain
\[
  \Wcal_e \approx \int_{y_p(\Omega)} \frac12\abs{\nabla\tilde{\eta}-r}_\Ebb^2 \dd z.
\]

Now we define the \textbf{linearized elastic distortion}
\[
  e := \nabla y_e - \Id-r,
\]
which satisfies
\[
  e=\nabla\tilde{\eta}-r, \qquad
  \curl e=-\curl r=-\curl P_\res.
\]
So, for the linearized energy we obtain the approximation
\begin{equation} \label{eq:small_distortion_We_approx}
  \Wcal_e \approx \int_{y_p(\Omega)} \frac12\abs{e}_\Ebb^2 \dd z.
\end{equation}

With a view to the Line-Tension Approximation to be introduced in the following section, our next goal is to adjust~\eqref{eq:small_distortion_We_approx} to separate out the contributions from the \enquote{free} elastic displacement and from the \enquote{geometrically necessary} distortion due to the presence of dislocations. So, define the \textbf{free distortion}
\[
  \fr[e] := e+\beta,
\]
which is a map defined on $y_p(\Omega)$ with values in $\R^{3\times3}$, where the \textbf{geometrically-necessary distortion} $\beta \colon y_p(\Omega) \to \R^{3\times3}$ is the solution to
\begin{equation} \label{eq:beta}
  \left\{ \begin{aligned}
    -\dive\Ebb\beta &= 0 &&\text{in $y_p(\Omega)$,} \\
    \curl\beta &= \curl r=-\curl e &&\text{in $y_p(\Omega)$,} \\
    (\Ebb\beta)n &= 0 &&\text{on $\partial y_p(\Omega)$.}
  \end{aligned} \right.
\end{equation}
Note that indeed $\fr[e]$ only depends on $e$. The boundary conditions here are the so-called \enquote{natural} ones, meaning that $-\dive\Ebb\beta=0$ in fact holds on \emph{all} of $\R^3$ if $\beta$ is extended by zero. This is in accordance with stress fields in elasticity over free boundaries. It is also the right choice to avoid any boundary terms in the orthogonality statement~\eqref{eq:fr_e_gradient_orthogonal} below. Via~\eqref{eq:curlP_yp_pushforward} and~\eqref{eq:curlP_consistency} we furthermore know that
\begin{equation} \label{eq:curl_r}
  \curl r(t)
  = \curl P_\res(t)
  = y_p(t)_*[\curl P]
  = \frac12\sum_{b\in\Bcal}b\otimes y_p(t)_*S^b(t).
\end{equation}

Now, as a consequence of~\eqref{eq:ye_tilde_ue},
\begin{align}
  \nabla y_e
  &= \Id+\nabla\tilde{\eta} \notag\\
  &= \Id+e+r \notag\\
  &= \Id+\fr[e]-\beta+r. \label{eq:nabla_ye_decomp_small_distortion}
\end{align}
Once the decomposition in the Small-Distortion Hypothesis and the rigid-motion normalization have been fixed, $r,\beta$ are uniquely determined by the state of the dislocations and $\fr[e]$ is the only part of $\nabla y_e$ that is free to change in order to accommodate external forces or boundary conditions. Thus, $\fr[e]$ constitutes the elastic degrees of freedom of our system under the Small-Distortion Hypothesis, explaining why we call it the \enquote{free} elastic distortion.

From the definition of $\beta$ in~\eqref{eq:beta} it also follows that
\begin{equation} \label{eq:curl_fr_0}
  \curl \fr[e] = 0.
\end{equation}
Thus, if $\Omega$ is simply connected then~\eqref{eq:curl_fr_0} immediately implies the existence of a map $\eta \colon y_p(\Omega) \to \R^3$ satisfying $\fr[e] = \nabla \eta$. However, if $\Omega$ is not simply connected, there are curl-free fields that are not gradients. Indeed, if many dislocations wrap around a \enquote{tunnel} in $\Omega$, then these dislocations will influence the elastic properties of the specimen. We assume that such \enquote{non-null-homologous} dislocations have been removed from the slip trajectory system:

\begin{hypothesis}
There is a \textbf{free elastic displacement} $\eta \colon y_p(\Omega) \to \R^3$ with
\[
  \fr[e] = \nabla \eta.
\]
\end{hypothesis}

It is then only $\eta$ (satisfying some imposed boundary conditions) that may be varied elastically. Moreover, $\fr[e]$ and $\beta$ are $\Ebb$-orthogonal, that is
\begin{align*}
  \int_{y_p(\Omega)} \Ebb \beta : \fr[e] \dd z
  &= \int_{y_p(\Omega)} \Ebb \beta : \nabla \eta \dd z \\
  &= - \int_{y_p(\Omega)} (\dive \Ebb \beta) \cdot \eta \dd z + \int_{\partial y_p(\Omega)} \Ebb \beta : (\eta \otimes n) \dd \Hcal^2 \\
  &= 0,
\end{align*}
so that the orthogonal decomposition
\begin{equation} \label{eq:fr_e_gradient_orthogonal}
  \int_{y_p(\Omega)} \abs{e}_\Ebb^2 \dd z
  = \int_{y_p(\Omega)} \abs{\nabla \eta - \beta}_\Ebb^2 \dd z
  = \int_{y_p(\Omega)} \abs{\nabla \eta}_\Ebb^2 \dd z + \int_{y_p(\Omega)} \abs{\beta}_\Ebb^2 \dd z
\end{equation}
holds. 

\begin{remark} \label{rem:r_LieAlgebras}
One could instead of~\eqref{eq:r} define $r(z)$ via the formula
\[
  P_\res(z) = \Exp(r(z)),  \qquad z \in y_p(\Omega).
\]
For $P_\res$ uniformly close to the identity, this equation is solvable (see, e.g.,~\cite{Hall15book}). This is almost the same as~\eqref{eq:r} since $\Exp(A) \approx \Id + A$ for small matrices $A$, but it preserves additional constraints better: For instance, if $\det P_\res \equiv 1$, it follows that $\tr r \equiv 0$. More generally, if $P_\res \in \Pfrak$ for a Lie group $\Pfrak$, then $r \in \pfrak$ with $\pfrak = \Lie(\Pfrak)$ the corresponding Lie algebra. In this way, one may transport restrictions on $P_\res$ to restrictions on $r$.
\end{remark}

\subsection{Line-Tension Approximation} \label{sc:line_tension_approximation}

From the Small-Distortion Hypothesis we derived in~\eqref{eq:small_distortion_We_approx} that the elastic energy at time $t$ may be approximated as
\[
  \Wcal_e \approx \int_{y_p(\Omega)} \frac12 \abs{e}_\Ebb^2 \dd z.
\]
Moreover,
\[
  e = \fr[e] - \beta,
\]
with $\beta$ solving (see~\eqref{eq:beta} in conjunction with~\eqref{eq:curl_r})
\[
  \left\{ \begin{aligned}
    - \dive \Ebb \beta &= 0          &&\text{in $y_p(\Omega)$,} \\
    \curl \beta &= \frac12 \sum_{b \in \Bcal} b \otimes y_{p*}S^b(t)  &&\text{in $y_p(\Omega)$,} \\
    (\Ebb \beta)n &= 0                 &&\text{on $\partial y_p(\Omega)$.}
  \end{aligned} \right.
\]

We have already seen the orthogonality relation~\eqref{eq:fr_e_gradient_orthogonal}, that is,
\[
  \Wcal_e \approx \int_{y_p(\Omega)} \frac12 \abs{\nabla \eta}_\Ebb^2 \dd z + \int_{y_p(\Omega)} \frac12 \abs{\beta}_\Ebb^2 \dd z.
\]
Evidently, the second term, namely the energy of the geometrically-necessary distortion, is completely independent of the first and only depends on the dislocations that are present. We want to approximate this term by the so-called \emph{line-tension} of the dislocation lines.

The basic idea is the following: Consider first the prototypical situation that there is only one Burgers vector $b$ and $y_{p*}S^b(t)$ consists of a single straight line $\ell = \spn\{e_3\}$ with (unit-length) direction $\tau = \ee_3$ and multiplicity $1$. Assume also, for the sake of illustration, that $\Ebb$ is the identity tensor, so that $(\Ebb A) : B = A : B$ for $A,B \in \R^{3 \times 3}$. Finally, we disregard the boundary conditions and consider a stress field on all of $\R^3$, thus neglecting effects sufficiently far away from the dislocation line. The PDE for $\beta$ then simplifies to
\begin{equation} \label{eq:beta_line_PDE}
  \left\{ \begin{aligned}
    - \dive \beta &= 0, \\
    \curl \beta &= b \otimes \tau \; \Hcal^1 \restrict \ell,\\
  \end{aligned} \right.
\end{equation}
with $\Hcal^1 \restrict \ell$ denoting the \enquote{measure} (distribution of order zero) representing the dislocation line $\ell = \spn\{e_3\}$.

This system has rotational symmetry and can be solved explicitly. The solution is best written in cylindrical coordinates $x = (r,\phi,z)$, representing a point $x$ as lying in a plane $z \ee_3 + \spn\{\ee_1, \ee_2\}$, with distance $r$ from the line $\ell$ and angle $\phi$ relative to the $\ee_2$-direction in this plane. Denoting the unit tangent vector to the circle with radius $r$ by $\hat{\phi}$ (with orientation according to the right-hand rule relative to the direction $\ee_3$), the solution to the above PDE turns out to be
\begin{equation} \label{eq:beta_line}
  \beta(r,\phi,z) = b \otimes \frac{\hat{\phi}}{2 \pi r}.
\end{equation}
To verify the second equation in~\eqref{eq:beta_line_PDE} in an elementary way, let
\[
  B_R(z) := z \ee_3 + B_R^{\R^2}
\]
denote the cross-sectional disk of radius $R>0$ at height $z$. Applying the Stokes theorem on $B_R(z)$, understood distributionally at the point $B_R(z) \cap \ell$, we obtain
\[
  \int_{B_R(z)} (\curl\beta)\ee_3 \dd\Hcal^2
  = \int_{\partial B_R(z)} \beta\hat\phi \dd\Hcal^1.
\]
Since the positively oriented unit tangent to $\partial B_R(z)$ is $\hat\phi$, the right hand side can be evaluated as
\[
  \int_{\partial B_R(z)} \beta\hat\phi \dd\Hcal^1
  = \frac{b\abs{\hat\phi}^2}{2\pi R}\,(2\pi R)
  = b.
\]
Integrating with respect to the axial variable $z$ over $(0,h)$, we therefore find
\[
  \int_0^h \int_{B_R(z)} (\curl\beta)\ee_3 \dd\Hcal^2 \dd z
  = \int_0^h \int_{\partial B_R(z)} \beta\hat\phi \dd\Hcal^1 \dd z
  = hb.
\]
Thus, for every $R,h>0$,
\[
  \int_{Z_{R,h}} (\curl\beta)\ee_3 \dd\Hcal^3 = hb,
\]
and hence, in the sense of distributions,
\[
  \curl\beta = b\otimes\ee_3\;\Hcal^1\restrict\ell.
\]

Moreover, expressing the divergence in cylindrical coordinates, we compute
\[
  \dive \biggl[ \frac{\hat{\phi}}{2 \pi r} \biggr]
  = \frac{1}{r} \frac{\partial}{\partial \phi} \biggl[ \frac{1}{2 \pi r} \biggr]
  = 0.
\]
From this we conclude that $\dive \beta = 0$, verifying that~\eqref{eq:beta_line} indeed is a solution for our PDE~\eqref{eq:beta_line_PDE}.

Note that $\beta$ is not square-integrable whenever the domain of integration contains a piece of the line. Indeed, we may integrate over any solid hollow cylinder $Z_{\eps,R,h}$ around the line with inner radius $\eps > 0$, outer radius $R > 0$, and height $h > 0$ as follows:
\begin{align*}
  \int_{Z_{\eps,R,h}} \abs{\beta}^2 \dd x
  &= \frac{\abs{b}^2}{4\pi^2} \int_0^h \int_0^{2\pi} \int_\eps^R \frac{1}{r^2} \; r \, \di r \dd \phi \dd z \\
  &= \frac{\abs{b}^2 h}{2\pi} \int_\eps^R \frac{1}{r} \dd r \\
  &= \frac{\abs{b}^2 h}{2\pi} \; \ln r \big|_{r=\eps}^R.
\end{align*}
So, the above energy for $\beta$ is infinite in the limit $\eps \todown 0$, $R \toup \infty$. In fact, both the region near the \enquote{core} (that is, for $\eps \todown 0$) and the \enquote{tail} (that is, for $R \toup \infty$) make the integral infinite. We will examine both of these effects in turn.

For the core region, the above computation, that is, for $\eps$ small, is in fact not physically valid since if $\eps$ gets to within the order of magnitude of the interatomic distances, the continuum assumption breaks down and a discrete approach to compute the so-called \textbf{core energy} is necessary. Concretely, if only elastic and not atomistic effects are to be considered, this means to limit the lower bound to a \emph{fixed} $\eps_\mathrm{min} > 0$. We compensate this neglect of the core by adding another term of the form $E^b_\mathrm{core} h$ with a line energy density $E^b_\mathrm{core}$, which may depend on the Burgers vector $b$. This must scale linearly with the length of the line segment by symmetry, and could also incorporate other effects like a chemical potential (i.e., the bonding energy).

For the tails, we just neglect the energy \enquote{far away} from the line since the material has a finite size and so there is an absolute upper bound, say $R_\mathrm{max}$ on $R$. Besides, the derivative of the logarithm goes to zero, so unless we go \enquote{all the way out to infinity}, the error made in this procedure is small (smaller anyway than the other modeling errors already accumulated). 

In conclusion, the preceding considerations entail that we may approximate
\[
  \Wcal_e 
  \approx \frac{\abs{b}^2 h}{4\pi} \; \ln r \big|_{r=\eps_\mathrm{min}}^{R_\mathrm{max}} + E^b_\mathrm{core} h
  = E^b_\mathrm{line} h,
\]
where $E^b_\mathrm{line}$ is a constant representing the total line-tension energy per unit length. The salient point here is that the energy scales \emph{linearly} with the length $h$ of the dislocation.

In the following we will need to write a number of integrals over $1$-currents, for which we introduce a convenient shorthand notation: For a $G \colon \R^3 \to [0,\infty)$ that is positively $1$-homogeneous (that is, $G(\alpha v) = \alpha G(v)$ for any $v \in \R^3$ and all $\alpha \geq 0$), and a $1$-current $T = \vec{T} \, \tv{T}$ on $\R^3$ set
\[
  \int G(T) := \int G(\vec{T}) \dd \tv{T}.
\]

We are now ready to state the \textbf{Line-Tension Approximation} assumption:

\begin{hypothesis}
The energy of the geometrically-necessary distortion
\[
  \int_{y_p(\Omega)} \frac12 \abs{\beta}_\Ebb^2 \dd z
\]
can be approximated by a \textbf{line-tension energy} $\Wcal_\linet$, which has the form of an anisotropic length for the dislocations in the plastically transformed system, that is,
\[
  \Wcal_\linet = \frac{1}{2} \sum_{b \in \Bcal} \int G^b\bigl( y_{p*} S^b(t) \bigr),
\]
with the absolutely $1$-homogeneous \textbf{line-tension densities} $G^b \colon \R^3 \to [0,\infty)$ ($b \in \Bcal$), which satisfy $G^{-b} = G^b$ and $G^b(\alpha \tau) = \abs{\alpha}G^b(\tau)$ for \emph{all} $\alpha \in \R$.
\end{hypothesis}

Note that the line-tension energy $\Wcal_\linet$ takes as its input the dislocation system $\Tb = (T^b)_b$, transported via pushforward under $y_p(t)$ at time $t$ and does not depend on $\fr[e]$. However, $\Wcal_\linet$ depends on $y_p$, so it is important to ensure that this is uniquely defined.

\begin{remark} \label{rem:atomistic}
The above form of the line-tension approximation has been made mathematically rigorous (under various technical assumptions) in a series of works~\cite{BourgainBrezis04,ContiGarroniOrtiz15,ContiGarroniMassaccesi15,Ginster19,Ginster19b,ContiGarroni21,GarroniMarzianiScala21,ContiGarroniMarziani23} for both linearized and nonlinear elasticity. On the other hand, it has been shown that for both the defect core and the far-field, atomistic effects in fact have to be taken into account to get a good approximation for the stresses, see for instance the recent articles~\cite{BraunBuzeOrtner19,BraunHudsonOrtner22,BraunOrtnerWangZhang25} and the references cited therein. Consequently, a body of evidence has been accumulated suggesting that any purely elastic continuum approximation -- be it linearized or fully nonlinear -- necessarily diverges from the true stress field in a significant (but predictable) way. However, our line-tension approximation hypothesis ultimately only assumes that $\Wcal_\linet$ behaves like a (possibly anisotropic) \enquote{length} of the dislocations. Reducing the interaction energy to a line tension is thus a reasonable approximation in cases where the chemical energy stored in the dislocation core is considerably larger than the energy stored by long-range elastic interactions between dislocations. Further refinements to the model could be introduced to include these.
\end{remark}

We also define the \textbf{(linearized) free elastic energy}
\begin{equation} \label{eq:Wfree}
  \Wcal_\free := \int_{y_p(\Omega)} \frac12 \abs{\fr[e]}_\Ebb^2 \dd z = \int_{y_p(\Omega)} \frac12 \abs{\nabla \eta}_\Ebb^2 \dd z.
\end{equation}
Given external forces $f = f(t,x) \in \R^3$, we then (re-)define the \textbf{total energy} as
\[
  \Ecal := \Wcal_\free + \Wcal_\linet - \dprb{f, \eta},
\]
where
\[
  \dprb{f(t),\eta} := \int_{y_p(\Omega)} f \cdot \eta \dd z.
\]
Also other forms of the external loading are admissible, of course.

The Euler--Lagrange equation for the free elastic part, which is necessary and sufficient for an $\eta$ to be a minimizer of $\Wcal_\free$ (holding all other quantities fixed), is the \textbf{linearized elastic equilibrium equation}
\[
  - \dive (\Ebb \nabla \eta) = f   \qquad \text{in $y_p(\Omega)$,}
\]
augmented with the relevant boundary conditions.

\subsection{Change of reference frame} \label{sc:reference_change} 

We expressed the state of our system via the variables
\[
  \eta, \; P, \; y_p, \; P_\res, \; (S^b)_b.
\]
These variables are of course not independent and various relations hold between them. The fields $P$ and $y_p$ are defined on $\Omega$, whereas $\eta$ and $P_\res$ are defined on $y_p(\Omega)$; the currents $S^b$ live in space-time over $\Omega$. In the present section we want to investigate how these variables transform if we change the reference domain via an orientation-preserving invertible smooth map
\[
  z \colon \Omega \to z(\Omega) := \setb{ z(x) }{ x \in \Omega },  \qquad\text{satisfying}\qquad
  \det \nabla z \equiv 1.
\]
This is to verify that our model is indeed \emph{independent} of the choice of reference domain in that all relations between the variables transform correctly. In this way we see that the choice of reference domain just fixes the \enquote{labeling} of material points, but has no bearing on the description of the physics. In particular, by setting $z := y_p$, we could change the description to a (semi-)Eulerian point of view, where the material points are indexed relative to the plastically deformed configuration (via $y_p$); we will come back to this case at the end of the present section.

The natural way to transform the above state variables under $z$ is as follows:
\begin{align*}
  \tilde{\eta} &:= \eta, \\
  \tilde{P} &:= (P \circ z^{-1}) \, \nabla z^{-1}, \\
  \tilde{y}_p &:= y_p \circ z^{-1}, \\
  \tilde{P}_\res &:= P_\res, \\
  \tilde{S}^b &:= (\id,z)_* S^b,
\end{align*}
where $\tilde{\eta}, \tilde{P}_\res$ are defined on $y_p(\Omega)$ and all other quantities are defined on the domain $z(\Omega)$ (and for times $t$, which are not transformed). Here we have denoted the inverse map of $z$ by $z^{-1} \colon z(\Omega) \to \Omega$ and its gradient by $\nabla z^{-1}$. As before, by the formula for the gradient of the inverse we have
\[
  \nabla z^{-1} \circ z = (\nabla z)^{-1},  \qquad
  (\nabla z \circ z^{-1})^{-1} = \nabla z^{-1}.
\]

First, we note that the transformation from $S^b$ to $\tilde{S}^b = (\id,z)_* S^b$ entails that
\[
  \tilde{S}^b(t) = z_* S^b(t) = (\DD z \circ z^{-1})[\pbf\vec{S}^b(t) \circ z^{-1}] \; \tv{S^b(t)} \circ z^{-1}.
\]
Here, $\tv{S^b(t)} \circ z^{-1}$ is the transport (measure-theoretic pushforward) of the \enquote{measure} $\tv{S^b(t)}$ to $z(\Omega)$, which acts by applying $\tv{S^b(t)}$ to the preimages under $z$ of any set it is applied to (that is, for a set $A \subseteq z(\Omega)$, apply $\tv{S^b(t)}$ to the set $z^{-1}(A)$).

Next, we have the multiplicative Helmholtz-type decomposition
\[
  P = (P_\res \circ y_p) \, \nabla y_p,
\]
see~\eqref{eq:multiplicative_Helmholtz} (this is how $y_p$ and $P_\res$ are defined). For $\tilde{P}, \tilde{P}_\res, \tilde{y}_p$, as defined above, it then holds that
\[
  (\tilde{P}_\res \circ \tilde{y}_p) \, \nabla \tilde{y}_p
  = (P_\res \circ y_p \circ z^{-1}) \, (\nabla y_p \circ z^{-1}) \, \nabla z^{-1}
  = (P \circ z^{-1}) \, \nabla z^{-1}
  = \tilde{P}.
\]
Moreover, by the same transformations as those for~\eqref{eq:curlP_yp_pushforward} we obtain
\[
  \curl \tilde{P}_\res = \tilde{y}_{p*} [\curl \tilde{P}]
\] 
This shows that also $\tilde{y}_p$, $\tilde{P}_\res$ give a multiplicative Helmholtz decomposition of $\tilde{P}$.

By the same computations as Section~\ref{sc:multiplicative_Helmholtz}, see~\eqref{eq:curlP_yp_pushforward}, we obtain that the consistency relation
\[
  \curl P(t) = \frac12 \sum_{b \in \Bcal} b \otimes S^b(t)
\]
transforms into
\[
  \curl \tilde{P}
  = z_* [\curl P] 
  = \frac12 \sum_{b \in \Bcal} b \otimes (z_* S^b(t))
  = \frac12 \sum_{b \in \Bcal} b \otimes \tilde{S}^b(t),
\]
as required.

A change of variables analogously to~\eqref{eq:We_transform} furthermore shows that the minimization problem~\eqref{eq:multiplicative_Helmholtz_minimum_linearized} transforms as expected:
\begin{align*}
  &\int_{z(\Omega)} \frac12 \absb{\nabla (\hat{y}_p \circ z^{-1}) \, \tilde{P}^{-1} - \Id}_\Ebb^2 \dd x \\
  &\qquad = \int_{z(\Omega)} \frac12 \absb{(\nabla \hat{y}_p \circ z^{-1}) \, \nabla z^{-1} \, (\nabla z^{-1})^{-1} \, (P \circ z^{-1})^{-1} - \Id}_\Ebb^2 \dd x \\
  &\qquad = \int_\Omega \frac12 \absb{\nabla \hat{y}_p P^{-1} - \Id}_\Ebb^2 \dd x.
\end{align*}
So, the minimizer $\tilde{y}_p$ in the transformed configuration equals $y_p\circ z^{-1}$, exactly as defined above. 

For the line energy $\tilde{\Wcal}_l$ in the transformed configuration we have
\begin{align*}
  \tilde{\Wcal}_l
  &= \frac{1}{2} \sum_{b \in \Bcal} \int G^b \bigl( \tilde{y}_{p*} \tilde{S}^b(t) \bigr) \\
  &= \frac{1}{2} \sum_{b \in \Bcal} \int G^b \bigl( \tilde{y}_{p*} z_* S^b(t) \bigr) \\
  &= \frac{1}{2} \sum_{b \in \Bcal} \int G^b \bigl( y_{p*} S^b(t) \bigr) \\
  &= \Wcal_\linet,
\end{align*}
so that we see that also the line energy transforms correctly.

For the free elastic energy $\tilde{\Wcal}_\free$ of the transformed system, we obtain
\begin{align*}
  \tilde{\Wcal}_\free
  &= \frac{1}{2} \int_{\tilde{y}_p(z(\Omega))} \abs{\nabla \tilde{\eta}}_\Ebb^2 \dd x \\
  &= \frac{1}{2} \int_{y_p(\Omega)} \abs{\nabla \eta}_\Ebb^2 \dd x \\
  &= \Wcal_\free.
\end{align*}
Consequently, for the total energy it holds that
\[
  \min \tilde{\Ecal} = \min \Ecal
\]
since there is a one-to-one correspondence of minimizers. Thus, the transformed system behaves exactly the same with respect to the energetics as the original system.

Consider now the plastic flow equation in the form~\eqref{eq:P_forward}, that is,
\[
  P(t) = P(0) + \frac12 \sum_{b \in \Bcal} b \otimes \hodge \pbf_*(S^b \restrict \{0 < \tbf < t\}).
\]
If we change the spatial argument via $z^{-1}$ and then apply $\nabla z^{-1}$ from the right on both sides, we see using the Hodge dual transformation rule~\eqref{eq:hodge_transform2} (that is, $\hodge (L\eta) = (\det L) \, L^{-\T} (\hodge \eta)$) that
\begin{align*}
  \tilde{P}(t)
  &= \tilde{P}(0) + \frac12 \sum_{b \in \Bcal} b \otimes \hodge \pbf_*\bigl( (\id, z)_* S^b \restrict \{0 < \tbf < t\} \bigr) \\
  &= \tilde{P}(0) + \frac12 \sum_{b \in \Bcal} b \otimes \hodge \pbf_*\bigl( \tilde{S}^b \restrict \{0 < \tbf < t\} \bigr),
\end{align*}
whereby also the plastic flow equation transforms in the correct way.

As a consequence of the above calculations, we conclude that for any volume-preserving invertible map $z \colon \Omega \to z(\Omega)$ we may fully transform our system to the transformed reference domain $z(\Omega)$, which indexes the points in a different way, without changing either the dynamics or energetics.

Finally, we consider the special case mentioned above: For $z := y_p$, which we assume to be (globally) invertible, we obtain
\begin{align*}
  \tilde{\eta} &= \eta, \\
  \tilde{P} &= (P \circ y_p^{-1}) \, \nabla y_p^{-1} = P_\res, \\
  \tilde{y}_p &= \id, \\
  \tilde{P}_\res &= P_\res, \\
  \tilde{S}^b &= (\id,y_p)_* S^b,
\end{align*}
which are now \emph{all} defined in the plastically deformed spatial domain $y_p(\Omega)$, so we have rewritten the system from a \textbf{(semi-)Eulerian point of view}.

While here we have only considered \emph{volume-preserving} changes of reference, one could also carry through this argument for general orientation-preserving diffeomorphisms, where, however, additional Jacobians appear due to the volumetric change.

\section{Summary of relations} \label{sc:summary_relations}

The model developed so far is summed up by the following relations:

\begin{enumerate}[(1)] \setlength\itemsep{8pt}
\item Domain: $\Omega \subseteq \R^3$ is open, bounded, and connected.

\item Fundamental variables: $y_p = y_p(t,x) \in \R^3$, for $(t,x) \in [0,\infty) \times \Omega$, the plastic deformation, $P = P(t,x) \in \R^{3 \times 3}$ the plastic distortion, and $\eta = \eta(t,\frarg) \in \R^3$ (defined on $y_p(t,\Omega)$) the free elastic displacement.

\item Plastic restriction: $\det P \equiv 1$, or if additional restrictions are present, $P \in \Pfrak$ for a Lie group $\Pfrak \subseteq \SL(3)$ with Lie algebra $\pfrak = \Lie(\Pfrak)$.

\item Burgers vectors: $b \in \Bcal$ for a finite set $\Bcal$ of lattice vectors.

\item Slip trajectories:
\[
  \Sb = (S^b)_b,
\]
with $S^b$ a $2$-current in $[0,\infty) \times \Omega$ for every Burgers vector $b \in \Bcal$, satisfying the symmetry relations
\[
  S^{-b} = -S^b,  \qquad
  (\partial S^b) \restrict (0,T) \times \Omega = 0.
\]

\item Geometric derivative (dislocation velocity):
\[
  \frac{\DD}{\DD t} S^b \bigg|_t(x)
  := \frac{\pbf\xi^b(t,x)}{\tbf\xi^b(t,x)},
\]
where
\[
  \vec{S}^b = \xi^b \wedge \vec{S}^b|_t, \qquad \xi^b \perp \vec{S}^b|_t, \qquad \abs{\vec{S}^b|_t} = \abs{\xi^b} = 1, \qquad \tbf\xi^b = \xi^b \cdot \ee_0 > 0.
\]

\item Plastic flow (forward) equation:
\[
  P(t) = P(0) + \frac12 \sum_{b \in \Bcal} b \otimes \hodge \pbf_*(S^b \restrict \{0 < \tbf < t\})
\]
or
\[
  \dot{P}(t) = \frac12 \sum_{b\in\Bcal} b \otimes \biggl[\frac{\DD}{\DD t} S^b \bigg|_t \times \tau^b(t) \biggr] \, \tv{S^b(t)}.
\]

\item Elasticity tensor:
\[
  \Ebb := 4\DD^2 \hat W_e(\Id) \qquad\text{($= \DD^2 W_e(\Id)$ on symmetric matrices)},
\]
which acts on a matrix $A = (A^i_j) \in \R^{3 \times 3}$ via $\Ebb A := [\Ebb^{ijkl} A^k_l]^i_j$ and satisfies
\[
  \Ebb^{ijkl} = \Ebb^{klij}, \qquad
  \Ebb^{ijkl} = \Ebb^{jikl},  \qquad
  \Ebb^{ijkl} = \Ebb^{ijlk}.
\]

\item Multiplicative Helmholtz decomposition:
\[
  P = (P_\res \circ y_p) \, \nabla y_p
  \quad\text{with}\quad
  \left\{\begin{aligned}
    &\det \nabla y_p \equiv 1,\\
    &\text{$y_p$ minimizes $\hat{y}_p \mapsto \int_\Omega W_e(\nabla \hat{y}_p P^{-1}) \dd x$,} \\
    &\curl P_\res = y_{p*}(\curl P),
  \end{aligned}\right.
\]
for $y_p \colon \Omega \to \R^3$ and $P_\res \colon y_p(\Omega) \to \R^{3 \times 3}$.

\item Small-Distortion assumption:
\[
  P_\res = \Id + r,
\]
with $r \colon y_p(\Omega) \to \R^{3 \times 3}$ small.

\item Linearized elastic distortion:
\[
  e := \nabla y_e - r - \Id.
\]

\item Free distortion:
\[
  \nabla \eta = \fr[e] := e + \beta,
\]
where the geometrically-necessary distortion $\beta \colon y_p(\Omega) \to \R^{3 \times 3}$ is the solution to
\[
  \left\{ \begin{aligned}
    - \dive \Ebb \beta &= 0 &&\text{in $y_p(\Omega)$,} \\
    \curl \beta &= \curl P_\res  &&\text{in $y_p(\Omega)$,} \\
    (\Ebb \beta)n &= 0        &&\text{on $\partial y_p(\Omega)$.}
  \end{aligned} \right.
\]

\item Free elastic energy:
\[
  \Wcal_\free := \int_{y_p(\Omega)} \frac12 \abs{\nabla \eta}_\Ebb^2 \dd z.
\]

\item Line-tension energy:
\[
  \Wcal_\linet = \frac{1}{2} \sum_{b \in \Bcal} \int G^b\bigl( y_{p*} S^b(t) \bigr),
\]

\item Total energy:
\[
  \Ecal := \Wcal_\free + \Wcal_\linet - \dprb{f, \eta},
\]
where
\[
  \dprb{f(t),\eta} := \int_{y_p(\Omega)} f \cdot \eta \dd z.
\]

\item Elastic equilibrium equation:
\[
  - \dive (\Ebb \nabla \eta) = f   \qquad \text{in $y_p(\Omega)$.}
\]
\end{enumerate}

In a concrete application, these relations have to be augmented by boundary conditions, e.g., the \textbf{boundary conditions of place}
\[
  y = g   \quad\text{on $\Gamma_\mathrm{Dir} \subseteq \partial \Omega$.}
\]

\section{Energetic formulation} \label{sc:energetic}

What is missing from the list of relations in the previous section is a suitable flow rule relating the dislocation flow to the plastic stresses. In the following we will discuss the so-called energetic formulation, which is also directly amenable to a mathematical analysis. 

Experiments show that plastic flow is rate-dependent, but only slightly so below absolute temperatures of approximately $0.35 \, \vartheta_m$, where $\vartheta_m$ is the melting temperature of the material, see~\cite[Section~78]{GurtinFriedAnand10book}. For instance, in a commonly used power viscosity law, see~\cite[Section~5.4]{LemaitreChaboche90}, the stress depends on the rate with exponent $1/N$. For example, steel with 35\% carbon at 450 \textdegree{}C has $N \approx 15$ and the titanium-aluminium alloy TA6V at 350 \textdegree{}C has $N \approx 120$; more values can be found in Table~6.2 of~\cite{LemaitreChaboche90}. Since all these values of $N$ are \enquote{large}, the flow stress varies only slightly with the flow rate. Thus, plasticity is often considered as being (nearly) rate-independent if the loading is slow enough, and we will consider only this regime in the present section.

\subsection{Abstract rate-independent systems}

We will first look at rate-independent systems abstractly and then apply this approach to our concrete situation in the next section.

Let $\Qcal$ be the \textbf{state space}, which contains all variables (vectors, tensors, or more complex objects) that are necessary to give a complete description of the system at an instant in time. Also assume the existence of a \textbf{total energy functional}
\[
  \Ecal \colon [0,T] \times \Qcal \to \R \cup \{+\infty\},
\]
in analogy to the one defined before. Assume for the moment that $\Ecal$ is \textbf{G\^ateaux differentiable} on its effective domain (where it is real-valued) with respect to the state variable. That means that
\[
  \DD\Ecal(t,q)[v] := \lim_{h \to 0} \frac{\Ecal(t,q+hv)-\Ecal(t,q)}{h},
\]
exists for all states $q \in \Qcal$ with $\Ecal(t,q) < \infty$, and all rates (velocities) $v \in \Qcal_0$, where $\Qcal_0$ may be different from $\Qcal$, but is such that $q+hv \in \Qcal$ for all sufficiently small $\abs{h}$. The functional $\DD\Ecal(t,q)$ is called the \textbf{G\^ateaux differential} of $\Ecal(t,\frarg)$ at $q$.

Further, let
\[
  \Rcal \colon \Qcal_0 \to [0,\infty]
\]
be a convex \textbf{dissipation potential}. 

We may then consider the \textbf{abstract Biot inclusion} for the functionals $\Ecal, \Rcal$ (which is sometimes also called an \textbf{evolutionary doubly-nonlinear (sub-)differential inclusion}), namely, for all $t \in [0,T]$,
\begin{equation} \label{eq:DI}
  -\DD\Ecal(t,q(t)) \in \partial \Rcal(\dot{q}(t)),
\end{equation}
as functionals on $\Qcal_0$, that is, with values in the \textbf{dual space} $\Qcal_0^*$ to $\Qcal_0$ (which by definition contains all continuous linear $\R$-valued functionals on $\Qcal_0$). Here, the~\textbf{subdifferential} $\partial \Rcal(v)$ is given as
\[
  \partial \Rcal(v) := \setb{ F \in \Qcal_0^* }{ \Rcal(v) + \dpr{F,\hat{v}-v} \leq \Rcal(\hat{v})  \quad\text{for all $\hat{v} \in \Qcal_0$}}.
\]
Consequently, we can rewrite~\eqref{eq:DI} as an \textbf{(abstract) evolutionary variational inequality}
\begin{equation} \label{eq:EVI}
  \DD\Ecal(t,q(t))[\hat{v}-\dot{q}(t)] + \Rcal(\hat{v}) - \Rcal(\dot{q}(t)) \geq 0
\end{equation}
for all $\hat{v} \in \Qcal_0$ and $t \in [0,T]$. In particular, $\DD\Ecal(t,q(t))$ takes the role of an \textbf{(abstract) configurational stress}.

The inclusion~\eqref{eq:DI} is nothing else than the balance between elastic and dissipative stresses (or forces). This can be made even clearer by assuming for the moment that $\Rcal$ is differentiable (which in the rate-independent regime it is not), so that
\[
  -\DD\Ecal(t,q(t)) = \DD\Rcal(\dot{q}(t)),
\]
that is, a balance between stresses. Further details on~\eqref{eq:DI}, including its interpretation as a \enquote{generalized gradient flow}, can be found in~\cite{MielkeRoubicek15book}.

The equation~\eqref{eq:DI}, or its equivalent form~\eqref{eq:EVI}, is complemented by an initial condition
\[
  q(0) = q_0 \in \Qcal.
\]

The property that makes the above inclusion \textbf{rate-independent} is the \textbf{positive $1$-homogeneity} of $\Rcal$, i.e.
\[
  \Rcal(\lambda v) = \lambda \Rcal(v)  \qquad\text{for all $\lambda \geq 0$ and $v \in \Qcal_0$,}
\]
which we henceforth assume. So, for all $v, \hat{v} \in \Qcal_0$, $F \in \Qcal_0^*$, and $\lambda > 0$,
\[
  \Rcal(\lambda v) + \dprb{F,\hat{v} - \lambda v} - \Rcal(\hat{v})
    = \lambda \left[ \Rcal(v) + \dprBB{F,\frac{\hat{v}}{\lambda} - v}
      - \Rcal\biggl(\frac{\hat{v}}{\lambda}\biggr) \right],
\]
and substituting $\hat{v}' := \hat{v}/\lambda$, one can cancel $\lambda$ to see that $F \in \partial \Rcal(\lambda v)$ if and only if $F \in \partial \Rcal(v)$. Thus, we arrive at 
\[
  \partial \Rcal(\lambda v) = \partial \Rcal(v).
\]
This expresses that $\partial\Rcal$ is $0$-homogeneous, which now indeed gives \enquote{rate-independence}: Since the time-derivative $\dot{q}$ of the process $q$ occurs inside $\partial \Rcal$ in~\eqref{eq:DI}, the energetic balance is independent of the speed of the solution (and so depends only on the direction $\abs{\dot{q}}^{-1} \dot{q}$ whenever $\dot{q} \neq 0$). This property has profound implications for the behavior of the system. 

Another consequence of rate-independence is that the dissipation can be defined without referring to a time derivative. To this end, let
\[
  \Dcal \colon \Qcal \times \Qcal \to [0,\infty]
\]
be a \textbf{dissipation distance}, satisfying
\[
  \Dcal(q,q)=0
  \qquad\text{and}\qquad
  \Dcal(q_0,q_2) \leq \Dcal(q_0,q_1)+\Dcal(q_1,q_2)
\]
for all $q,q_0,q_1,q_2\in\Qcal$. Symmetry of $\Dcal$ is not required. The \textbf{total dissipation} of a process $q\colon[0,T]\to\Qcal$ on the interval $[s,t]\subseteq[0,T]$ is defined by
\[
  \Diss(q;[s,t]) := \sup\setBB{ \sum_{j=1}^{N} \Dcal\bigl(q(t_{j-1}),q(t_j)\bigr) }{ s = t_0 < t_1 < \cdots < t_N = t }.
\]
If $\Qcal$ is a linear space and the dissipation distance is induced by the positively $1$-homogeneous dissipation potential $\Rcal$ through
\[
  \Dcal(q_0,q_1):=\Rcal(q_1-q_0),
\]
then, for every absolutely continuous process $q$, one has
\begin{equation} \label{eq:total_dissipation_AC}
  \Diss(q;[s,t])
  = \int_s^t \Rcal(\dot q(\tau))\dd\tau.
\end{equation}

We will now define a different notion of solution to~\eqref{eq:DI} that does not require differentiability of either the energy or the process. A process $q\colon[0,T]\to\Qcal$ is called an \textbf{energetic solution} if, for every $t\in[0,T]$ and all $0\leq s<t\leq T$, the following conditions hold:
\[
\left\{
  \begin{aligned}
    &\text{\textbf{(C) Consistency:}} \\
    &\quad\qquad q(t)\in\Qcal, \\[8pt]
    &\text{\textbf{(S) Stability:} For all $\hat q\in\Qcal$,} \\
    &\quad\qquad \Ecal(t,q(t))
      \leq \Ecal(t,\hat q)+\Dcal(q(t),\hat q), \\[8pt]
    &\text{\textbf{(E) Energy balance:}} \\
    &\quad\qquad \Ecal(t,q(t))+\Diss(q;[s,t])
      = \Ecal(s,q(s))
      +\int_s^t \partial_\sigma\Ecal(\sigma,q(\sigma))\dd\sigma.
  \end{aligned}
\right.
\]
This notion was introduced by Mielke and Theil in~\cite{MielkeTheil99,MielkeTheilLevitas02}.

The corresponding \textbf{stability sets} are
\[
  \Scal(t)
  := \setb{q\in\Qcal}{
    \Ecal(t,q)<\infty\text{ and }
    \Ecal(t,q)\leq\Ecal(t,\hat q)+\Dcal(q,\hat q)
    \text{ for all }\hat q\in\Qcal
  },
\]
so that condition~(S) can equivalently be written as
\[
  \text{(S')}\quad q(t)\in\Scal(t).
\]

In the following, assume additionally that $\Qcal=\Qcal_0$ is a linear space, that
\[
  \Dcal(q_0,q_1)=\Rcal(q_1-q_0),
\]
that $q$ is absolutely continuous, and that $\Ecal$ has the differentiability properties required below. Under these assumptions,~\eqref{eq:total_dissipation_AC} holds, and the energetic formulation can be compared with the differential inclusion~\eqref{eq:DI}, or equivalently with the evolutionary variational inequality~\eqref{eq:EVI}.

In the following, we will show that if $\Ecal$ is G\^ateaux differentiable and \emph{convex}, which necessitates $\Qcal$ ($= \Qcal_0$) to be a linear space, the energetic formulation indeed is an extension of the traditional formulations via the differential inclusion~\eqref{eq:DI} or, equivalently, via the evolutionary variational inequality~\eqref{eq:EVI}. So, let $q \colon [0,T] \to \Qcal$ be an energetic solution. The stability condition~(S) for $\hat{q} := q(t) + h\hat{v}$ with $\hat{v} \in \Qcal_0$ and $h > 0$ gives for all $t \in [0,T]$ that
\[
  \frac{\Ecal(t,q(t)+h\hat{v}) - \Ecal(t,q(t))}{h} + \frac{\Rcal(h\hat{v})}{h} \geq 0.
\]
Using the $1$-homogeneity of $\Rcal$ and letting $h \todown 0$, we then obtain
\begin{equation} \label{eq:Sloc}
  \DD\Ecal(t,q(t))[\hat{v}] + \Rcal(\hat{v}) \geq 0.
\end{equation}
On the other hand, writing out the energy balance~(E) with the definition of the dissipation $\Diss(q;[0,t])$, we have
\[
  \Ecal(t,q(t)) + \int_0^t \Rcal(\dot{q}(\tau)) \dd \tau = \Ecal(0,q(0))
    + \int_0^t \partial_t \Ecal(\tau,q(\tau)) \dd \tau.
\]
Differentiating with respect to $t$ and cancelling terms, we obtain by the chain rule that
\[
  \DD\Ecal(t,q(t))[\dot{q}(t)] + \Rcal(\dot{q}(t)) = 0.
\]
Subtracting this from~\eqref{eq:Sloc}, we arrive at
\[
  \DD\Ecal(t,q(t))[\hat{v}-\dot{q}(t)] + \Rcal(\hat{v}) - \Rcal(\dot{q}(t)) \geq 0,
\]
which is~\eqref{eq:EVI}.

We now turn to the converse implication, for which we additionally assume that $\Ecal(t,\frarg)$ is convex. Then,~\eqref{eq:EVI} with $v := h^{-1}\hat{v}$, where $\hat{v} \in \Qcal_0$ and $h > 0$, reads after multiplication with $h$ as
\[
  \DD\Ecal(t,q(t))[\hat{v}-h\dot{q}(t)] + \Rcal(\hat{v}) - h\Rcal(\dot{q}(t)) \geq 0
\]
for almost all $t \in [0,T]$. Letting $h \todown 0$,
\begin{equation} \label{eq:Sloc_w}
  \DD\Ecal(t,q(t))[\hat{v}] + \Rcal(\hat{v}) \geq 0
\end{equation}
and the convexity of $\Ecal(t, \frarg)$ implies
\[
  \Ecal(t,v) - \Ecal(t,q(t)) + \Rcal(v-q(t)) \geq \DD\Ecal(t,q(t))[v-q(t)] + \Rcal(v-q(t)) \geq 0,
\]
where the last inequality follows from~\eqref{eq:Sloc_w} with $\hat{v} := v-q(t)$ for $v \in \Qcal_0$. This is~(S). 

Next,~\eqref{eq:EVI} with $v := 0$ and~\eqref{eq:Sloc_w} with $\hat{v} := \dot{q}(t)$ are the \enquote{$\leq$}- and the \enquote{$\geq$}-part, respectively, of
\[
  \DD\Ecal(t,q(t))[\dot{q}(t)] + \Rcal(\dot{q}(t)) = 0.
\]
Adding $\partial_t \Ecal(t,q(t))$ on both sides and using the chain rule backwards, the equality
\[
  \frac{\mathrm{d}}{\mathrm{d}t} \Ecal(t,q(t)) + \Rcal(\dot{q}(t)) = \partial_t \Ecal(t,q(t))
\]
is established. Integrating this from $0$ to $t \in [0,T]$ yields~(E).

Summing up, the energetic formulation uses neither $\DD\Ecal$, $\partial\Rcal$, nor $\dot q$ in conditions~(C), (S), and~(E), and no linear structure of $\Qcal$ is needed. Under the additional differentiability, convexity, linearity, and absolute-continuity assumptions imposed above, it is equivalent to the traditional formulation via~\eqref{eq:DI}, or equivalently via~\eqref{eq:EVI}.

\subsection{The rate-independent Small-Distortion Geometric Model} \label{sc:SDGM_ERIS}

In this section we will describe the energetic formulation of the Small-Distortion Geometric Model in the rate-independent regime. We use the following variables:
\begin{align*}
  P &\colon [0,T] \times \Omega \to \R^{3 \times 3} \quad \text{the plastic distortion}, \\
  \eta(t) &\colon y_p[P(t)](\Omega) \to \R^3   \quad \text{the elastic displacement}, \\
  \Sb &= (S^b)_b  \quad \text{a slip trajectory system}
\end{align*}
satisfying
\[
  S^{-b} = -S^b, \qquad
  (\partial S^b) \restrict (0,T) \times \Omega = 0
\]
for every Burgers vector $b \in \Bcal$. Here, we assume that the multiplicative Helmholtz decomposition~\eqref{eq:multiplicative_Helmholtz} admits a unique (or otherwise fixed) solution, denoted by
\[
  y_p[P] \colon \Omega \to \R^3, \qquad
  P_\res[P] \colon y_p[P](\Omega) \to \R^{3 \times 3}.
\]
Thus,
\begin{equation} \label{eq:SDGM_Helmholtz_reconstruction}
  P = \bigl(P_\res[P] \circ y_p[P]\bigr) \, \nabla y_p[P],
\end{equation}
where $\det \nabla y_p[P] \equiv 1$, the map $y_p[P]$ minimizes the functional in~\eqref{eq:multiplicative_Helmholtz_minimum} among the admissible plastic deformations, and
\[
  \curl P_\res[P] = y_p[P]_*[\curl P].
\]
If the minimizer is not unique, the notation $y_p[P]$ refers to a fixed admissible selection.

We also require the initial conditions
\[
  P(0) = P_0, \qquad
  \eta(0) = \eta_0,
\]
and, as before,
\[
  S^b(0+) = T_0^b, \qquad\text{where}\qquad
  S^b(0+) := -\pbf_*\bigl[(\partial S^b) \restrict \{0\}\bigr],
\]
for a given admissible initial plastic distortion $P_0 \colon \Omega \to \R^{3 \times 3}$, a given initial elastic displacement $\eta_0 \colon y_p[P_0](\Omega) \to \R^3$, and a given initial dislocation system $\Tb_0 = (T_0^b)_b$ satisfying
\[
  T_0^{-b} = -T_0^b, \qquad
  \partial T_0^b = 0, \qquad
  \curl P_0 = \frac{1}{2} \sum_{b \in \Bcal} b \otimes T_0^b
\]
for every $b \in \Bcal$.

For the internal state
\[
  Z = (P,\Tb),
\]
consisting of plastic distortion and a dislocation system, the total energy of the system is defined to be
\[
  \Ecal(t,\eta,Z) := \Wcal_\free(P,\eta) + \Wcal_\linet(P,\Tb) - \dprb{f(t),\eta},
\]
where
\begin{align*}
  \Wcal_\free(P,\eta) &:= \int_{y_p[P](\Omega)} \frac{1}{2} \abs{\nabla\eta}_\Ebb^2 \dd x', \\
  \Wcal_\linet(P,\Tb) &:= \frac{1}{2} \sum_{b \in \Bcal} \int_{y_p[P](\Omega)} G^b\bigl(\vec{T}_{y_p[P]}^b\bigr) \dd \tv{T_{y_p[P]}^b},
\end{align*}
and
\[
  T_{y_p[P]}^b := y_p[P]_*T^b, \qquad b \in \Bcal.
\]
Here, $G^b \colon \R^3 \to [0,\infty)$ ($b \in \Bcal$) are the absolutely $1$-homogeneous line-tension densities. To make the loading term meaningful for every competitor, we assume that $f(t,\frarg)$ is defined on a fixed ambient set containing every admissible intermediate configuration $y_p[P](\Omega)$.

We next define the forward operator on the joint internal state. Let $\hat{\Sb} = (\hat{S}^b)_b$ be a slip trajectory system on the interval $[0,1]$, starting from the dislocation system $\Tb = (T^b)_b$. This initial condition means that, for suitable terminal currents $\hat T^b$,
\begin{equation} \label{eq:SDGM_competitor_boundary}
  \partial \hat{S}^b = \delta_1 \times \hat T^b - \delta_0 \times T^b, \qquad b \in \Bcal.
\end{equation}
The updated plastic distortion and terminal dislocation system are defined by
\begin{align*}
  \hat{P} &:= P + \frac{1}{2} \sum_{b \in \Bcal} b \otimes \hodge\pbf_* \bigl(\hat{S}^b \restrict \{0 < \tbf < 1\}\bigr), \\
  \hat{\Tb} &:= (\hat T^b)_b.
\end{align*}
We then define the \textbf{joint forward operator} by
\[
  \hat{\Sb}_\ff Z := (\hat P,\hat\Tb).
\]
Thus, the first component records the evolved plastic distortion, whereas the second component is the terminal dislocation system. The competitor plastic deformation is $y_p[\hat P]$, obtained from the multiplicative Helmholtz reconstruction~\eqref{eq:SDGM_Helmholtz_reconstruction}; it is not varied independently. The forward operator is defined only for trajectories satisfying~\eqref{eq:SDGM_competitor_boundary} for which the updated distortion $\hat P$ belongs to the admissible class of the multiplicative Helmholtz problem.

For the dissipation we use $\Diss = \Diss_1$, defined in~\eqref{eq:Diss1}, that is, with $R^b_1 \colon \R^3 \to [0,\infty]$ the ($1$-homogeneous, convex) rate-independent dissipation potential,
\begin{align*}
  \Diss([s,t]) 
  &:= \frac12 \sum_{b\in\Bcal} \int_s^t \int_\Omega R^b_1 \biggl( P(\tau)^{-\T} \frac{\hodge \pbf \vec{S}^b}{\tbf\xi^b} \biggr) \dd \tv{S^b(\tau)} \dd \tau\\
  &\phantom{:}= \frac12 \sum_{b\in\Bcal} \int_{[s,t] \times \Omega} R^b_1 \bigl( P(\tau)^{-\T} \hodge \pbf \vec{S}^b \bigr) \dd \tv{S^b},
\end{align*}
where we also have used~\eqref{eq:gb_geometric_nowedge} to express the geometric slip rate.

Bundling
\[
  Z(t) := (P(t),\Sb(t)),
\]
which has initial value
\[
  Z_0 := (P_0,\Tb_0),
\]
our notion of energetic solution is the following:
\[
\left\{
  \begin{aligned}
    &\text{\textbf{(C) Consistency:}} \\
    &\quad\qquad
      S^{-b} = -S^b, \quad
      (\partial S^b) \restrict (0,T) \times \Omega = 0, \\
    &\quad\qquad
      \curl P(t) = \frac{1}{2} \sum_{b \in \Bcal} b \otimes S^b(t), \quad
      \det P(t) = 1 \\[8pt]
    &\text{\textbf{(S) Stability:} For all admissible $\hat{\Sb}$ on $[0,1]$ starting from $\Sb(t)$} \\
    &\text{\phantom{\textbf{(S) Stability:}} and all $\hat\eta \colon y_p[\hat P](\Omega) \to \R^3$, where $(\hat P,\hat\Tb) = \hat{\Sb}_\ff Z(t)$,} \\
    &\quad\qquad
      \Ecal(t,\eta(t),Z(t)) \leq \Ecal(t,\hat\eta,\hat{\Sb}_\ff Z(t)) + \Diss(\hat{\Sb}) \\[8pt]
    &\text{\textbf{(E) Energy balance:}} \\
    &\quad\qquad
      \Ecal(t,\eta(t),Z(t)) = \Ecal(0,\eta_0,Z_0) - \Diss(\Sb;[0,t]) - \int_0^t \dprb{\dot f(\tau),\eta(\tau)} \dd \tau \\[8pt]
    &\text{\textbf{(P) Plastic flow:}} \\
    &\quad\qquad
      P(t) = P_0 + \frac{1}{2} \sum_{b \in \Bcal}  b \otimes \hodge\pbf_* \bigl(S^b \restrict \{0 < \tbf < t\}\bigr).
  \end{aligned}
\right.
\]

\section{Small-strain linearization} \label{sc:SDGM_linearization}

We now consider the small-strain linearization of the Small-Distortion Geometric Model, which entails that the plastic deformation $y_p$ is close to the identity map.

\subsection{Linearization}

The \textbf{small-strain hypothesis} is the following:

\begin{hypothesis}
For the plastic deformation $y_p$ (determined from the Small-Distortion Hypothesis in Section~\ref{sc:small_distortion_hypothesis}) it holds that
\[
  y_p(x) = x + u_p(x),  \qquad x \in \Omega,
\]
with a small plastic displacement $u_p \colon \Omega \to \R^3$, and $\nabla y_p$ is close to the identity field everywhere. 
\end{hypothesis}

As expected, this additional assumption leads to significant simplifications, most notably that the multiplicative Kr\"{o}ner--Lee decomposition linearizes into an additive decomposition. Of course, in general it is not true that $y_p$ (and its gradient) is close to the identity; however, this is a reasonable assumption for a short amount of time after the start of plastic flow from the reference configuration. 

For the derivation let us augment our previous notation as follows: Quantities defined on $y_p(\Omega)$ are identified by the overset symbol \enquote{$\check{\phantom{m}}$}. With this convention, the Small-Distortion Hypothesis in Section~\ref{sc:small_distortion_hypothesis} reads as
\[
  P = (\Id + \check{r} \circ y_p) \, \nabla y_p
\]
for a smoothly invertible map $y_p \colon \Omega \to \R^3$ and a small $\check{r} \colon y_p(\Omega) \to \R^{3 \times 3}$. We know from~\eqref{eq:nabla_ye_decomp_small_distortion} that for the corresponding elastic deformation $\check{y}_e \colon y_p(\Omega) \to \R^3$ it then holds that
\begin{equation} \label{eq:ye_lin}
  \nabla \check{y}_e \approx \Id + \check{e} + \check{r},  \qquad \check{e} = \fr[\check{e}] - \check{\beta},
\end{equation}
with the free elastic distortion $\fr[\check{e}] \colon y_p(\Omega) \to \R^{3 \times 3}$ and the geometrically-necessary distortion $\check{\beta} \colon y_p(\Omega) \to \R^{3 \times 3}$, which solves
\[
  \left\{ \begin{aligned}
    - \dive \Ebb \check{\beta} &= 0 &&\text{in $y_p(\Omega)$,} \\
    \curl \check{\beta} &= \curl \check{r}  &&\text{in $y_p(\Omega)$,} \\
    (\Ebb \check{\beta})n &= 0        &&\text{on $\partial y_p(\Omega)$.}
  \end{aligned} \right.
\]
Plugging~\eqref{eq:ye_lin} into~\eqref{eq:y_ye_yp}, that is, $\nabla y = (\nabla \check{y}_e \circ y_p) \, \nabla y_p$, and neglecting lower-order terms when multiplying out (using that $\Id + \nabla u_p \approx \Id$), we obtain
\begin{align*}
  \nabla y 
  &\approx \bigl(\Id + \check{e} \circ y_p + \check{r} \circ y_p \bigr) \, \nabla y_p \\
  &= \bigl(\fr[\check{e}] \circ y_p - \check{\beta} \circ y_p \bigr) \, (\Id + \nabla u_p) + P \\
  &\approx \Id + \fr[\check{e}] \circ y_p - \check{\beta} \circ y_p + p,
\end{align*}
where we have defined the \textbf{linearized plastic distortion} as before via $p := P - \Id$.

We retain throughout this section the rigid-motion normalization fixed in Section~\ref{sc:small_distortion_hypothesis}; in particular, the elastic deformation is understood in the corresponding co-rotated frame.

Since $y_p$ is assumed to be close to the identity by the small-strain hypothesis, we then have that
\[
  \check{\beta} \approx \beta \circ y_p^{-1}
\]
for $\beta \colon \Omega \to \R^{3 \times 3}$ solving
\[
  \left\{ \begin{aligned}
    - \dive \Ebb \beta &= 0 &&\text{in $\Omega$,} \\
    \curl \beta &= \curl p  &&\text{in $\Omega$,} \\
    (\Ebb \beta)n &= 0        &&\text{on $\partial \Omega$.}
  \end{aligned} \right.
\]
Hence, for the \textbf{total displacement} $u \colon \Omega \to \R^3$ defined as
\[
  u(x) := y(x) - x,  \qquad x \in \Omega,
\]
we obtain the relation
\begin{equation} \label{eq:nabla_u_linearization_approx}
  \nabla u \approx p + \fr[\check{e}] \circ y_p - \beta.
\end{equation}

Defining the \textbf{linearized elastic distortion} via
\[
  e := \nabla u - p,
\]
the \textbf{linearized Kr\"{o}ner--Lee decomposition}
\[
  \nabla u = e + p
\]
holds (trivially). Moreover, the \textbf{(linearized) free distortion}
\begin{equation} \label{eq:fr_e_linearized}
  \bar{\fr}[\nabla u - p] := \nabla u - p + \beta
\end{equation}
is curl-free (hence a gradient if $\Omega$ is simply connected), depends only on $\nabla u - p$ (since $\beta$ depends only on $\curl p = -\curl (\nabla u - p)$), and satisfies (via~\eqref{eq:nabla_u_linearization_approx})
\[
  \bar{\fr}[\nabla u - p] \approx \fr[\check{e}] \circ y_p.
\]
Thus, from the definition of the free elastic energy in~\eqref{eq:Wfree} and since also $\det \nabla y_p \approx 1$,
\begin{align*}
  \Wcal_\free
  &= \int_{y_p(\Omega)} \frac12 \abs{\fr[\check{e}]}_\Ebb^2 \dd z  \\
  &\approx \int_{y_p(\Omega)} \frac12 \abs{(\nabla u - p + \beta) \circ y_p^{-1}}_\Ebb^2 \dd z  \\
  &\approx \int_{\Omega} \frac12 \abs{\nabla u - p + \beta}_\Ebb^2 \dd x  \\
  &= \int_{\Omega} \frac12 \abs{\bar{\fr}[\nabla u - p]}_\Ebb^2 \dd x.
\end{align*}
In particular, the \textbf{linearized elastic energy}
\begin{equation} \label{eq:Wfree_linearized}
  \overline{\Wcal}_\free
  := \int_{\Omega} \frac12 \abs{\bar{\fr}[\nabla u - p]}_\Ebb^2 \dd x
\end{equation}
depends only on the (linearized) free distortion $\bar{\fr}[\nabla u - p]$.

For the \textbf{linearized line-tension energy} we set
\begin{equation} \label{eq:Wline_linearized}
  \overline{\Wcal}_\linet := \frac{1}{2} \sum_{b \in \Bcal} \int G^b(S^b(t)),
\end{equation}
that is,
\[
  \overline{\Wcal}_\linet = \frac{1}{2} \sum_{b \in \Bcal} \int G^b(\vec{S}^b(t)) \dd \tv{S^b(t)}.
\]

We now turn to the plastic flow equation. Since $\dot{P} = \dot{p}$, the linearized form of~\eqref{eq:dotP_geometric_nowedge} is the same, namely
\[
  \dot{p}(t) = \frac12 \sum_{b \in \Bcal} b \otimes \biggl(\frac{\hodge \pbf \vec{S}^b(t)}{\tbf\xi^b(t)} \, \tv{S^b(t)}\biggr).
\]
As before we may integrate this over time to arrive at an analogous expression to~\eqref{eq:P_forward}, namely
\[
  p(t) = p(0) + \frac12 \sum_{b \in \Bcal} b \otimes \hodge \pbf_*(S^b \restrict \{0 < \tbf < t\}). 
\]

\begin{remark} \label{rem:linearization_singularities}
Our derivation of the plastic small-strain limit never assumes that \emph{gradients} of quantities with non-zero curl, such as $e,p,r,\beta$, are small. This is a technical, but important point since these gradients in general may indeed not be small: For instance, in~\eqref{eq:beta_line} we saw for a $\beta$ with curl concentrated in a (dislocation) line that $\abs{\beta} \approx \abs{b} r^{-1}$ with $r$ the distance to the dislocation line. This approximation is valid for $r \geq k \eps$ with $\eps$ denoting the lattice size and $k$ a constant (on the order of $10$). If also the Burgers vector $b$ scales with the lattice spacing, $\abs{b} \approx \eps$, then we obtain
\[
  \abs{\beta} \lessapprox \frac{\eps}{k\eps} \ll 1
\]
away from the singularity and the smallness assumption is valid. For the gradient of $\beta$, however, one has $\abs{\nabla \beta} \approx \abs{b} r^{-2}$ and so
\[
  \abs{\nabla \beta} \approx \frac{\eps}{k^2\eps^2} \gg 1
\]
for $\eps$ small.
\end{remark}

\subsection{Energetic formulation in the small-strain rate-independent regime}

In analogy to Section~\ref{sc:SDGM_ERIS} we now give an energetic formulation of the Linearized Small-Distortion Geometric Model. For this we will use the following variables:
\begin{align*}
  u  &\colon [0,T] \times \Omega \to \R^3 \quad \text{the total displacement}, \\
  p  &\colon [0,T] \times \Omega \to \R^{3 \times 3} \quad \text{the (linearized) plastic distortion}, \\
  \Sb &= (S^b)_b \quad \text{a slip trajectory system}
\end{align*}
satisfying
\[
  S^{-b} = -S^b,  \qquad
  (\partial S^b) \restrict (0,T) \times \Omega = 0
\]
for every Burgers vector $b \in \Bcal$. We also require the initial conditions
\[
  u(0) = u_0, \qquad
  p(0) = p_0,
\]
and, as before,
\[
  S^b(0+) = T^b_0,  \qquad\text{where}\qquad
  S^b(0+) := - \pbf_* \bigl[(\partial S^b) \restrict \{0\}\bigr],
\]
for a given initial total displacement $u_0 \colon \Omega \to \R^3$, a given initial plastic distortion $p_0 \colon \Omega \to \R^{3 \times 3}$, and a given initial dislocation system $\Tb_0 = (T^b_0)_b$ (satisfying $T^{-b}_0 = -T^b_0$ and $\partial T^b_0 = 0$ for all $b \in \Bcal$).

Then, the total energy of the system is now defined to be
\[
  \Ecal(t,u,p,\Tb) := \overline{\Wcal}_\free(u,p) + \overline{\Wcal}_\linet(\Tb) - \dprb{f(t), u}
\]
where
\begin{align*}
  \overline{\Wcal}_\free(u,p) &:= \frac{1}{2} \int_\Omega \abs{\bar{\fr}[\nabla u - p]}_\Ebb^2 \dd x, \\
  \overline{\Wcal}_\linet(\Tb) &:= \frac{1}{2} \sum_{b \in \Bcal} \int_\Omega G^b(\vec{T}^b) \dd \tv{T^b},
\end{align*}
and $f = f(t,x)$ is the external loading (which now acts on $u$ in accordance with the small-strain hypothesis). Here, $G^b \colon \R^3 \to [0,\infty)$ ($b \in \Bcal$) are the absolutely $1$-homogeneous (that is, $G^b(\alpha v) = \abs{\alpha} G^b(v)$ for all $\alpha \in \R$) line-tension densities. The dissipation and forward operators are exactly as before.

For the dissipation we now use
\[
  \Diss([s,t]) = \frac12 \sum_{b\in\Bcal} \int_{[s,t] \times \Omega} R^b_1 \bigl( \hodge \pbf \vec{S}^b \bigr) \dd \tv{S^b},
\]
with $R^b_1 \colon \R^3 \to [0,\infty]$ the ($1$-homogeneous, convex) rate-independent dissipation potential.

Bundling
\[
  Z(t) := (p(t),\Sb(t)),
\]
where $\Sb(t) := (S^b(t))_b$ is the dislocation system at time $t$, which has initial value
\[
  Z_0 := (p_0,\Tb_0),
\]
we extend the forward operator to the joint internal variable as follows: Let $\hat{\Sb}=(\hat{S}^b)_b$ be a slip trajectory system on $[0,1]$ starting from $\Tb$. We define the \textbf{joint forward operator} by
\[
  \hat{\Sb}_\ff Z
  := \biggl(p + \frac12 \sum_{b\in\Bcal} b \otimes \hodge\pbf_*\bigl( \hat{S}^b \restrict \{0<\tbf<1\} \bigr), \hat{\Sb}_\ff\Tb \biggr).
\]
Thus, the first component records the evolved plastic distortion, whereas the second component is the terminal dislocation system. As before, this operator is defined only when $\hat{\Sb}$ starts from $\Tb$, that is, when for each $b\in\Bcal$ there exists a terminal current $\hat T^b$ such that
\[
  \partial\hat{S}^b=\delta_1\times\hat T^b-\delta_0\times T^b.
\]
We also use the shorthand
\[
  \Ecal(t,u,Z):=\Ecal(t,u,p,\Tb) \qquad\text{for $Z=(p,\Tb)$.}
\]
Then, our notion of energetic solution is the following:
\[
\left\{
  \begin{aligned}
    &\text{\textbf{(C) Consistency:}} \\
    &\quad\qquad  S^{-b} = -S^b, \quad (\partial S^b) \restrict (0,T) \times \Omega = 0, \quad \curl p(t) = \frac12 \sum_{b \in \Bcal} b \otimes S^b(t)  \\[8pt]
    &\text{\textbf{(S) Stability:} For all $\hat{u} \colon \Omega \to \R^3$ and all $\hat{\Sb}$ (on $[0,1]$) starting from $\Sb(t)$,} \\
    &\quad\qquad  \Ecal(t,u(t),Z(t)) \leq \Ecal(t,\hat{u},\hat{\Sb}_\ff Z(t)) + \Diss(\hat{\Sb}) \\[8pt]
    &\text{\textbf{(E) Energy balance:}} \\
    &\quad\qquad  \Ecal(t,u(t),Z(t)) = \Ecal(0,u_0,Z_0) - \Diss(\Sb;[0,t]) - \int_0^t \dprb{\dot{f}(\tau),u(\tau)} \dd \tau. \\[8pt]
    &\text{\textbf{(P) Plastic flow:}} \\
    &\quad\qquad  p(t) = p_0 + \frac12 \sum_{b \in \Bcal} b \otimes \hodge\pbf_* \bigl(S^b \restrict \{0 < \tbf < t\}\bigr)
  \end{aligned}
\right.
\]
Thus, no additional differential equation for the plastic distortion is required in this formulation. The value of $p(t)$ is determined directly by the initial plastic distortion $p_0$ and the slip trajectory accumulated up to time $t$.

\begin{remark}
Also constraints on $p$, such as the infinitesimal incompressibility constraint $\tr p \equiv 0$, can be incorporated by either constraining motions to respect this constraint or by projecting on the (linear) constraint space. We have not done this here to keep the formulas simpler.
\end{remark}

\subsection{The Geometric Transport Equation}

So far, we have specified the evolution of dislocations in the space-time framework, which allows us to express the effect of \enquote{applying} a slip trajectory to the dislocation system, see in particular~\eqref{eq:S_forward}. However, to derive a PDE-version of our equations it is necessary to formulate a transport equation for currents.

To derive such a Geometric Transport Equation, we first consider a simple form of motion: Let $v \colon \R^d \to \R^d$ be a smooth, globally bounded \emph{autonomous} (i.e., time-independent) vector field. We recall that in this situation there exists a unique flow of $v$, that is, a family of smooth maps $\Phi \colon \R \times \R^d \to \R^d$ such that 
\[
  \left\{
  \begin{aligned}
	\frac{\di}{\di t} \Phi(t,x) &= v(\Phi(t,x)), \\
	\Phi(0,x) &= x.
  \end{aligned}
  \right.
\]
We will write
\[
  \Phi^t(x) := \Phi(t,x)
\]
and call the family $(\Phi^t)_{t \in \R}$ the \textbf{flow} of $v$. Standard facts ensure that the family $(\Phi^t)_{t \in \R}$ is a family of smoothly invertible maps (diffeomorphisms) of $\R^d$ to itself and that the group properties
\begin{equation}\label{eq:flow_group}
  \Phi^t \circ \Phi^{s} = \Phi^{t+s} = \Phi^s \circ \Phi^t,  \qquad \Phi^0 = \id
\end{equation}
hold for all $s,t \in \R$.

The natural extension of this flow to a \mbox{$1$-current} $T$ is the pushforward $\Phi^t_* T$, as defined in Section~\ref{sc:operations_on_currents}. According to~\eqref{eq:pushforward_def}, we have the expression
\[
  \dprb{\Phi^t_* T, \omega}
  = \int \dprb{\DD \Phi^t[\vec{T}], \omega \circ \Phi^t} \dd \tv{T}.
\]
for any smooth $1$-form $\omega$. 

To express a derivative of $T$ \emph{along the flow of $v$}, we now define the \textbf{Lie derivative} $\Lcal_v T$ of $T$ in direction $v$ as
\[
  \Lcal_v T := \lim_{h \to 0} \frac{1}{h} \bigl[ \Phi^h_* T - T \bigr].
\]
However, this is not a rigorous notion, since we would need to first introduce a suitable notion of convergence for currents (and even if we did, it would not be clear that this limit exists). We will see below how to furnish a better definition following~\cite{BonicattoDelNinRindler25} and proceed on an intuitive level for now. Fix $h > 0$ and first consider the cylinder \mbox{$2$-current} $\dbr{0,h} \times T$ obtained from $T$ by taking the Cartesian product with the interval $[0,h]$, which is oriented by the $2$-vector $\ee_0 \wedge \vec{T}$. Pushing forward this cylinder with $\Phi$ gives a distorted cylinder $\Phi_* (\dbr{0,h} \times T)$. To compute its boundary we use the rules in Section~\ref{sc:operations_on_currents} as well as the cylinder boundary formula
\[
  \partial(\dbr{0,h} \times T) = \delta_h \times T - \delta_0 \times T - \dbr{0,h} \times \partial T,
\]
where $\delta_0, \delta_h$ are the point mass $0$-currents at $0$ and $h$, respectively. So,
\begin{equation} \label{eq:Phi_homotopy}
  \partial \Phi_*(\dbr{0,h} \times T)
  = \Phi_* \partial(\dbr{0,h} \times T)
  = \Phi^h_* T - T - \Phi_*(\dbr{0,h} \times \partial T).
\end{equation}
Applying~\eqref{eq:Phi_homotopy} in rearranged form to the above definition of Lie derivative, we obtain
\[
  \Lcal_v T = \lim_{h \to 0} \frac{1}{h} \bigl[ \partial \Phi_*(\dbr{0,h} \times T) + \Phi_*(\dbr{0,h} \times \partial T) \bigr].
\]
Now observe that $\Phi_*(h^{-1} \dbr{0,h} \times T)$ converges (in a suitable sense) to $v \wedge \vec{T} \, \tv{T}$. While we cannot make this fully rigorous without much more theory at this point, it should be intuitively clear since $\Phi$ is the flow of the vector field $v$ and so looks like the map $(t,x) \mapsto x + tv$ near $t = 0$, whereby the cylinder $\Phi_*(h^{-1} \dbr{0,h} \times T)$ looks roughly like the translation of $T$ along the line with direction $v$ starting from the origin. Taking this (and the corresponding fact relating to $\partial T$) for granted, we arrive at
\begin{equation} \label{eq:Lie_current_Cartan}
  \Lcal_v T = \partial (v \wedge T) + v \wedge \partial T.
\end{equation}
We call this the \textbf{(dual) Cartan formula} since it is the analogue for currents of the classical \emph{Cartan \enquote{magic} formula} for the Lie derivative of a differential form. From now on we take $\Lcal_v T$ to be \emph{defined} through~\eqref{eq:Lie_current_Cartan}.

\begin{remark}
A more rigorous derivation of~\eqref{eq:Lie_current_Cartan} proceeds as follows: The pushforward $\Phi^t_* T$, as defined in Section~\ref{sc:operations_on_currents} satisfies
\[
  \dprb{\Phi^t_* T, \omega} = \dprb{T, (\Phi^t)^* \omega}
\]
for any smooth $1$-form $\omega$. Here, $(\Phi^t)^* \omega$ is the \textbf{pullback} of $\omega$ under $\Phi^t$, which is defined via
\[
  \dprb{\xi, ((\Phi^t)^* \omega)(x)} := \dprb{\DD \Phi^t(x)[\xi], \omega(\Phi^t(x))},  \qquad \xi \in \Wedge_1 \R^d.
\]
The point of this notational shifting is that $(\Phi^t)^* \omega$ is still a smooth $1$-form and $(\Phi^t)^* \omega$ can be computed by taking classical derivatives of $\omega$, which is always well-defined (even if $T$ is very singular). We then compute, using also the group property~\eqref{eq:flow_group},
\begin{align*}
  \frac{\di}{\di t} \dprb{\Phi^t_* T, \omega}
  &= \frac{\di}{\di t} \dprb{T, (\Phi^t)^* \omega} \\
  &= \lim_{h \to 0} \frac{1}{h} \dprb{T, (\Phi^{t+h})^* \omega - (\Phi^t)^* \omega} \\
  &= \lim_{h \to 0} \frac{1}{h} \dprb{\Phi^t_* T, (\Phi^h)^* \omega - \omega} \\
  &= \dprBB{\Phi^t_* T, \lim_{h \to 0} \frac{(\Phi^h)^* \omega - \omega}{h}}.
\end{align*}
The expression on the right in the duality bracket is a well-known quantity in differential geometry, namely the \textbf{Lie derivative} $\Lcal_v \omega$ of the form $\omega$ in direction $v$, that is,
\[
  \Lcal_v \omega := \lim_{h \to 0} \frac{(\Phi^h)^* \omega - \omega}{h}.
\]
This is again a smooth $1$-form. The work~\cite{BonicattoDelNinRindler25} then introduced the \textbf{Lie derivative} $\Lcal_v T$ of a current $T$ by duality, that is,
\begin{equation} \label{eq:Lie_current_duality}
  \dprb{\Lcal_v T, \omega} := \dprb{T, \Lcal_v \omega}.
\end{equation}
Note, however, that we use the opposite sign convention here to that work, which seems to fit better with differential geometric notions. The well-known \textbf{Cartan \enquote{magic} formula} (see, e.g., Theorem~14-35 in~\cite{Lee13book}) says that
\[
  \Lcal_v \omega = \iota_v (\di\omega) + \di (\iota_v \omega),
\]
where $\di \omega$ is the exterior differential of $\omega$ (see Section~\ref{sc:smooth_currents} for the definition) and $\iota_v$ acts on a $2$-covector $\alpha$ as $\dpr{\xi, \iota_v \alpha} := \dpr{v \wedge \xi, \alpha}$ for any $2$-vector $\xi$. Combining this with~\eqref{eq:Lie_current_duality} and using that $\partial$ is dual to $\di$ (see~\eqref{eq:boundary_Stokes}), we obtain
\[
  \dprb{\Lcal_v T, \omega}
  = \dprb{T, \iota_v (\di\omega) + \di (\iota_v \omega)}
  = \dprb{\partial(v \wedge T) + v \wedge \partial T, \omega},
\]
that is,~\eqref{eq:Lie_current_Cartan}.
\end{remark}

The above derivation thus yields that for the special flow along the autonomous vector field $v$ it holds that
\[
  \frac{\di}{\di t} \dprb{\Phi^t_* T, \omega} = \dprb{\Lcal_v[\Phi^t_* T], \omega}
\]
for all smooth $1$-forms $\omega$. We now generalize this discussion as follows: Call a time-indexed family $(T_t)_{t \geq 0}$ of \mbox{$1$-currents} a solution to the \textbf{Geometric Transport Equation}
\begin{equation} \label{eq:GTE}
  \frac{\di}{\di t} T_t - \Lcal_v T_t = 0
\end{equation}
where $v = v(t,x) \colon \R^{1+3} \to \R^3$ is a non-autonomous (time-dependent) vector field, if
\[
  \frac{\di}{\di t} \dprb{T_t, \omega} 
  = \dprb{\Lcal_v T_t, \omega}
  = \dprb{\partial (v \wedge T_t) + v \wedge \partial T_t, \omega}
\]
holds for all smooth $1$-forms $\omega$ (in space). This formulation has no ill-defined space-derivatives any longer, so even makes sense if $T_t$ is supported on singular sets such as a collection of curves.

We now consider the Geometric Transport Equation~\eqref{eq:GTE} in coordinates. So, assume that we have a family of 1-currents $(T_t)_{t \geq 0}$ satisfying the Geometric Transport Equation~\eqref{eq:GTE}. In the following, we will take derivatives of $T_t$. This can be understood in two ways: Either we suppose that the $T_t$ are densities, so that they can be identified with a time-dependent (1-)vector field $\tau = (\tau_1, \tau_2, \tau_3)$, or we take derivatives in the distributional sense. Both approaches are equivalent if the $T_t$ are densities, but distributional derivatives also make sense for very singular objects.

In this way, for every smooth $1$-form $\omega$, which we write as
\[
  \omega = \sum_{i=1}^3 \hat{\omega}_i \, \di x^i,  \qquad
  \hat{\omega} := \begin{bmatrix} \hat{\omega}_1 \\ \hat{\omega}_2 \\ \hat{\omega}_3 \end{bmatrix},
\]
it holds that (with $\dive$ as usual denoting the row-wise divergence) 
\[
  - \dprb{\partial (v \wedge T_t), \omega}
  = \int [\dive(T_t \otimes v - v\otimes T_t)] \cdot \hat{\omega} \dd x
\]
and 
\[
  - \dprb{v\wedge \partial T_t, \omega }
  = \int \dive(T_t) \, v \cdot \hat{\omega} \dd x.
\]
We conclude that~\eqref{eq:GTE} in this case reads as
\[
  \frac{\di}{\di t} T_t + \dive(T_t \otimes v - v\otimes T_t) + v \dive T_t = 0. 
\]
If we further assume that every $T_t$ is boundaryless (equivalently, $\dive T_t = 0$ for every $t$) and use the classical vector calculus identity  
\[
  \dive(f \otimes g - g \otimes f) = - \curl(g \times f), 
\]
we obtain that in coordinates~\eqref{eq:GTE} reads as
\begin{equation} \label{eq:GTE_1_coordinates}
  \frac{\di}{\di t} T_t - \curl(v \times T_t) = 0.
\end{equation}
This transport equation for curves is to be understood in a weak sense: For any smooth and compactly supported test vector field $\psi \colon \R^{1+3} \to \R^3$ it holds that
\[
  \int_{\R^{1+3}} - T_t(x) \cdot \frac{\di}{\di t} \psi(t,x) - (v(t,x) \times T_t(x)) \cdot \curl_x \psi(t,x) \dd x \dd t = 0.
\]
This follows via an integration by parts and using that $\curl$ is self-adjoint (since $\psi$ is zero outside a bounded set). The weak formulation makes sense both for individual curves, with $T_t$ being interpreted as a measure on $\R^3$ for all times $t$, as well as for fields of lines, with $T_t$ a vector field. Moreover, also time-dependent driving velocity fields $v = v(t,x)$ can be handled analogously.

\begin{remark}
We note in passing that if one writes~\eqref{eq:GTE} in coordinates for densities $\rho = \rho(t,x)$, considered as $0$-currents (mass densities), one obtains instead the \emph{continuity equation}
\[
  \frac{\di}{\di t} \rho + \dive(v \rho) = 0.
\]
Furthermore, for scalar fields $\alpha = \alpha(t,x)$ in $\R^3$, considered as $3$-currents (field strengths), the Geometric Transport Equation~\eqref{eq:GTE} reads as the classical \emph{transport equation}
\[
  \frac{\di}{\di t} \alpha + v \cdot \nabla \alpha = 0.
\]
\end{remark}

The final task in this section is to relate the Geometric Transport Equation to our existing approach with space-time currents. For this, we will show that for a space-time \mbox{$2$-current} $S^b$ with boundaryless slices $S^b(t) := \pbf_*(S^b|_t)$ these slices are advected in the sense of~\eqref{eq:GTE} by the geometric derivative
\[
  v^b := \frac{\DD}{\DD t} S^b,
\]
that is,
\begin{equation} \label{eq:GTE_Sbt}
  \frac{\di}{\di t} S^b(t) - \Lcal_{v^b} S^b(t)
  = \frac{\di}{\di t} S^b(t) - \partial(v^b \wedge S^b(t)) = 0.
\end{equation}
The arguments in Section~\ref{sc:plastic_flow}, in particular see~\eqref{eq:P_via_S_part1},~\eqref{eq:P_via_S_part2},~\eqref{eq:P_via_S_part3}, show that for any $h > 0$ we have
\begin{align*}
  \int_t^{t+h} v^b \wedge S^b(s) \dd s
  &= \int_t^{t+h} \frac{\pbf\xi^b}{\tbf\xi^b} \wedge \pbf\vec{S}^b|_s \dd s \\
  &= \int_t^{t+h} \pbf \vec{S}^b \; \tv{S^b(s)} \; \frac{\di s}{\abs{\tbf\xi^b}} \\
  &= \pbf_*(S^b \restrict \{t < \tbf < t+h\}).
\end{align*}
Thus,
\[
  \int_t^{t+h} \partial(v^b \wedge S^b(s)) \dd s
  = \partial \pbf_*(S^b \restrict \{t < \tbf < t+h\})
  = S^b(t+h) - S^b(t).
\]
Dividing by $h$ and letting $h \to 0$, as well as a similar argument for $h < 0$, yields that
\[
  \Lcal_{v^b} S^b(t)
  = \partial(v^b \wedge S^b(t))
  = \frac{\di}{\di t} S^b(t).
\]
Hence,~\eqref{eq:GTE_Sbt} holds.

Let us finally observe how the Geometric Transport Equation evolves the mass of an advected current: For this, let $T$ be a smooth absolutely continuous boundaryless $1$-current ($\partial T = 0$) with weighted tangent field $\tau$, that is, $T = \tau \, \Lcal^3$, and let $v$ be an autonomous velocity field. Assume that $\tau$ is sufficiently smooth and satisfies $\abs{\tau(x)} > 0$ for every $x\in\R^3$, with sufficient decay at infinity to justify the following integrations by parts. Using the Geometric Transport Equation in coordinate form, see~\eqref{eq:GTE_1_coordinates}, we obtain for the \textbf{mass}
\[
  \Mbf(\Phi^t_* T) = \int \di \tv{\Phi^t_* T}
\]
of the current $T$ advected by the field $v$ that
\begin{equation} \label{eq:dMdt_GTE}
  \frac{\di}{\di t} \Mbf(\Phi^t_* T) \bigg|_{t=0}
  = \int \frac{\tau}{\abs{\tau}} \cdot \curl(v \times \tau) \dd x.
\end{equation}
If $T$ (and hence $\Phi^t_* T$) is smooth, the mass is just the $\Lrm^1$-norm and the above calculation is rigorous; if $T$ is not smooth it needs to be understood in a generalized sense.

On the other hand, the following expression is classical (see, e.g., Chapter~1 in~\cite{Tonegawa19book}):
\begin{equation} \label{eq:dMdt_Brakke}
  \frac{\di}{\di t} \Mbf(\Phi^t_* T) \bigg|_{t=0}
  = \int (\dive_\tau v) \, \abs{\tau} \dd x,
\end{equation}
with $\dive_\tau v$ the tangential divergence of $v$ over the tangent space given by $\tau$, defined as
\[
  \dive_\tau v 
  = \frac{\tau \cdot (\nabla v \, \tau)}{\abs{\tau}^2}
  = \frac{\tau^\T \nabla v \, \tau}{\abs{\tau}^2} = \frac{\tau \otimes \tau}{\abs{\tau}^2} : \nabla v.
\]
Note that here we have normalized $\tau$ first before defining $\dive_\tau v$. This is because we allow for non-unit density (or weight). It is also well known~\cite{Tonegawa19book} that one can rewrite~\eqref{eq:dMdt_Brakke} as
\[
  \frac{\di}{\di t} \Mbf(\Phi^t_* T) \bigg|_{t=0}
  = - \int h \cdot v \, \abs{\tau} \dd x,
\]
with $h$ the (generalized) \emph{mean curvature vector} (of the $1$-varifold associated with $T$, see~\cite{Tonegawa19book}), assuming that this exists.

Let us observe that (in the whole space) the two expressions~\eqref{eq:dMdt_Brakke},~\eqref{eq:dMdt_GTE} in fact agree: Starting from~\eqref{eq:dMdt_GTE} and using $\curl(v \times \tau) = \dive(v \otimes \tau) - \dive(\tau \otimes v)$ as well as 
\[
  [\dive(f \otimes g)]^i
  = \sum_j \partial_j (f_i g_j)
  = \sum_j \partial_j f_i \cdot g_j + \sum_j f_i \cdot \partial_j g_j
  = \bigl[ \nabla f \, g + f \, \dive g \bigr]^i,
\]
we compute (at least for smooth fields $v$ and somewhat formally):
\begin{align*}
  \frac{\di}{\di t} \Mbf(\Phi^t_* T)
  &= \int \frac{\tau}{\abs{\tau}} \cdot \curl(v \times \tau) \dd x \\
  &= \int \frac{\tau}{\abs{\tau}} \cdot \bigl[ \nabla v \, \tau + v \, \dive \tau - \nabla \tau \, v - \tau \, \dive v \bigr] \dd x \\
  &= \int \frac{\tau^\T \nabla v \, \tau}{\abs{\tau}} 
  		+ \frac{\tau \cdot v}{\abs{\tau}} \dive \tau 
  		- \frac{\tau^\T \nabla \tau \, v}{\abs{\tau}} 
  		- \abs{\tau} \dive v \dd x.
\end{align*}
Note that here, $\tau$ is implicitly evaluated at $t$, but this is omitted for ease of notation. The first term is precisely the expression in~\eqref{eq:dMdt_Brakke}, the second term is zero since $\tau$ is boundaryless (divergence-free) and for the remaining terms we observe using integration by parts, assuming that $\abs{\tau} \cdot \abs{v}$ is asymptotically vanishing, that
\[
  \int \abs{\tau} \dive v \dd x
  = - \int \frac{\tau^\T \nabla \tau \, v}{\abs{\tau}} \dd x.
\]
Hence, these terms cancel each other and we conclude that~\eqref{eq:dMdt_GTE} and~\eqref{eq:dMdt_Brakke} agree.

If we instead do the above computations on a fixed domain $\Omega$, we pick up a boundary term:
\[
  \int_\Omega \abs{\tau} \dive v \dd x
  = - \int_\Omega \frac{\tau^\T \nabla \tau \, v}{\abs{\tau}} \dd x
  - \int_{\partial \Omega} \abs{\tau} \, v \cdot n \dd \Hcal^2
\]
where $n$ is the inner normal to $\partial \Omega$. On the other hand, if in~\eqref{eq:dMdt_GTE} we consider the mass to remain conserved, we must also move the domain by $v$, so set $\Omega(t) := \Phi^t(\Omega)$. Recall the Reynolds transport theorem, that is, 
\[
  \frac{\di}{\di t} \int_{\Omega(t)} f \dd x
  = \int_{\Omega(t)} \frac{\partial f}{\partial t} \dd x - \int_{\partial \Omega(t)} (v \cdot n) \, f \dd \Hcal^2
\]
for any $f \colon \R \times \R^3 \to \R$. Then, omitting the time argument for $\tau$ as before,
\begin{align*}
  \frac{\di}{\di t} \Mbf_{\Omega(t)}(\tau)
  &= \int_{\Omega(t)} \frac{\di}{\di t} \abs{\tau} \dd x - \int_{\partial \Omega(t)} (v \cdot n) \, \abs{\tau} \dd \Hcal^2 \\
  &= \int_{\Omega(t)} \frac{\tau}{\abs{\tau}} \cdot \curl(v \times \tau) \dd x - \int_{\partial \Omega(t)} (v \cdot n) \, \abs{\tau} \dd \Hcal^2
\end{align*}
and the additional term precisely cancels the one above. In conclusion, the approach via the Geometric Transport Equation agrees with classical arguments whenever both make sense.

\section{PDE formulation of the small-strain model} \label{sc:PDE}

In this section we will rewrite the Small-Distortion Geometric Model for small strains in the form of a system of PDEs. This formulation applies to dislocation-density fields in the rate-dependent regime only, since the derivation uses classical derivatives.

\subsection{Field formulation}

To consider a PDE formulation in the small plastic strain regime, as derived in Section~\ref{sc:SDGM_linearization}, we first need to posit that all quantities are given by \emph{fields}. So, for a slip trajectory $S = \vec{S} \, \tv{S}$ in $[0,T] \times \Omega \subseteq \R^{1+3}$ with $\partial (S|_t) = 0$ for all times $t$, we assume that
\[
  \vec{S} = \xi \wedge \tau_1,  \qquad
  \tv{S} = m \, \Lcal^{1+3}
\]
with
\[
  \xi \in \R^{1+3}, \quad 
  \tau_1 \in \{0\} \times \R^3, \quad  
  \xi \cdot \tau_1 = 0, \quad
  \abs{\xi} = \abs{\tau_1} = 1, \quad
  \tbf\xi = \xi \cdot \ee_0 > 0,
\]
and \enquote{$\Lcal^{1+3}(\di x)$} is the volume (Lebesgue) measure on space-time $\R^{1+3}$, meaning that $\tv{S}$ is a Lebesgue-absolutely continuous density. Then,
\[
  \pbf \vec{S} = (v \wedge \tau_1) \cdot \tbf\xi,
\]
with $v$ the geometric derivative of $S$,
\[
  v := \frac{\DD}{\DD t}S := \frac{\pbf\xi}{\tbf\xi}.
\]
As before (see Section~\ref{sc:plastic_flow}), $\frac{\DD}{\DD t}S$ is the normal velocity of the moving dislocation. Setting
\[
  \tau := \tau_1 \cdot \tbf\xi \cdot m \in \{0\} \times \R^3 \sim \R^3,
\]
where $m$ is the multiplicity, we observe from the property that the slices are boundaryless, that is, $\partial (S|_t) = 0$ for all times $t$, via~\eqref{eq:partial_div} that
\[
  \dive \tau = 0  \qquad \text{in $\Omega$.}
\]
Here, and in all of the following, $\dive = \dive_x$ is the divergence in space only.

Thus, a field slip trajectory system $\Sb = (S^b)_b$ with respect to the set $b \in \Bcal$ of Burgers vectors as defined in Section~\ref{sc:slip_trajectories}, can be described by the variables
\[
  \taub = (\tau^b)_{b \in \Bcal}, \; \vb := (v^b)_{b \in \Bcal},  \qquad\text{where}\qquad
  \tau^b, v^b \colon [0,T] \times \Omega \to \R^3,
\]
satisfying
\[
  \tau^{-b} = -\tau^b, \qquad
  v^{-b} = v^b, \qquad
  v^b \cdot \tau^b = 0,
\]
as well as
\[
  \dive \tau^b = 0  \qquad \text{in $\Omega$.}
\]
Here, the symmetry relations $\tau^{-b} = -\tau^b$, $v^{-b} = v^b$ follow from $S^{-b} = -S^b$. Note also that the condition $\tbf\xi > 0$ is no longer required. Moreover, the evolution of the $\tau^b$ has been expressed via the Geometric Transport Equation in the previous section, see in particular~\eqref{eq:GTE_1_coordinates}, so that $\tau$ satisfies the line transport equation
\begin{equation} \label{eq:GTE_tau}
  \frac{\di}{\di t} \tau^b = \curl (v^b \times \tau^b).
\end{equation}

For the linearized kinematic variables we use
\[
  u \colon [0,T] \times \Omega \to \R^3, \qquad
  p \colon [0,T] \times \Omega \to \R^{3 \times 3}.
\]
The evolution of the (linearized) plastic distortion $p = P - \Id$ can be expressed in terms of $\tau^b,v^b$ as follows: Combine the plastic flow equation~\eqref{eq:plastic_flow} with~\eqref{eq:gb_cross_currents}, that is, the expression
\[
  g^b(t) = v^b \times \tau^b
\]
for the slip normal rate, to obtain
\begin{equation} \label{eq:pdot_linearized}
  \dot{p} = \frac12 \sum_{b \in \Bcal} b \otimes (v^b \times \tau^b).
\end{equation}
Recall also the plastic consistency condition~\eqref{eq:curlP_consistency}, which here reads as
\[
  \curl p = \frac12 \sum_{b \in \Bcal} b \otimes \tau^b.
\]
The arguments leading to~\eqref{eq:curlP_consistency} show that this relation is carried along the flow.

\begin{remark}
We can also derive the plastic consistency condition directly, without recourse to the geometric theory, as follows: Taking the (row-wise) curl on both sides and applying the line transport equation, we see that
\[
  \curl \dot{p} 
  = \frac12 \sum_{b \in \Bcal} b \otimes \curl (v^b \times \tau^b)
  = \frac12 \sum_{b \in \Bcal} b \otimes \frac{\di}{\di t} \tau^b,
\]
whereby
\[
  \frac{\di}{\di t} (\curl p) = \frac{\di}{\di t} \biggl( \frac12 \sum_{b \in \Bcal} b \otimes \tau^b \biggr).
\]
Integrating from $0$ to a time $t$, we obtain the plastic consistency condition, assuming it is satisfied at the initial time.
\end{remark}

Finally, for the total energy we set
\[
  \Ecal(t,u,p,\taub) := \overline{\Wcal}_\free(u,p) + \overline{\Wcal}_\linet(\taub) - \dprb{f, u},
\]
where (see~\eqref{eq:Wfree_linearized},~\eqref{eq:Wline_linearized})
\begin{align*}
  \overline{\Wcal}_\free(u,p)   &:= \int_\Omega \frac12 \abs{\bar{\fr}[\nabla u - p]}_\Ebb^2 \dd x, \\
  \overline{\Wcal}_\linet(\taub) &:= \frac12 \sum_{b \in \Bcal} \int_\Omega G^b(\tau^b) \dd x, \\
  \dprb{f(t), u} &:= \int_\Omega f \cdot u \dd x.
\end{align*}
Here, $\Ebb = \Ebb^{ijkl}$ is the (linearized) elasticity tensor as defined in Section~\ref{sc:We_linearize}, and $G^b \colon \R^3 \to [0,\infty)$ ($b \in \Bcal$) are the absolutely $1$-homogeneous line-tension densities with $G^{-b} = G^b$ (which are here assumed to be smooth away from the origin).

\subsection{Balance laws formulated via PDEs}

In the following formal PDE derivation, we assume that all fields are sufficiently smooth and that there exists a constant $c > 0$ such that
\[
  \abs{\tau^b(t,x)} \geq c
  \qquad \text{for all $(t,x) \in [0,T] \times \overline{\Omega}$, $b \in \Bcal$.}
\]
Then, all the following expressions involving $\abs{\tau^b}^{-1}$, as well as $\DD G^b(\tau^b)$ and $\curl \DD G^b(\tau^b)$, are well-defined throughout $\Omega$.

We now impose versions of the elastic equilibrium equation and the flow rule that are adapted to the present situation. So, we require the (linearized) elastic equilibrium equation
\[
  - \dive (\Ebb \bar{\fr}[\nabla u - p]) = f   \qquad \text{in $\Omega$,}
\]
together with boundary conditions. Here, we have specified $f$ on $\Omega$, as this is more natural in the linearized setting. In the same vein, and restricting to the rate-dependent setting only, we obtain the following flow rule:
\begin{equation} \label{eq:flow_for_PDE}
  -\DD_{\vb} \Ecal(t,u,p,\taub) = \DD_{\vb} \Rcal(\taub,\vb),
\end{equation}
where
\[
  \Rcal(\taub,\vb) := \frac12 \sum_{b \in \Bcal} \int_\Omega R^b(v^b, \tau^b) \dd x,
\]
with $R^b \colon \R^3 \times \R^3 \to [0,\infty)$ a dissipation potential that is continuously differentiable with respect to $v$, satisfies $R^b(0,\frarg) = 0$ and $R^{-b} = R^b$, and is absolutely $1$-homogeneous in $\tau$, that is, $R^b(v,\alpha \tau) = \abs{\alpha} R^b(v, \tau)$ for all $\alpha \in \R$. The reason for the latter condition is that the multiplicity of the dislocations must be taken into account linearly. For instance, twice as many moving dislocations must dissipate exactly twice as much energy.

Here, $\DD_{\vb} \Ecal(t,u,p,\taub)$ is to be understood as the change of $\Ecal$ with respect to moving the dislocations via $\vb$ as follows: Let $\hat{\vb} = (\hat{v}^b)_b$ be test dislocation velocities that have compact support inside $\Omega$ and satisfy $\hat{v}^{-b} = \hat{v}^b$ for all $b \in \Bcal$. Thus, for each pair $\{b,-b\}$, both components are varied simultaneously. Since $\tau^b$ is transported via~\eqref{eq:GTE_tau}, we have
\[
  \delta \tau^b = \curl(\hat{v}^b \times \tau^b),
  \qquad
  \delta \tau^{-b} = \curl(\hat{v}^b \times \tau^{-b})
  = -\curl(\hat{v}^b \times \tau^b).
\]
In the following, $\DD_{v^b}$ denotes this paired variation. In particular, the factor $\frac12$ in the definition of the plastic flow cancels against the two equal contributions:
\[
  \DD_{v^b} \dot p[\hat{v}^b]
  = \frac12 \bigl[b \otimes (\hat{v}^b \times \tau^b) + (-b) \otimes (\hat{v}^{-b} \times \tau^{-b})\bigr]
  = b \otimes (\hat{v}^b \times \tau^b).
\]

We will first compute the derivative for the free elastic energy $\overline{\Wcal}_\free$. For this, we first need to make the dependence of this functional on $v^b$ explicit. According to the derivation in Section~\ref{sc:SDGM_linearization}, we have for the total displacement $u \colon [0,T] \times \Omega \to \R^3$, in particular see~\eqref{eq:fr_e_linearized}, that
\[
  \bar{\fr}[\nabla u - p] = \nabla u - \Hcal p,  \qquad\text{where}\qquad
  \Hcal p := p - \beta
\]
with $\beta \colon \Omega \to \R^{3 \times 3}$ solving
\[
  \left\{ \begin{aligned}
    - \dive \Ebb \beta &= 0 &&\text{in $\Omega$,} \\
    \curl \beta &= \curl p  &&\text{in $\Omega$,} \\
    (\Ebb \beta)n &= 0        &&\text{on $\partial \Omega$.}
  \end{aligned} \right.
\]
From the linearity of the operator $\Hcal$ it follows that
\[
  \DD_p (\Hcal p)[\hat{p}] = \Hcal \hat{p}
\]
for any $\hat{p} \colon \Omega \to \R^{3 \times 3}$.

Holding the total displacement $u$ constant, we observe that
\[
  \DD_p (\bar{\fr}[\nabla u - p])[\hat{p}] = - \DD_p (\Hcal p)[\hat{p}] = - \Hcal \hat{p}.
\]
Furthermore, by~\eqref{eq:pdot_linearized},
\[
  \DD_{v^b} (\bar{\fr}[\nabla u - p])[\hat{v}^b]
  = \DD_p (\bar{\fr}[\nabla u - p]) \bigl[ b \otimes (\hat{v}^b \times \tau^b) \bigr]
  = - \Hcal \bigl[ b \otimes (\hat{v}^b \times \tau^b) \bigr].
\]
Thus,
\begin{align*}
  \DD_{v^b} \overline{\Wcal}_\free(u,p)[\hat{v}^b]
  &= \int_\Omega \Ebb(\bar{\fr}[\nabla u - p]) : \DD_{v^b} (\bar{\fr}[\nabla u - p])[\hat{v}^b] \dd x \\
  &= - \int_\Omega \Ebb(\bar{\fr}[\nabla u - p]) : \Hcal \bigl[ b \otimes (\hat{v}^b \times \tau^b) \bigr] \dd x \\
  &= - \int_\Omega \bigl(b^\T \Hcal^* \bigl[\Ebb(\bar{\fr}[\nabla u - p])] \bigr)^\T \cdot (\hat{v}^b \times \tau^b) \dd x \\
  &= - \int_\Omega \bigl( \tau^b \times \bigl(b^\T \Hcal^* \bigl[\Ebb(\bar{\fr}[\nabla u - p])] \bigr)^\T \bigr) \cdot \hat{v}^b \dd x,
\end{align*}
where $\Hcal^*$ denotes the adjoint operator of $\Hcal$ and in the last line we have used the elementary equality $z \cdot (v \times \tau) = (\tau \times z) \cdot v$.

For a test velocity $\hat{v}^b$, we compute the variation of the linearized line-tension energy as follows:
\begin{align*}
  \DD_{v^b} \overline{\Wcal}_\linet(\taub)[\hat{v}^b]
  &= \DD_{\tau^b} \overline{\Wcal}_\linet(\taub) \bigl[ \curl(\hat{v}^b \times \tau^b) \bigr] \\
  &= \int_\Omega \DD G^b(\tau^b) \cdot \curl(\hat{v}^b \times \tau^b) \dd x \\
  &= \int_\Omega \curl \DD G^b(\tau^b) \cdot (\hat{v}^b \times \tau^b) \dd x \\
  &= \int_\Omega \bigl( \tau^b \times \curl \DD G^b(\tau^b) \bigr) \cdot \hat{v}^b \dd x.
\end{align*}
Here we have used the self-adjointness of the curl operator (and that $\hat{v}^b$ has compact support). The factor $\frac12$ in $\overline{\Wcal}_\linet$ cancels because the paired variations of the $b$- and $-b$-components give equal contributions (note that we assume $G^{-b} = G^b$ and $G^b(\alpha \tau) = \abs{\alpha}G^b(\tau)$ for all $\alpha \in \R$).

For the dissipation we likewise have
\[
  \DD_{v^b} \Rcal(\taub,\vb)[\hat{v}^b] = \int_\Omega \DD_v R^b(v^b, \tau^b) \cdot \hat{v}^b \dd x,
\]
where again the factor $\frac12$ cancels because both components of the pair are varied.

Hence, combining the variations of the free energy, the line-tension energy, and the dissipation in~\eqref{eq:flow_for_PDE}, we obtain
\[
  \int_\Omega \Bigl\{ \DD_v R^b(v^b, \tau^b) + \tau^b \times \Bigl[\curl \DD G^b(\tau^b) - \bigl(b^\T \Hcal^*\bigl[\Ebb(\bar{\fr}[\nabla u-p])\bigr]\bigr)^\T \Bigr] \Bigr\} \cdot \hat{v}^b \dd x = 0.
\]
Then, since $\hat{v}^b$ was arbitrary, we arrive at the PDE-formulation of the flow rule
\[
  \DD_v R^b(v^b, \tau^b) + \tau^b \times \Bigl[\curl \DD G^b(\tau^b) - \bigl(b^\T \Hcal^*\bigl[\Ebb(\bar{\fr}[\nabla u-p])\bigr]\bigr)^\T \Bigr] = 0 \qquad \text{in $\Omega$,}
\]
for every $b \in \Bcal$.

\begin{example}
For instance, one could use $R^b(v^b, \tau^b) := \bar{r} \abs{v^b}^q \abs{\tau^b}$, corresponding to a rate-dependent dissipation if $q > 1$. Note that $\abs{v^b \times \tau^b} = \abs{v^b} \abs{\tau^b}$ since $v^b \cdot \tau^b = 0$, so that we could write the above $R^b$ also as depending on the speed $\abs{v^b}$ and the (weighted) slip surface normal $v^b \times \tau^b$ instead, which is more in line with the physical understanding of dislocation transport.
\end{example}

In conclusion, we have reformulated the linearized Small-Distortion Geometric Model in the variables
\begin{align*}
  u &\colon [0,T] \times \Omega \to \R^3, \\
  p &\colon [0,T] \times \Omega \to \R^{3 \times 3}, \\
  \vb &= (v^b)_{b \in \Bcal} \quad\text{with}\quad v^b \colon [0,T] \times \Omega \to \R^3, \\
  \taub &= (\tau^b)_{b \in \Bcal} \quad\text{with}\quad \tau^b \colon [0,T] \times \Omega \to \R^3,
\end{align*}
as follows:
\[
\left\{
  \begin{aligned}
    &\text{\textbf{(C) Consistency:}} \\
    &\quad\qquad  v^{-b} = v^b, \qquad \tau^{-b} = -\tau^b, \qquad v^b \cdot \tau^b = 0, \qquad \dive \tau^b = 0 \\[8pt]
    &\text{\textbf{(B) Elastic balance:}} \\
    &\quad\qquad  - \dive \Ebb(\bar{\fr}[\nabla u - p]) = f \\[8pt]
   	&\text{\textbf{(V) Velocity equations:} For $b \in \Bcal$,}\\
   	&\quad\qquad \DD_v R^b(v^b, \tau^b) + \tau^b \times \bigl[ \curl \DD G^b(\tau^b) - \bigl(b^\T \Hcal^* \bigl[\Ebb(\bar{\fr}[\nabla u-p])\bigr]\bigr)^\T \bigr] = 0 \\[8pt]
    &\text{\textbf{(P) Plastic flow:}} \\
    &\quad\qquad  \frac{\di}{\di t} \tau^b = \curl (v^b \times \tau^b),  \qquad
                  \frac{\di}{\di t} p = \frac12 \sum_{b \in \Bcal} b \otimes (v^b \times \tau^b)
  \end{aligned}
  \right.
\]
Note that there is also an additional condition, $\curl p = \frac12 \sum_{b \in \Bcal} b \otimes \tau^b$, that is automatically preserved along the flow.

\begin{remark}
It holds that
\[
  \left\{\begin{aligned}
  - \dive \Ebb \Hcal p &= - \dive \Ebb p, \\
  \curl \Hcal p &= 0
  \end{aligned}\right.
\]
by construction of $\Hcal$, so in particular $\Hcal p$ is a gradient itself if $\Omega$ is simply connected. Moreover, $\Hcal$ maps $\Wrm^{1,1}$ (or $\BV$) to $\Lrm^{3/2}$ by~\cite[Proposition~4.2]{ContiGarroniOrtiz15} (based on~\cite{BourgainBrezis04}) and the Gagliardo--Nirenberg embedding. In particular, one may not assume $\Hcal p$ to be square-integrable, which requires careful arguments in the rigorous theory~\cite{BonicattoRindlerTurnbull26c?}.
\end{remark}

\subsection{Comparison to Field Dislocation Mechanics}

Consider the following simplifying assumptions:
\[
  G^b(\tau) = \abs{\tau},  \qquad
  R^b(v,\tau) = \frac12 \abs{v}^2 \abs{\tau}.
\]
This means that the line-tension energy equals the Euclidean length, and the dissipation is the easiest possible rate-dependent one. Then, our equations simplify to the following:
\[
\left\{
  \begin{aligned}
    &\text{\textbf{(C) Consistency:}} \\
    &\quad\qquad  v^{-b} = v^b, \qquad \tau^{-b} = -\tau^b, \qquad v^b \cdot \tau^b = 0, \qquad \dive \tau^b = 0 \\[8pt]
    &\text{\textbf{(B) Elastic balance:}} \\
    &\quad\qquad  - \dive \Ebb(\bar{\fr}[\nabla u - p]) = f \\[8pt]
  	&\text{\textbf{(V) Velocity equations:} For $b \in \Bcal$,}\\
  	&\quad\qquad v^b \abs{\tau^b} = - \tau^b \times \bigl[ \curl (\abs{\tau^b}^{-1} \tau^b) - \bigl(b^\T \Hcal^* \bigl[\Ebb(\bar{\fr}[\nabla u - p])] \bigr)^\T \bigr] \\[8pt]
    &\text{\textbf{(P) Plastic flow:}} \\
    &\quad\qquad  \frac{\di}{\di t} \tau^b = \curl (v^b \times \tau^b),  \qquad
                  \frac{\di}{\di t} p = \frac12 \sum_{b \in \Bcal} b \otimes (v^b \times \tau^b)
  \end{aligned}
  \right.
\]
Note that $v^b \cdot \tau^b = 0$ is automatically satisfied by the velocity equation. Moreover, care needs to be taken with the term $\tau^b \times \curl (\abs{\tau^b}^{-1} \tau^b)$ when defining this rigorously to make sure this does not degenerate for $\tau^b \to 0$ (which we excluded here), either by a regularization or by a subdifferential interpretation.

We want to compare our model to Acharya's Field Dislocation Mechanics (FDM) model that was introduced in~\cite{Acharya01,Acharya03,Acharya04} (also see~\cite{AcharyaTartar11,AroraAcharya20} and the references cited therein). We compare the present model to the small-strain version reproduced in~\cite{AcharyaTartar11}. We only carry out this comparison for the model with the above simplifying assumptions $G^b(\tau) = \abs{\tau}$ and $R^b(v,\tau) = \frac12 \abs{v}^2 \abs{\tau}$ in order to keep the notation manageable.

Let us consider first the case of only one Burgers vector (that is, one active slip system), so assume that
\[
  \Bcal = \{\pm b\}.
\]
Simplifying notation $\tau^b$ to $\tau$ and $v^b$ to $v$, we then have the following system:
\[
\left\{
  \begin{aligned}
    &\text{\textbf{(C) Consistency:}} \\
    &\quad\qquad  \dive \tau = 0 \\[8pt]
    &\text{\textbf{(B) Elastic balance:}} \\
    &\quad\qquad  - \dive \Ebb(\bar{\fr}[\nabla u - p]) = f \\[8pt]
  	&\text{\textbf{(V) Velocity equation:}}\\
  	&\quad\qquad v \abs{\tau} = -\tau \times \bigl[ \curl (\abs{\tau}^{-1} \tau) - \bigl(b^\T \Hcal^* \bigl[\Ebb(\bar{\fr}[\nabla u - p])] \bigr)^\T \bigr] \\[8pt]
    &\text{\textbf{(P) Plastic flow:}} \\
    &\quad\qquad  \frac{\di}{\di t} \tau = \curl (v \times \tau),  \qquad
                  \frac{\di}{\di t} p = b \otimes (v \times \tau)
  \end{aligned}
  \right.
\]

In this setting, the small-strain FDM considered here reads as follows: The variables
\begin{align*}
  u &\colon [0,T] \times \Omega \to \R^3, \\
  \alpha &\colon [0,T] \times \Omega \to \R^{3 \times 3}, \\
  \chi &\colon [0,T] \times \Omega \to \R^{3 \times 3}, \\
  V &\colon [0,T] \times \Omega \to \R^3, \\
  \nabla z &\colon [0,T] \times \Omega \to \R^{3 \times 3}
\end{align*}
are specified to satisfy the following equations (in our sign convention for Burgers vectors, see Remark~\ref{rem:sign}):
\begin{enumerate}[(1)]
  \item $\curl \chi = -\alpha$,
  \item $\dive \chi = 0$,
  \item $\dive (\nabla \dot{z}) = -\dive (\alpha \times V)$,
  \item $\dive T[\nabla u - \nabla z + \chi] = 0$,
  \item $\dot{\alpha} = - \curl (\alpha \times V)$,
  \item $V = X[-T + (\curl \alpha)^\T, \alpha]$.
\end{enumerate}
Here $T$ is the stress, that is, the derivative of the elastic energy density, and
\[
  X[A,B] := \bigl[ \eps_{ijk} A^j_r B^r_k \bigr]^i.
\]

We now translate our quantities into the FDM variables. Set
\begin{align*}
  \alpha &:= b \otimes \tau, \\
  V &:= v, \\
  \nabla z &:= p + \chi,
\end{align*}
and $\chi$ the solution to 
\begin{equation} \label{eq:FDM_chi}
  \left\{ \begin{aligned}
    \dive \chi &= 0         &&\text{in $\Omega$,} \\
    \curl \chi &= -\curl p  &&\text{in $\Omega$,} \\
    \chi n &= 0       &&\text{on $\partial \Omega$.}
  \end{aligned} \right.
\end{equation}
Note that $\chi$ does not solve the same system as $\beta$, the difference being that for $\chi$ we use the identity elasticity tensor. Then, applying~(C)--(P),
\begin{enumerate}[(1)]\setlength{\itemsep}{6pt}
  \item $\curl \chi = -\curl p = -b \otimes \tau = -\alpha$.
  \item $\dive \chi = 0$.
  \item $\dive (\nabla \dot{z}) = \dive \dot{p} + \frac{\di}{\di t} [\dive \chi] = \dive [ b \otimes (v \times \tau)] = - \dive [ b \otimes (\tau \times v)] = -\dive (\alpha \times V)$.
  \item $\dive T[\nabla u - \nabla z + \chi] = \dive T[\nabla u - p] = 0$.
  \item $\dot{\alpha} = b \otimes \dot{\tau} = \curl [b \otimes (v \times \tau)] = -\curl [b \otimes (\tau \times V)] = - \curl (\alpha \times V)$.
\end{enumerate}
For~(6) we compute:
\begin{align*}
  \bigl(X[-T,\alpha]\bigr)^i
  &= -\eps_{ijk}T^j_r b^r\tau^k \\
  &= -\eps_{ijk}(Tb)^j\tau^k \\
  &= \bigl(-(Tb)\times\tau\bigr)^i \\
  &= \bigl(\tau\times(b^\T T^\T)^\T\bigr)^i.
\end{align*}
Moreover, since $\alpha = b\otimes\tau$, we have $\curl\alpha=b\otimes\curl\tau$. Consequently,
\begin{align*}
  \bigl(X[(\curl\alpha)^\T,\alpha]\bigr)^i
  &= \bigl(X[(\curl\tau)\otimes b,b\otimes\tau]\bigr)^i \\
  &= \eps_{ijk}(\curl\tau)^j b_r b^r\tau^k \\
  &= \abs{b}^2\bigl((\curl\tau)\times\tau\bigr)^i \\
  &= -\abs{b}^2\bigl(\tau\times(\curl\tau)\bigr)^i.
\end{align*}
Thus, under the constitutive specialization considered here,
\begin{align*}
  V &= \tau\times(b^\T T^\T)^\T -\abs{b}^2\tau\times(\curl\tau) \\
  &= -\tau\times\bigl\{\abs{b}^2\curl\tau-(b^\T T^\T)^\T \bigr\}.
\end{align*}
This is almost the same as our velocity equation~(V), that is,
\[
  v = -\abs{\tau}^{-1}\tau \times\Bigl[ \curl(\abs{\tau}^{-1}\tau) - \bigl(b^\T\Hcal^*\bigl[\Ebb(\bar{\fr}[\nabla u-p])\bigr]\bigr)^\T \Bigr],
\]
interpreted suitably. We conclude that Field Dislocation Mechanics has a similar structure as the PDE version of the linearized Small-Distortion Geometric Model. There are, however, some differences:

\begin{enumerate}[(a)]
  \item In the particular small-strain constitutive specialization considered here, the FDM velocity law contains the term $\tau \times (\curl \tau)$, which scales quadratically under $\tau \mapsto m\tau$, whereas in our system we have the mean-curvature type term $\abs{\tau}^{-1} \tau \times \curl (\abs{\tau}^{-1} \tau)$. The normalization reflects our modeling convention that multiplicity represents superposed, infinitesimally close dislocations whose individual velocity is independent of that multiplicity. This differs from the constitutive dependence on dislocation density in the FDM specialization considered above. Indeed, our velocity equation gives a velocity that is \emph{independent} of the dislocation multiplicity.
  \item The velocity equation~(V) uses the nonlocal energy term $b^\T \Hcal^*[\Ebb(\nabla u - p + \beta)]$, whereas FDM has the term $b^\T T^\T$. The two models also produce different elastic driving forces: Our expression contains the nonlocal operator $\Hcal^*$ arising from the free-distortion construction, whereas the FDM specialization couples the dislocation velocity directly to the stress. This reflects the different kinematic and energetic formulations of the two theories. On the other hand, the Small-Distortion Geometric Model imposes the Line-Tension Approximation, which treats lines as infinitesimally small geometric objects.
\end{enumerate}

The situation is quite different when considering multiple Burgers vectors. In this case, we set, analogously to before,
\begin{align*}
  \alpha &:= \frac12 \sum_{b \in \Bcal} b \otimes \tau^b, \\
  \nabla z &:= p + \chi,
\end{align*}
and $\chi$ from~\eqref{eq:FDM_chi}. However, for independently moving dislocations, there is in general no single field $V$ satisfying both
\[
  \dot{\alpha} = - \curl (\alpha \times V)  \qquad\text{and}\qquad
  \alpha \times V = \frac12 \sum_b b \otimes (\tau^b \times v^b).
\]
Indeed, 
\begin{align*}
  \dot{\alpha}
  &= \frac12 \sum_{b \in \Bcal} b \otimes \dot{\tau}^b \\
  &= \frac12 \sum_{b \in \Bcal} \curl [b \otimes (v^b \times \tau^b)] \\
  &= - \curl \biggl[\frac12 \sum_{b \in \Bcal} b \otimes (\tau^b \times v^b)\biggr].
\end{align*}
In general, however, the tensor
\[
  \frac12 \sum_{b\in\Bcal} b\otimes(\tau^b\times v^b)
\]
cannot be factorized pointwise as $\alpha\times V$ for a single velocity field $V$. This reflects the fact that dislocations with different Burgers vectors can move independently, whereas $V$ represents a single velocity field for $\alpha$.

\begin{remark} \label{rem:commutators}
The preceding discussion is also the reason for the appearance of \enquote{commutators} when passing to a meso- or macroscale, see the Appendix to~\cite{AroraAcharya20}. In particular, the commutator quantity
\[
  L^p := \overline{\alpha \times V} - \overline{\alpha} \times \overline{V},
\]
with the bar denoting averaging, defined in eq.~(30) of \emph{loc.~cit.}, is in general non-zero when passing from the micro- to the mesoscale. In the Small-Distortion Geometric Model, the oriented space-time area, and hence the corresponding plastic-flow contribution, is retained directly under the chosen convergence in the linear space of wedge products (recall that $\alpha \times V = \hodge(\alpha \wedge V)$). This provides a structural distinction between the two formulations, namely that here the plastic-flow contribution is retained directly rather than reconstructed from separately averaged fields.
\end{remark}

\begin{example} \label{ex:wedge_better}
To illustrate the point of the preceding remark further, consider the lines
\[
  \ell_{N,k} := \R \ee_2 + \frac{k}{N} \ee_1,
  \qquad k=1,\ldots,N,
\]
and the dislocation current
\[
  T_N := \frac{1}{N} \sum_{k=1}^N (-1)^k \dbr{\ell_{N,k}}.
\]
Thus, $T_N$ is a collection of $N$ lines with weight $1/N$, all of which are translates of the line $\dbr{\R \ee_2}$, which goes through the origin and has direction $\ee_2$, but with alternating orientations. We assume that all lines have the same Burgers vector $b$ and set
\[
  \alpha_N := b \otimes T_N.
\]
Suppose that the lines with positive direction (that is, in direction $\ee_2$) move with velocity $\ee_3$ and those with negative direction move in direction $-\ee_3$. More precisely, the tangent and velocity of the $k$th line are
\[
  \tau_{N,k} := (-1)^k \ee_2,
  \qquad
  v_{N,k} := (-1)^k \ee_3.
\]
Now, in the limit $N \to \infty$, the lines cancel each other due to their alternating orientation, so that $\overline{\alpha}=0$; likewise one may see that $\overline V=0$. This is, however, undesirable since
\[
  v_{N,k} \times \tau_{N,k}
  = (-1)^{2k} \ee_3 \times \ee_2
  = -\ee_1
\]
does \emph{not} change sign between the lines and hence the associated plastic effects of all the line motions are \emph{additive} (and do \emph{not} cancel). In the Space-Time Framework, the relevant plastic-flow contribution is retained directly under the chosen convergence: The area elements
\[
  v_{N,k} \wedge \tau_{N,k}
  = (-1)^{2k} \ee_3 \wedge \ee_2
  = \ee_3 \wedge \ee_2
\]
are the same for \emph{all} lines. In particular,
\[
  \hodge(v_{N,k} \wedge \tau_{N,k})
  = v_{N,k} \times \tau_{N,k}
  = -\ee_1.
\]
Thus, the corresponding plastic-flow contributions are all equal to $-b \otimes \ee_1$ and therefore add rather than cancel. Hence, no commutator appears.
\end{example}

A precise discrete-to-field homogenization result for the Small-Distortion Geometric Model in the small-strain regime is to be found in~\cite{BonicattoRindlerTurnbull26c?}.

\bibliography{SmallDistortionPlast.bib}
\bibliographystyle{amsplain}


\end{document}